\documentclass[lettersize,journal]{IEEEtran}
\usepackage{amsmath,amsfonts,amsthm,amssymb}
\usepackage[ruled,vlined,linesnumbered]{algorithm2e}
\usepackage{array}
\usepackage[caption=false,font=normalsize,labelfont=sf,textfont=sf]{subfig}
\usepackage{textcomp}
\usepackage{stfloats}
\usepackage{url}
\usepackage{verbatim}
\usepackage{graphicx}
\usepackage{cite}
\usepackage{hyperref}
\usepackage{enumitem}
\usepackage{adjustbox}

\usepackage{centernot}
\usepackage{comment}
\usepackage{multirow}
\usepackage{makecell}
\usepackage{xcolor,soul}

\newcommand{\julien}[1]{#1}

\newcommand{\revision}[1]{#1}

\graphicspath{{figures/}}

\newcommand{\highDim}{h}
\newcommand{\lowDim}{\ell}
\newcommand{\homologyDim}{k}

\newcommand{\inputPointCloud}{X}
\newcommand{\latentPointCloud}{Z}
\newcommand{\projection}{\phi}

\newcommand{\diameterThreshold}{r}

\newcommand{\emst}{\mathsf{MST}}
\newcommand{\rng}{\mathsf{RNG}}
\newcommand{\ug}{\mathsf{UG}}
\newcommand{\del}{\mathsf{DEL}}
\newcommand{\mml}{\mathsf{MML}}
\newcommand{\critical}[1]{\pi^{#1}}

\newcommand{\bbn}{\mathbb{N}}
\newcommand{\bbr}{\mathbb{R}}
\newcommand{\bbk}{\mathbb{K}}
\newcommand{\bbz}{\mathbb{Z}}

\newcommand{\calb}{\mathcal{B}}
\newcommand{\calc}{\mathcal{C}}
\newcommand{\cald}{\mathcal{D}}
\newcommand{\calh}{\mathcal{H}}
\newcommand{\calk}{\mathcal{K}}
\newcommand{\call}{\mathcal{L}}
\newcommand{\calo}{\mathcal{O}}
\newcommand{\calp}{\mathcal{P}}

\newcommand{\calw}{\mathcal{W}}
\newcommand{\calx}{\mathcal{X}}
\newcommand{\calz}{\mathcal{Z}}

\newcommand{\dist}[2]{\lVert{#1}-{#2}\rVert_2}

\DeclareMathOperator{\dr}{Del}
\DeclareMathOperator{\lens}{Lens}
\DeclareMathOperator{\llens}{LeftLens}
\DeclareMathOperator{\rlens}{RightLens}
\DeclareMathOperator{\rank}{rank}
\DeclareMathOperator{\rips}{Rips}
\DeclareMathOperator{\enc}{\mathrm{enc}}
\DeclareMathOperator{\dec}{\mathrm{dec}}

\DeclareMathOperator{\skelcasc}{\mathsf{GCS}}

\DeclareMathOperator{\TA}{TA}
\DeclareMathOperator{\LC}{LC}
\DeclareMathOperator{\Cont}{Cont}
\DeclareMathOperator{\Trust}{Trust}

\DeclareMathOperator{\RMSE}{RMSE}

\DeclareMathOperator{\metwasser}{PD\calw^1} 
\DeclareMathOperator{\metwasserzero}{PD\calw^0}
\DeclareMathOperator{\metdistor}{\mathfrak{D}}
\newcommand{\timing}{t\:(s)}

\newcommand{\dgm}[1]{\cald^{#1}}
\newcommand{\dgmrips}[1]{\cald_{\rips}^{#1}}

\newcommand{\augmented}[1]{\overline{#1}}

\newcommand{\tAE}{TopoAE}
\newcommand{\ltopoae}[1]{\mathcal{L}_{\mathrm{TAE}}^{#1}}
\newcommand{\lcascae}[1]{\mathcal{L}_{\mathrm{CD}}^{#1}}

\DeclareMathOperator{\Ima}{Im}
\newcommand{\chain}[1]{\calc_{#1}(\calk)}
\newcommand{\cycle}[1]{\calz_{#1}(\calk)}
\newcommand{\bound}[1]{\calb_{#1}(\calk)}
\newcommand{\homol}[1]{\calh_{#1}(\calk)}
\newcommand{\betti}[1]{\beta_{#1}}
\newcommand{\filt}[1]{\calk_{#1}}
\newcommand{\HG}[1]{\calh_{\homologyDim}(\filt{#1})}
\newcommand{\PHG}[2]{\calh_{\homologyDim}^{{#1},{#2}}(\bbk)}
\newcommand{\morphism}[2]{f_{\homologyDim}^{{#1},{#2}}}
\newcommand{\PHC}{\gamma}
\newcommand{\skeleton}[2]{{#1}^{({#2})}}
\newcommand{\simplicesofdim}[2]{{#1}^{{#2}}}

\newcommand{\PH}[1]{\ensuremath{\text{PH}^{#1}}}

\DeclareMathOperator{\cascade}{\mathrm{cascade}}
\DeclareMathOperator{\partner}{\mathrm{partner}}

\newcommand{\ripsKiller}{e_{\mml}}
\newcommand{\DRKiller}{e_{\dr}}
\newcommand{\edgeMML}{e_{\mml}}
\newcommand{\deathValue}{\delta_{\mml}}
\newcommand{\circumcircle}{\calc}

\newtheorem{lemma}{Lemma}

\newcommand{\carriereMethod}{\ensuremath{\text{\tAE+}\calw^1}}

\newcommand{\datathreeblobs}{\texttt{3Clusters}}
\newcommand{\datatwist}{\texttt{Twist}}
\newcommand{\datakfour}{\texttt{K\_4}}
\newcommand{\datakfive}{\texttt{K\_5}}
\newcommand{\datacoil}{\texttt{COIL-20-1}}
\newcommand{\datamocap}{\texttt{MoCap}}
\newcommand{\datasinglecell}{\texttt{SingleCell}}

\newcommand{\methodText}[2]{\rotatebox{90}{#1}\mbox{#2}}
\newcommand{\datasetText}[1]{\rotatebox{90}{#1}}

\begin{document}

\title{Topological Autoencoders++:
Fast and Accurate Cycle-Aware Dimensionality Reduction}

\author{Mattéo Clémot, Julie Digne, Julien Tierny
\thanks{Mattéo Clémot is with University Claude Bernard Lyon 1. E-mail: matteo.clemot@univ-lyon1.fr}
\thanks{Julie Digne is with the CNRS and University Claude Bernard Lyon 1. E-mail: julie.digne@cnrs.fr}
\thanks{Julien Tierny is with the CNRS and Sorbonne University. E-mail: julien.tierny@sorbonne-universite.fr}
}

\maketitle

\begin{abstract}
This paper presents
a novel topology-aware dimensionality
reduction approach aiming at accurately visualizing the cyclic
patterns present in high dimensional
data.
To that end, we build on the \emph{Topological Autoencoders} (TopoAE) \cite{moor2020topological} formulation.
First, we provide a novel theoretical analysis
of its associated loss and show that a zero loss indeed
induces identical persistence pairs (in high and low dimensions)
for the $0$-dimensional persistent homology (\PH{0}) of the Rips
filtration.
We also provide a counter example showing that this property no longer holds
for a naive extension of TopoAE to
\PH{d} for $d\ge 1$. Based on this observation, we introduce a novel generalization
of TopoAE to $1$-dimensional persistent homology
(\PH{1}), called TopoAE++, for the accurate
generation of cycle-aware planar embeddings, addressing the above failure case.
This generalization is based on
the notion of \emph{cascade  distortion}, a new penalty term favoring an isometric embedding of the $2$-chains filling persistent $1$-cycles, hence resulting in more faithful geometrical reconstructions of the $1$-cycles in the plane. 
We further
introduce
a novel, fast algorithm for the exact computation of
\PH{} for Rips filtrations in the plane, yielding
improved runtimes
over previously documented topology-aware methods. Our method also achieves a
better balance between the topological accuracy, as measured by the Wasserstein distance, and the visual preservation of
the cycles in low dimensions.
Our C++ implementation is available at
\url{https://github.com/MClemot/TopologicalAutoencodersPlusPlus}.


\end{abstract}


\begin{figure*}
	\adjustbox{width=1.013\linewidth,center}{
		\scriptsize{
		\begin{tabular}{|p{0cm}r||rr|rrr|rrr||r|}
			\hline
			
			\multicolumn{2}{|c||}{\multirow{2}{*}{Input}}&
			\multicolumn{2}{c|}{Global methods}&
			\multicolumn{3}{c|}{Locally topology-aware methods}&
			\multicolumn{4}{c|}{Globally topology-aware methods}\\
			
			\cline{3-11}
			\multicolumn{2}{|c||}{}
			&\multicolumn{1}{c}{PCA\cite{pearson1901liii}}
			&\multicolumn{1}{c|}{MDS\cite{torgerson1952multidimensional}}
			&\multicolumn{1}{c}{Isomap\cite{tenenbaum_global_2000}}
			&\multicolumn{1}{c}{t-SNE\cite{van2008visualizing}}
			&\multicolumn{1}{c|}{UMAP\cite{mcinnes2018umap}}
			&\multicolumn{1}{c}{TopoMap\cite{doraiswamy2020topomap}}
			&\multicolumn{1}{c}{\tAE\cite{moor2020topological}}
			&\multicolumn{1}{c||}{\carriereMethod\cite{carriere2021optimizing}}
			&\multicolumn{1}{c|}{\tAE++}\\
			
			\hline \raisebox{1.55cm}{\multirow{4}{*}{\rotatebox{90}{\datathreeblobs{} $(n=800)$}}} &
			\raisebox{-.5mm}{\includegraphics[width=.1\linewidth]{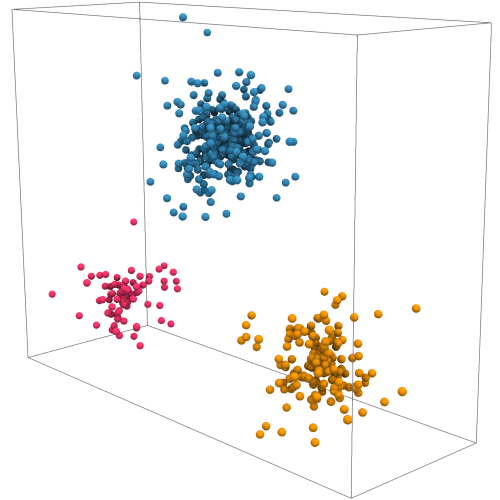}}&
			\includegraphics[width=.1\linewidth]{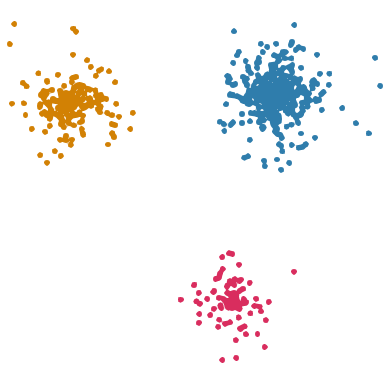}&
			\includegraphics[width=.1\linewidth]{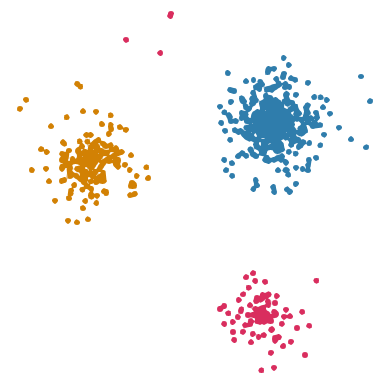}&
			\includegraphics[width=.1\linewidth]{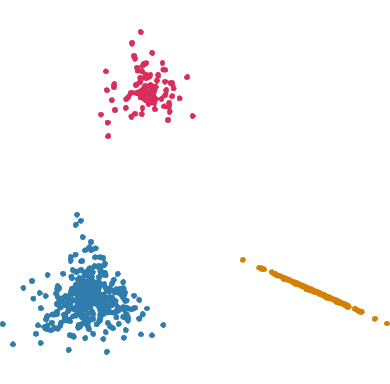}&
			\includegraphics[width=.1\linewidth]{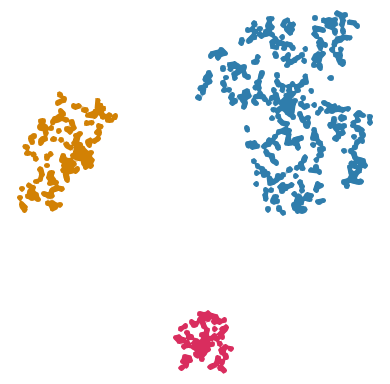}&
			\includegraphics[width=.1\linewidth]{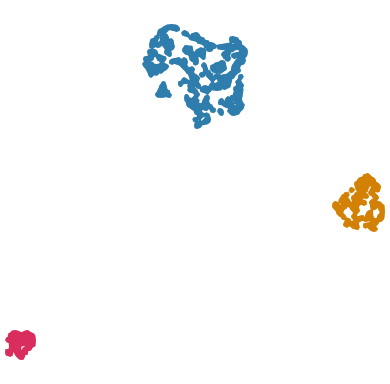}&
			\includegraphics[width=.1\linewidth]{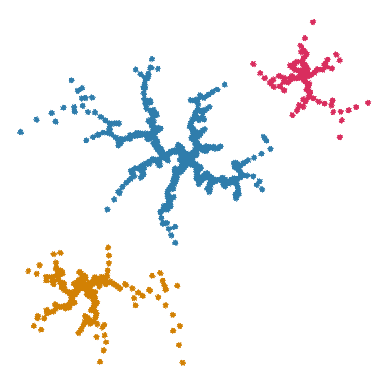}&
			\includegraphics[width=.1\linewidth]{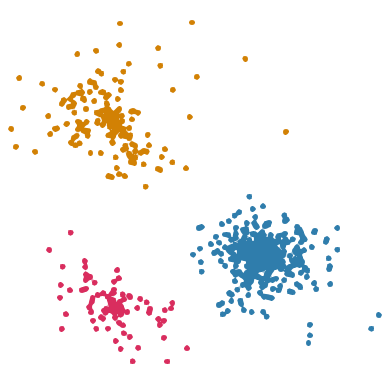}&
			\includegraphics[width=.1\linewidth]{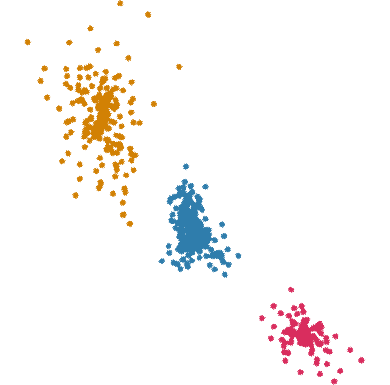}&
			\includegraphics[width=.1\linewidth]{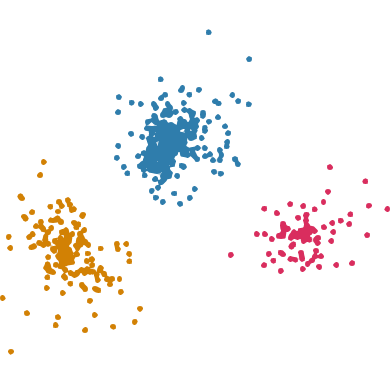}\\

			& $\metwasser(\inputPointCloud,\latentPointCloud)$ & 4.5e-01 & 5.0e-01 & 4.7e-01 & 2.3e+01 & 7.2e-01 & 4.3e+00 & 5.8e-01 & \textbf{1.1e-01} & \underline{4.4e-01}\\
			& $\metdistor(\inputPointCloud,\latentPointCloud)$ & \textbf{2.7e-01} & \underline{4.1e-01} & 1.1e+00 & 3.5e+01 & 8.5e+00 & 7.0e+00 & 1.3e+00 & 1.5e+00 & 6.4e-01\\
			& $\timing$ & 0.00049 & 2.5 & 0.16 & 0.76 & 2.6 & 0.086 & 81 & 830 & 5.0\\
			
			\hline \raisebox{1.25cm}{\multirow{4}{*}{\rotatebox{90}{\datatwist{} $(n=100)$}}} &
			\raisebox{-.5mm}{\includegraphics[width=.1\linewidth]{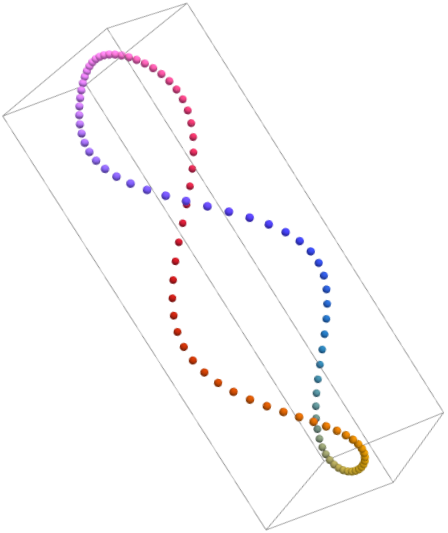}}&
			\includegraphics[width=.1\linewidth]{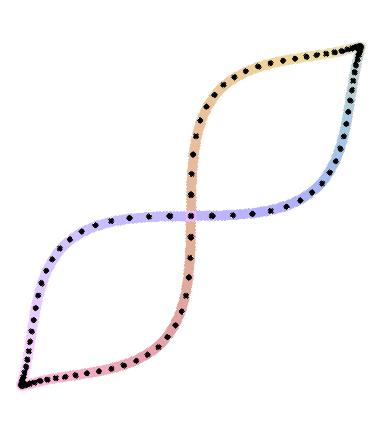}&
			\includegraphics[width=.1\linewidth]{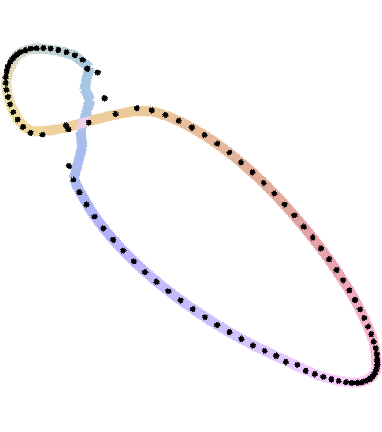}&
			\includegraphics[width=.1\linewidth]{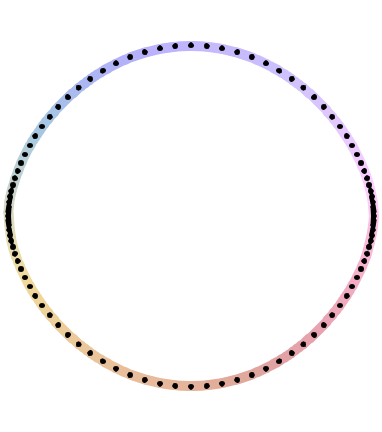}&
			\includegraphics[width=.1\linewidth]{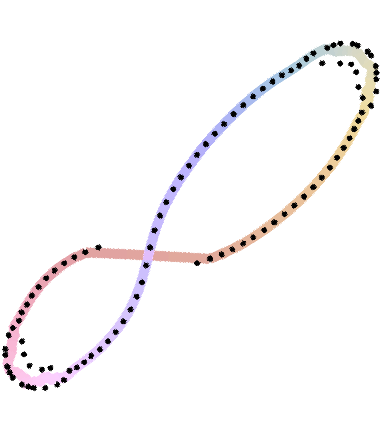}&
			\includegraphics[width=.1\linewidth]{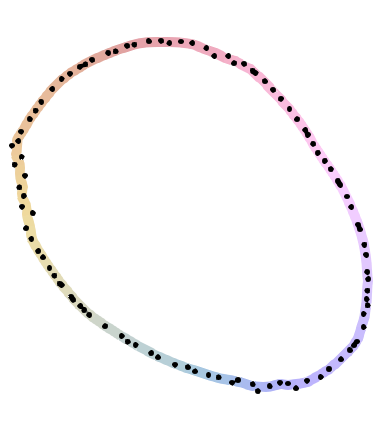}&
			\includegraphics[width=.1\linewidth]{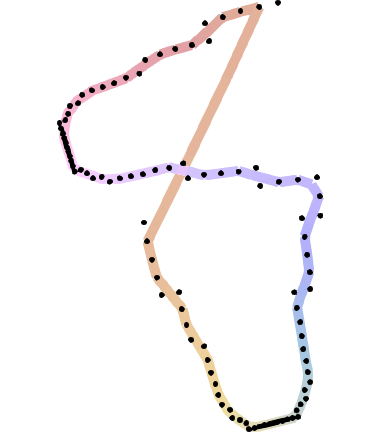}&
			\includegraphics[width=.1\linewidth]{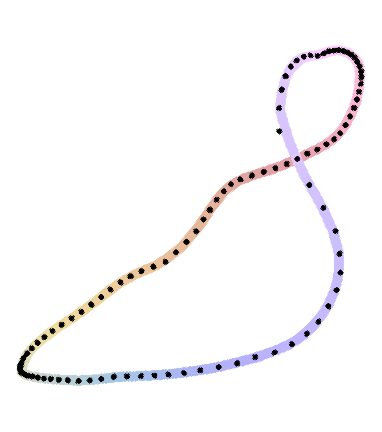}&
			\includegraphics[width=.1\linewidth]{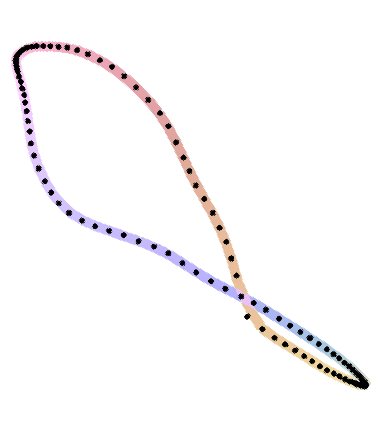}&
			\includegraphics[width=.1\linewidth]{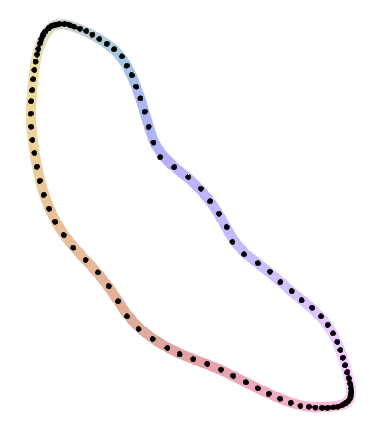}\\

			& $\metwasser(\inputPointCloud,\latentPointCloud)$ & 1.1e+00 & 2.2e-01 & 1.9e+01 & 3.3e+00 & 1.5e+01 & 1.8e-01 & 1.6e-01 & \textbf{1.8e-05} & \underline{1.7e-03}\\
			& $\metdistor(\inputPointCloud,\latentPointCloud)$ & \underline{2.2e-01} & \textbf{2.1e-01} & 2.4e+00 & 3.9e+00 & 2.8e+00 & 8.2e-01 & 3.8e-01 & 7.3e-01 & 7.8e-01\\
			& $\timing$ & 0.00049 & 0.073 & 0.0072 & 0.20 & 2.3 & 0.74 & 2.2 & 6.3 & 2.5\\
			
			\hline \raisebox{1.1cm}{\multirow{4}{*}{\rotatebox{90}{\datakfour{} $(n=300)$}}} &
			\raisebox{-.5mm}{\includegraphics[width=.1\linewidth]{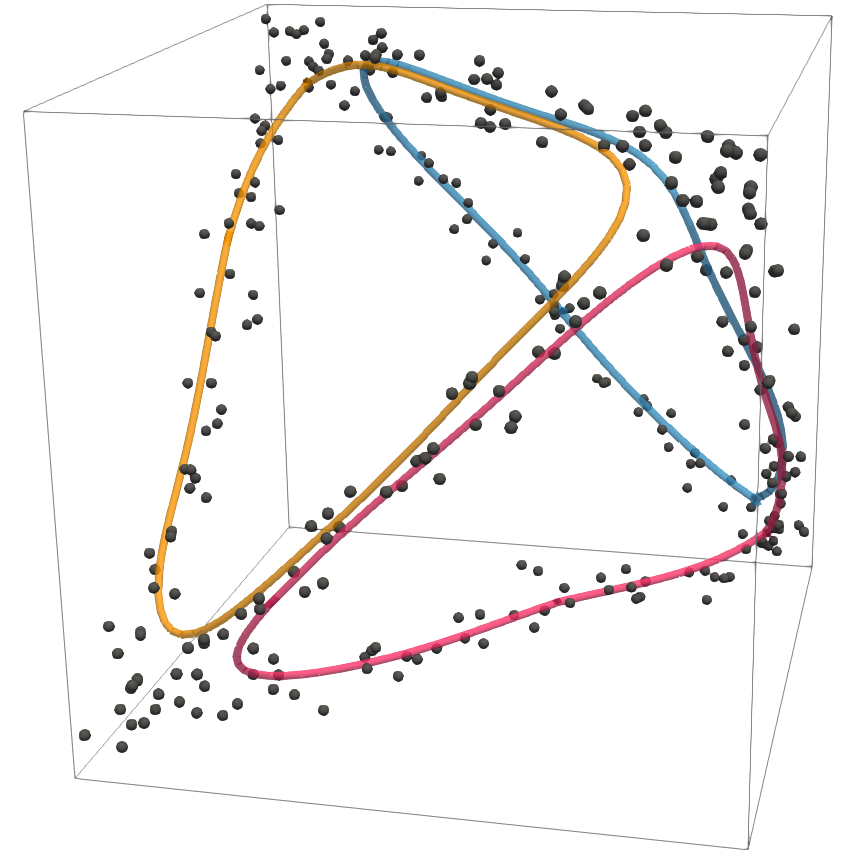}} &
			\includegraphics[width=.1\linewidth]{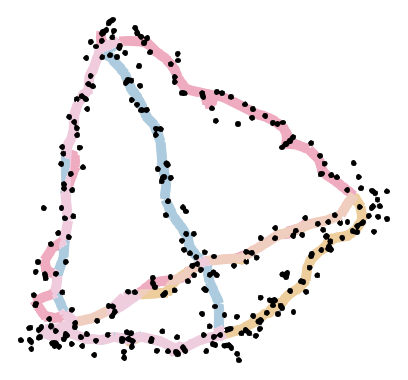}&
			\includegraphics[width=.1\linewidth]{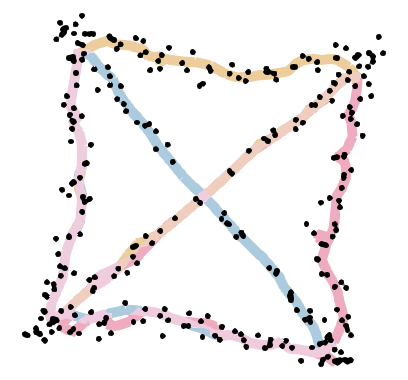}&
			\includegraphics[width=.1\linewidth]{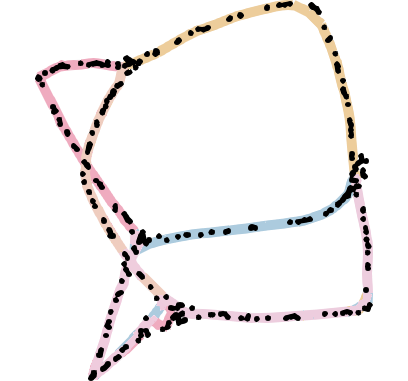}&
			\includegraphics[width=.1\linewidth]{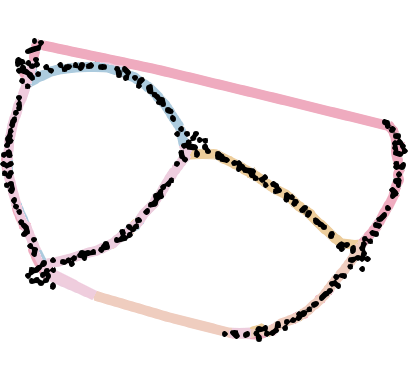}&
			\includegraphics[width=.1\linewidth]{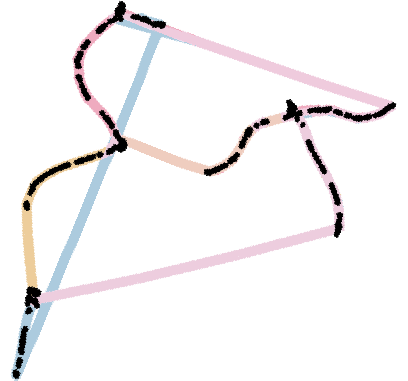}&
			\includegraphics[width=.1\linewidth]{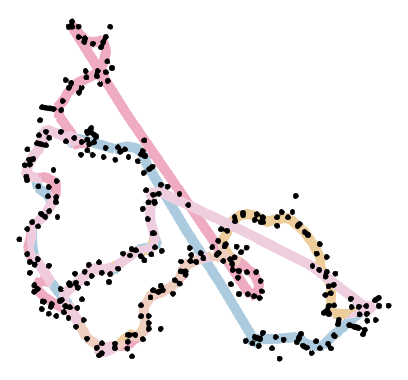}&
			\includegraphics[width=.1\linewidth]{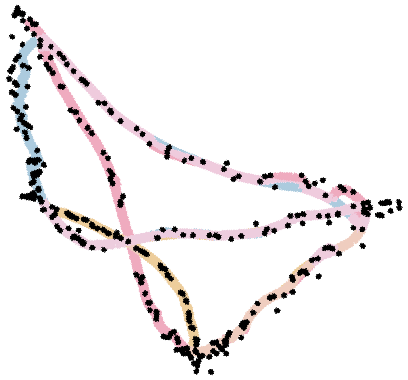}&
			\includegraphics[width=.1\linewidth]{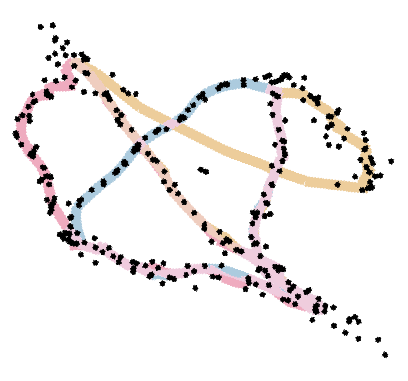}&
			\includegraphics[width=.1\linewidth]{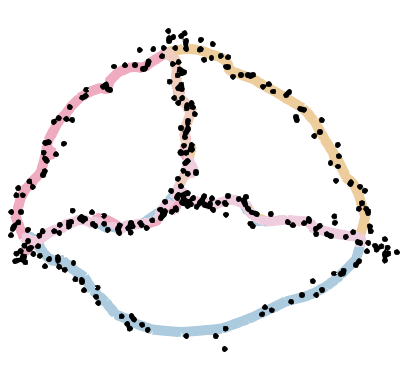}\\

			& $\metwasser(\inputPointCloud,\latentPointCloud)$ & 2.7e-01 & 2.8e-01 & 3.8e-01 & 4.7e+01 & 3.9e-01 & 1.4e-01 & 6.5e-02 & \textbf{7.9e-04} & \underline{2.4e-02}\\
			& $\metdistor(\inputPointCloud,\latentPointCloud)$ & \underline{2.7e-01} & \textbf{1.8e-01} & 5.1e-01 & 1.6e+01 & 9.1e+00 & 5.7e-01 & 1.0e+00 & 6.8e-01 & 3.4e-01\\
			& $\timing$ & 0.00048 & 0.22 & 0.026 & 0.46 & 2.4 & 1.7 & 11 & 63 & 6.2\\
			\hline
		\end{tabular}
		}
	}
	\caption{Comparison of DR methods on three synthetic 3D point clouds (see
	\autoref{sec:test_data} for a description), along with the metric distortion
	$\metdistor(\inputPointCloud,\latentPointCloud)$ between the input $\inputPointCloud$ and
	its planar embedding $\latentPointCloud$, the Wasserstein distance
	$\metwasser(\inputPointCloud,\latentPointCloud)=
	\calw_2\bigl(\dgmrips{1}(\inputPointCloud),\dgmrips{1}(\latentPointCloud)\bigr)$
	between their respective 1-dimensional persistence diagrams, and the running
	time $t$ in seconds. The best value for an indicator is written in bold, the second best is underlined.
	In the second line, a generator of the most persistent \PH{1} pair in the high-dimensional input
	is projected in the 2D embedding in
	transparency and slightly smoothed for visualization purposes (to
	distinguish it from the point cloud). The generator color map depicts its
	arc-length parameterization. In the third line, the input -- that is sampled
	around the edges of a tetrahedron in 3D -- features 3 significantly persistent \PH{1}
	pairs, and a generator for each is represented both in the input and in the 2D
	embeddings with a specific color. Quantitatively, our approach (TopoAE++) generates
	planar embeddings with competitive Wasserstein distances. Qualitatively, it produces
	planar embeddings yielding less crossings of the projected high-dimensional
	persistent generators (colored curves), thereby producing visualizations that
	depict more faithfully the topological handles present in high dimensions.}
	\label{fig:tableSyntheticData}
\end{figure*}

\section{Introduction}
\IEEEPARstart{D}{imensionality} reduction (DR)~\cite{borg97,dimensionReductionBook}
is a fundamental tool in data science for apprehending high-dimensional data.
Such datasets are typically represented as a set of $n$ points $\inputPointCloud = \{x_0, x_1, 
\dots, x_{n - 1}\}$ living in a high-dimensional Euclidean space 
$\mathbb{R}^{\highDim}$. Then, the goal of DR is to compute an embedding 
$\latentPointCloud = \projection(\inputPointCloud)$ of this input point cloud 
in a lower-dimensional Euclidean space $\mathbb{R}^{\lowDim}$, 
with $\lowDim \ll \highDim$. This process can be motivated in practice by 
the need to either \emph{(i)} simplify the input data (to facilitate further 
processing and analysis) or to \emph{(ii)} generate a visualization of the 
data that can be interpreted by a human observer (e.g., $\lowDim=2$ yields planar views of the input point cloud).

Unless $\inputPointCloud$ is already \emph{flat} (e.g., living in an
$\lowDim$-dimensional hyperplane), its projection to $\mathbb{R}^{\lowDim}$
will necessarily involve a form of \emph{distortion}. Depending on the
application needs, several distortion criteria have been introduced in the
literature.
Then, a vast corpus of DR algorithms~\cite{surveyDimensionReduction2,
surveyDimensionReduction1, NonatoA19} have been proposed, either
\emph{(i)} to minimize a specific distortion criterion,
or \emph{(ii)} to preserve interesting geometrical features (thereby
indirectly minimizing the associated distortion criterion).
For instance, the seminal multidimensional scaling algorithm~\cite{dimensionReductionBook} aims at minimizing metric distortion,
which is achieved in practice by trying to preserve pairwise distances.

In many use-cases, it may be desirable to preserve \emph{topological}
characteristics when projecting down to a visual space. For instance, the
preservation of the \emph{clusters} present in high-dimensions enables a quick
visual interpretation of the main trends in the dataset.
\emph{TopoMap}~\cite{doraiswamy2020topomap}, for example, guarantees that, after its
projection, the hierarchical clustering of the dataset is identical 
in high and low dimensions. This provides users with strong confidence 
regarding the structural interpretation of the projected data. 
Technically, the preservation of hierarchical clusters
relates to the preservation of the $0$-dimensional persistent homology (\PH{0}) of the Rips filtration of $\inputPointCloud$ (see \autoref{sec:persistentHomology} for  a 
technical description). 
In this context, several topology-aware DR 
methods~\cite{moor2020topological, doraiswamy2020topomap,
trofimov2023learning} have been proposed, but these mostly focused in 
practice on preserving PH$^0$, i.e., the data point clusters (see
\autoref{sec:relatedWorkTopology}).

In this work, we investigate the preservation of more 
advanced topological features than clusters. 
In particular, we focus on the preservation of the cyclic structures,
the \emph{topological handles}
captured by \PH{1}, when projecting to a planar visual space (i.e.,
$\lowDim = 2$).
\revision{Some datasets indeed include a periodic nature (e.g., time-series), yielding cycles in the high-dimensional space. Then, preserving these cyclic patterns in the visualization is important to visually convey and characterize the periodicity of the studied phenomenon. Examples of such datasets are provided in \autoref{sec:test_data}, with the \datamocap{} dataset capturing a recurring periodic gesture or with the \datasinglecell{} dataset, capturing gene expression at different stages of the cell life cycle.}

However, preserving \PH{1} while projecting to the plane is a 
significantly more challenging problem than preserving \PH{0}.
First, from a computational point of view, the $0$-dimensional case is a very 
specific instance of PH computation, which can be efficiently computed
 with a union-find data structure~\cite{cormen}
 (with a runtime complexity that is virtually linear with the number of
considered edges in the Rips complex).
As of dimension $1$, generic PH computation algorithms need to be considered, with a worst case runtime complexity that is cubic with the number
of considered simplices.
Second, as discussed by several authors~\cite{moor2020topological, doraiswamy2020topomap}, PH$^0$ preservation can be obtained by finding a planar layout of $\inputPointCloud$ which preserves its minimum spanning tree (MST).
Since trees can always be embedded in the plane, the existence of (at least) one 
solution (i.e., a projection preserving \PH{0}) is guaranteed.
Unfortunately, this observation no longer holds for higher dimensional PH.
As discussed later in \autoref{sec:results:analysis},
it is easy to design examples of
high-dimensional point clouds whose \PH{1} cannot be preserved through planar
projections (e.g., a dense sampling of a non-planar
graph, see \autoref{fig:tableK5}).
The existence of such cases, for which an exact preservation of
\PH{1} cannot be obtained,
illustrates the difficulty of the problem, and the need for efficient optimization methods for preserving \PH{1} as accurately as possible, which is the topic of this paper.

Specifically, this paper revisits the \emph{Topological Autoencoders} 
(TopoAE), a prominent technique in topology-aware dimensionality reduction,
which we extend into TopoAE++, for the accurate visualization of the cyclic
patterns present in the data. 
We make the following new contributions:

\begin{enumerate}
\item \emph{A novel theoretical analysis of TopoAE's loss
				($\ltopoae{0}$):}
\begin{itemize}
	\item We show that $\ltopoae{0}$ is an upper bound of the
		Wasserstein distance between the $0$-dimensional persistence diagrams in high
		and low dimensions, with regard to the Rips filtration. In particular, when $\ltopoae{0} = 0$,
		the persistence pairs are identical in high and low
		dimensions for the $0$-dimensional persistent homology (\PH{0}) of
		the Rips filtration.
	
	\item We provide a counter example showing that the above property does not hold for a naive extension of TopoAE to $1$-dimensional persistent homology (\PH{1}), with a loss noted $\ltopoae{1}$.
\end{itemize}

 \item \emph{A generalization of TopoAE's loss for $1$-dimensional
persistent homology (\PH{1}):} We introduce a new loss called
\emph{cascade distortion} (CD), noted $\lcascae{1}$, which addresses the above
counter example. This term favors an isometric embedding of the $2$-chains
filling persistent $1$-cycles. We show that for
\PH{0} this loss generalizes TopoAE's loss (i.e., $\lcascae{0} = \ltopoae{0}$).
Extensive experiments show the practical performance
of this loss for the accurate preservation of
generators through the projection.
 \item \emph{An efficient computation algorithm for TopoAE++:}
     To accelerate runtime, we provide a new,
fast algorithm for the exact computation of \PH{} for Rips filtrations in the
plane. This geometric algorithm relies on the
fast identification of cycle-killing triangles via local minmax triangulations of
the relative neighborhood graph, leveraging duality
for the efficient computation of \PH{1}. We believe this contribution to be of independent interest.
 \item \emph{Implementation:} We provide a C++ implementation of our
algorithms that can be used for reproducibility.
\end{enumerate}

\section{Related work}

The literature related to our work can be classified into two
categories: general purpose DR techniques
(\autoref{sec:relatedWorkGeneralPurpose}) and topology-aware techniques
(\autoref{sec:relatedWorkTopology}).

\subsection{General purpose dimensionality reduction}
\label{sec:relatedWorkGeneralPurpose}

Numerous DR techniques have been proposed and the related literature has been
summarized in several books~\cite{borg97, dimensionReductionBook} and surveys
\cite{surveyDimensionReduction2, surveyDimensionReduction1, NonatoA19}.
Principal Component Analysis (PCA)~\cite{pearson1901liii} is by far the most
popular linear DR technique.
Although it is an indispensable tool for data analysis,
its linear nature does not always allow it to apprehend complex non-linear
phenomena. One of the first non linear DR methods is the multidimensional
scaling (MDS)~\cite{torgerson1952multidimensional}. It aims at preserving as far
as possible the pairwise distances in the high- and low-dimensional point
clouds.
Another approach, particularly related to our work,
consists in optimizing an autoencoder neural network~\cite{hinton_reducing_2006}.
The \textit{encoder} is used to represent the explicit projection map from the
high-dimensional input space to the low-dimensional representation
space, while the \textit{decoder} tries to reconstruct the input data
from its encoded representation.
We will refer to these methods as \emph{global} methods.

Global methods have been used successfully in many applications, but
they do not take into
account the possible distribution of the input points over some implicit,
unknown manifold. This may lead to the unwanted preservation of distances
between points that are close in the ambient space but far apart on this
manifold. \emph{Locally topology-aware} methods have therefore been
introduced to address this issue. For instance,
Isomap~\cite{tenenbaum_global_2000}
preserves geodesic distances on a captured manifold structure of the
input data.
Other methods also feature neighborhood preservation objectives.
For example, Local Linear Embedding (LLE)~\cite{roweis2000nonlinear} relies
on linear reconstructions of local neighborhoods.
Other methods leverage additional landmarks~\cite{silva2003global} or user-provided
control points~\cite{joia:tvcg:2011}.

All these methods try to preserve local
Euclidean distances when projecting to a lower dimension.
However, this can sometimes lack relevance in the applications,
especially with high-dimensional datasets for which
the distribution of pairwise Euclidean distances tend to be uniform.
For such cases, local distance preservation fails at characterizing
well relevant local relations.
To alleviate this issue, SNE~\cite{hinton2002stochastic} and later
t\nobreakdash-SNE~\cite{van2008visualizing} use a conditional probability
formulation to represent similarities between points and try to
have similar distributions both in high- and low-dimension thanks to a
Kullback--Leibler divergence minimization.
More recently UMAP has been introduced~\cite{mcinnes2018umap} along a
theoretical foundation on category theory.
It provides results that are similar visually to t-SNE, but in a more
scalable way.
Variants were later introduced to better preserve the global structure in the embedding, such as TriMAP~\cite{amid2022trimap} that constrains the proximity order within triplets of points, or PaCMAP~\cite{wang_understanding_2021} that adds constraints on more distant point pairs.
Although these methods succeed in preserving the local topology, they are not
explicitly aware of the global structure
of the input, which can lead to the loss of noteworthy global,
topological features.

\subsection{Globally topology-aware dimensionality reduction}
\label{sec:relatedWorkTopology}

Topology-based methods have become popular over the last
two decades in data analysis and
visualization~\cite{heine16} and have been applied to various areas:
astrophysics~\cite{sousbie11, shivashankar2016felix},
biological imaging~\cite{beiBrain18, carr04, topoAngler},
quantum chemistry~\cite{chemistry_vis14,harshChemistry, D2CP05893F},
fluid dynamics~\cite{kasten_tvcg11, NauleauVBBT22},
material sciences~\cite{gyulassy_vis07, gyulassy_vis15, SolerPDPCT19},
turbulent combustion~\cite{gyulassy_ev14, laney_vis06}. They leverage tools that
define concise signatures of the data based on its topological properties and
that have been summarized in topological data analysis reference
books ~\cite{edelsbrunner_computational_2010, zomorodian_computational_2010}
and surveys~\cite{chazal_introduction_2021}.

Several DR methods have been proposed
by the visualization community to preserve specific topological signatures
of the input data. For instance, terrain metaphors have been
investigated for the visualization of an input high-dimensional scalar
field, in the form of a three-dimensional terrain, whose elevation yields an
identical contour tree~\cite{Weber:2007} or an identical set of separatrices
\cite{gerber2010, gerber2013}.
This framework has been extended to density
estimators~\cite{OesterlingHJS10,
OesterlingSTHKEW10, OHJSH11, Oesterling0WS13} for the support of
high-dimensional point clouds. However, such metaphors completely discard
the metric information of the input space~\cite{OesterlingHJS10}, possibly
placing next to each other points which are arbitrarily far
in the input space (and reciprocally). Yan et al.~\cite{abs-1806-08460}
introduced a DR approach driven by the Mapper structure~\cite{SinghMC07}, an
approximation of the Reeb graph~\cite{reeb46}, which can capture in practice
large handles in the data, however without guarantees, since the number of handles in the considered manifold is only an upper bound on the number of loops in the Reeb graph~\cite{edelsbrunner_computational_2010}.

To incorporate the metric information from the input data while
preserving at the same time some of their topological characteristics, several
approaches have focused on driving the projection by
the \emph{persistence diagram}
of the Rips filtration of the point cloud (see \autoref{sec:persistentHomology}
for  a technical description).
Carriere et al.~\cite{carriere2021optimizing} presented a generic persistence
optimization framework with an application to dimensionality reduction.
Their approach explicitly minimizes the Wasserstein distance
(\autoref{sec:persistentHomology}) between the $1$-dimensional persistence
diagrams in high and low dimensions. However, this approach solely focuses on
this penalization term. As a result,
although the number and persistence of cycles  may be well-preserved,
the solver tends to produce cycles in low dimensions which involve arbitrary
points (e.g., which were not necessarily located along the cycles in high
dimensions), which challenges visual interpretation, as later
detailed in \autoref{sec:results:analysis}.

To enforce a correspondence between the topological
features at the data point level, additional structures need to be preserved.
For the specific case of $0$-dimensional persistent homology (\PH{0}),
Doraiswamy et al. introduced \emph{TopoMap}~\cite{doraiswamy2020topomap}, an
algorithm which constructively preserves the \emph{persistence pairs}
(\autoref{sec:persistentHomology}) through the preservation of the minimum
spanning tree of the data. An accelerated version, with improved layouts, has
recently been proposed~\cite{guardieiro2024topomap++}.
Alternative approaches have considered the usage of an optimization framework
(typically based on an autoencoder neural network
\cite{hinton_reducing_2006}), with the integration of specific topology-aware
losses~\cite{moor2020topological,barannikov2021representation,
nelson2022topology,trofimov2023learning,schonenberger2020witness}. Among them,
a prominent approach is the \emph{Topological Autoencoders}
(TopoAE)~\cite{moor2020topological}. Its loss aims at preserving
the diameter of the simplices involved in
persistence pairs, when going from high to low dimensions and reciprocally.
However, the above techniques focused in practice on the
preservation of \PH{0} and did not, to our knowledge, report experiments
regarding the preservation of higher dimensional PH.
Specifically, we show in \autoref{sec:analysis} that, while a zero
TopoAE loss indeed implies a preservation of the persistence pairs for \PH{0},
it is not the case for higher dimensional PH. We provide a counter example for
\PH{1}, which is addressed by our novel, generalized loss.

\section{Background}
\label{sec:background}

This section presents the technical background to our work. We refer the
reader to textbooks~\cite{edelsbrunner_computational_2010,
zomorodian_computational_2010} for comprehensive introductions to computational
topology.

\subsection{Geometric structures}
\label{sec:geometry}

We introduce below geometric concepts related to persistent homology (\autoref{sec:geometricPH}) that are relevant to our algorithm.

\begin{figure}
	\centering
	\def\svgwidth{\linewidth}
\begingroup%
  \makeatletter%
  \providecommand\color[2][]{%
    \errmessage{(Inkscape) Color is used for the text in Inkscape, but the package 'color.sty' is not loaded}%
    \renewcommand\color[2][]{}%
  }%
  \providecommand\transparent[1]{%
    \errmessage{(Inkscape) Transparency is used (non-zero) for the text in Inkscape, but the package 'transparent.sty' is not loaded}%
    \renewcommand\transparent[1]{}%
  }%
  \providecommand\rotatebox[2]{#2}%
  \newcommand*\fsize{\dimexpr\f@size pt\relax}%
  \newcommand*\lineheight[1]{\fontsize{\fsize}{#1\fsize}\selectfont}%
  \ifx\svgwidth\undefined%
    \setlength{\unitlength}{994.84631348bp}%
    \ifx\svgscale\undefined%
      \relax%
    \else%
      \setlength{\unitlength}{\unitlength * \real{\svgscale}}%
    \fi%
  \else%
    \setlength{\unitlength}{\svgwidth}%
  \fi%
  \global\let\svgwidth\undefined%
  \global\let\svgscale\undefined%
  \makeatother%
  \begin{picture}(1,0.52441875)%
    \lineheight{1}%
    \setlength\tabcolsep{0pt}%
    \put(0,0){\includegraphics[width=\unitlength,page=1]{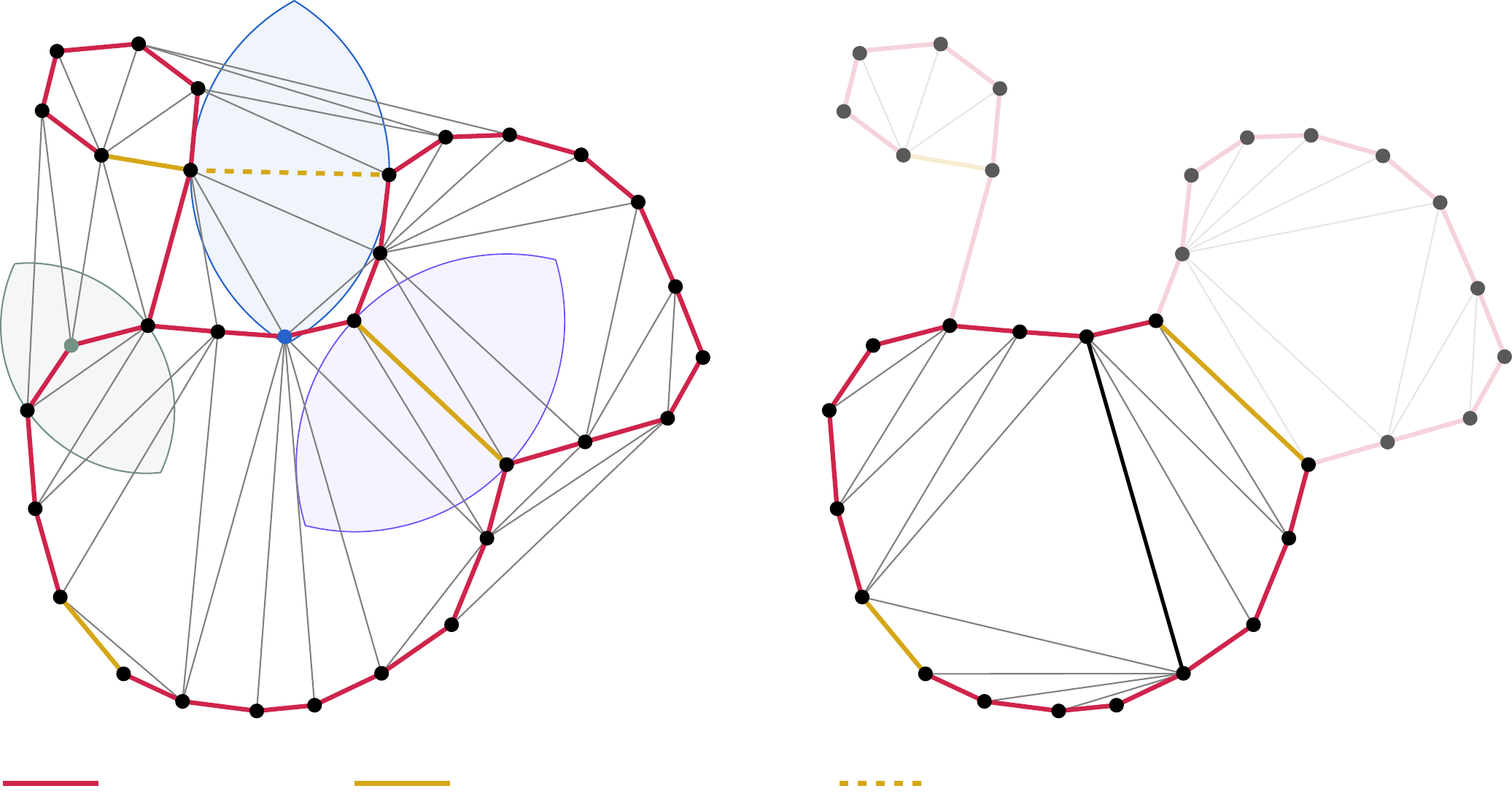}}%
    \put(0.07676666,-0.00331138){\color[rgb]{0,0,0}\makebox(0,0)[lt]{\lineheight{1.25}\smash{\begin{tabular}[t]{l}\footnotesize$\emst$\end{tabular}}}}%
    \put(0.31414994,-0.00308045){\color[rgb]{0,0,0}\makebox(0,0)[lt]{\lineheight{1.25}\smash{\begin{tabular}[t]{l}\footnotesize$\rng\setminus\emst$\end{tabular}}}}%
    \put(0.63044541,-0.00394023){\color[rgb]{0,0,0}\makebox(0,0)[lt]{\lineheight{1.25}\smash{\begin{tabular}[t]{l}\footnotesize$\ug\setminus\rng$\end{tabular}}}}%
  \end{picture}%
\endgroup%

	\caption{Left: geometric graphs for a point cloud $\inputPointCloud$ in the plane. Its $\emst$ is
depicted in red. The edges from the $\rng$ which are not in the $\emst$ are shown in continuous yellow, while those which are in the $\ug$ but not in the $\rng$ are shown in dashed yellow. The
remaining Delaunay edges are shown in gray.
Three lenses are shown:
the purple one is devoid of points of $\inputPointCloud$, therefore the associated yellow edge belongs to the $\rng$;
the blue one -- associated with the dashed yellow edge -- is devoid of points of the link of that edge but contains another point of $\inputPointCloud$ (the blue one), therefore this edge is in the $\ug$ but not in the $\rng$;
the green one contains the green point and the associated gray edge is not in $\rng$ nor in $\ug$.
Right: within the highlighted bottom $\rng$-polygon, a $\mml$ triangulation has
replaced the Delaunay triangulation, with its longest edge highlighted in
black. Note that for the two other polygons, the Delaunay triangulation is
already a $\mml$ triangulation within these polygons.}
	\label{fig:geometric_graphs}
\end{figure}

We consider in this section a point cloud $\inputPointCloud$ containing $n$ points, in a
$d$-dimensional space $\bbr^d$ endowed with the ambient Euclidean metric. The
lens of an edge $(a,b)\in \inputPointCloud^2$ is defined as the points of
$\inputPointCloud$ whose distance to either $a$ or $b$ is smaller than $\dist{a}{b}$.
The relative neighborhood graph $\rng(\inputPointCloud)$
\cite{toussaint_relative_1980} is the graph on $\inputPointCloud$ that features
every edge with an empty lens, i.e., $\rng(\inputPointCloud)=\{e\in
\inputPointCloud^2\mid\lens(e)=\varnothing\}$.
The $\rng$ is known to be a subset of the Delaunay triangulation
\cite{toussaint_relative_1980}. In the plane,
it can be computed in time $\calo(n\log n)$~\cite{supowit_relative_1983,
jaromczyk_note_1987, lingas_linear-time_1994}.
The Urquhart graph $\ug(\inputPointCloud)$ \cite{urquhart_algorithms_1980} is
another subgraph of the Delaunay triangulation that is obtained by removing
the longest edge of each triangle. It can also be thought as the
subgraph of $\del(\inputPointCloud)$ that only keeps edges whose lens does not
contain any points of its link%
\footnote{In an abstract simplicial complex $\calk$ (i.e., a family of sets that
is closed under taking subsets, where sets represent simplices and their
subsets represent their faces), the link of a simplex $\sigma\in\calk$ is the set
$\{\tau\in\calk\mid\sigma\cap\tau=\varnothing\text{ and }\sigma\cup\tau\in\calk\}$.
It provides a notion of neighborhood boundary for $\sigma$.}.
Finally, we write $\emst(\inputPointCloud)$ the Euclidean minimum spanning tree
of $\inputPointCloud$, i.e., the tree spanning $\inputPointCloud$ that minimizes
the sum of the Euclidean distances of its edges. Under general position
hypothesis\footnote{i.e., we suppose that no $d+2$ points lie on a
common $(d-1)$-sphere (for the Delaunay triangulation
\cite{edelsbrunner_computational_2010}), and that pairwise distances are unique
for the uniqueness of the $\emst$, which can be enforced
via small
perturbations.},
we have the following inclusions
\cite{toussaint_relative_1980} (\autoref{fig:geometric_graphs}, left, shows a case where the three inclusions are strict):
\begin{equation}
\emst(\inputPointCloud)\subseteq\rng(\inputPointCloud)
\subseteq\ug(\inputPointCloud)\subseteq\del(\inputPointCloud).
\end{equation}

Our algorithm also relies on the minmax length ($\mml$)
triangulations of a planar point cloud $\inputPointCloud\subset\bbr^2$. It is defined as
a triangulation that minimizes the length of its longest edge
(\autoref{fig:geometric_graphs}, right). Such a triangulation can be computed in quadratic time
$\calo(n^2)$ with an algorithm~\cite{edelsbrunner_quadratic_1993}
that breaks down the construction of such a triangulation over
each polygon formed by the $\rng$ and the convex hull of $\inputPointCloud$.

\begin{figure*}
	\begin{tabular}{@{}c@{}@{}c@{}@{}c@{}@{}c@{}@{}c@{}@{}c@{}}
	\includegraphics[width=.166\textwidth]{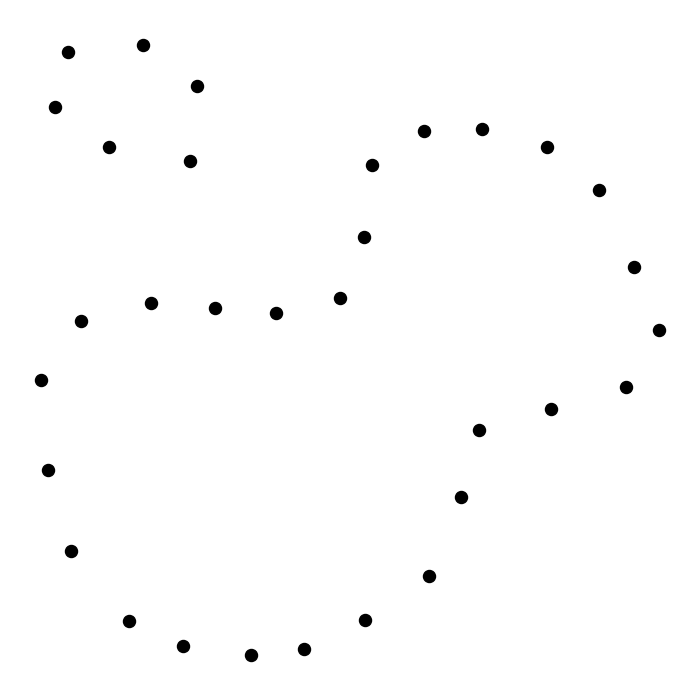}&
	\includegraphics[width=.166\textwidth]{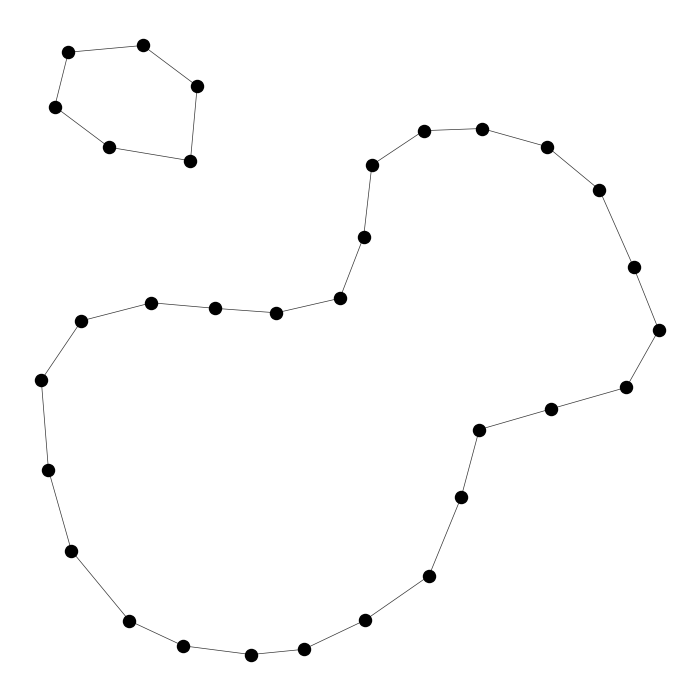}&
	\includegraphics[width=.166\textwidth]{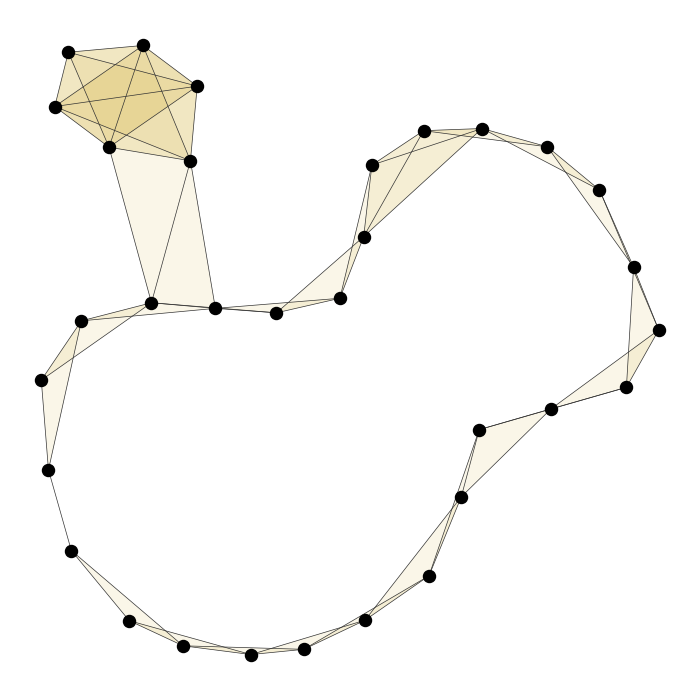}&
	\includegraphics[width=.166\textwidth]{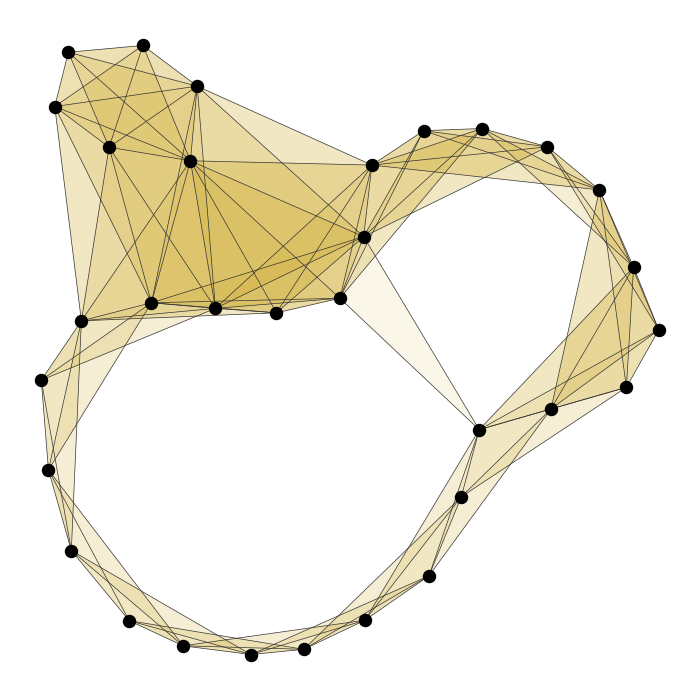}&
	\includegraphics[width=.166\textwidth]{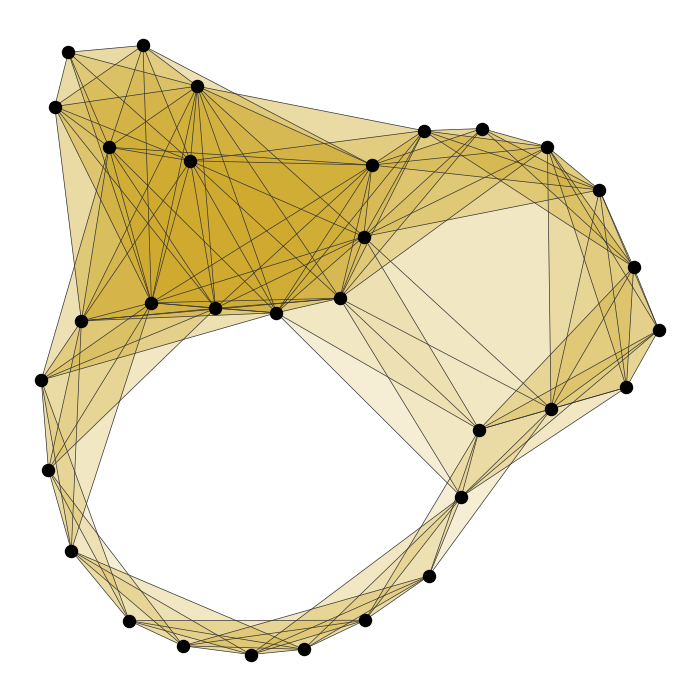}&
	\includegraphics[width=.166\textwidth]{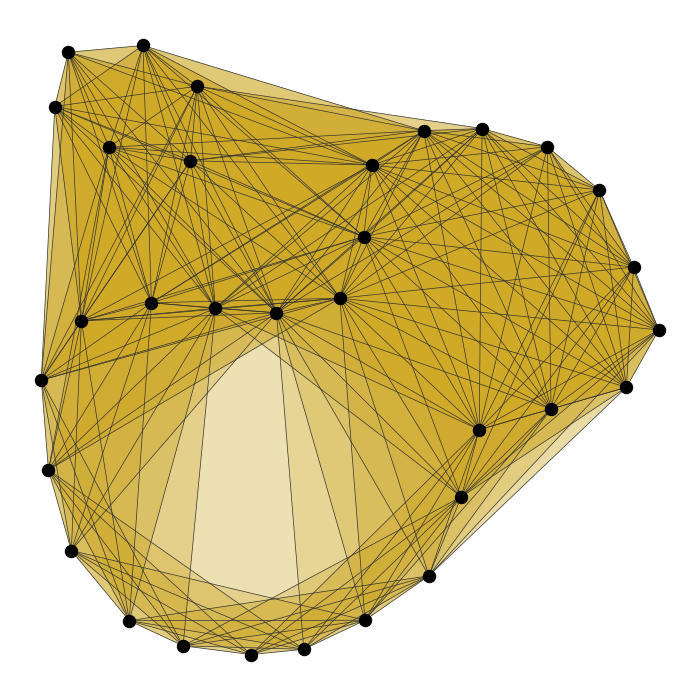}\\
	$\betti{0}=n,\;\betti{1}=0$&
	$\betti{0}=2,\;\betti{1}=2$&
	$\betti{0}=1,\;\betti{1}=1$&
	$\betti{0}=1,\;\betti{1}=2$&
	$\betti{0}=1,\;\betti{1}=1$&
	$\betti{0}=1,\;\betti{1}=0$\\
	\end{tabular}
	\caption{Rips complexes of the same planar point cloud $\inputPointCloud$,
	for increasing diameter thresholds $\diameterThreshold$
	(only the 2-skeletons, i.e., the vertices, edges and triangles, of the
	simplicial complexes are shown). This threshold increase
	induces a sequence of nested simplicial complexes,
	whose topology varies along the process. From left to right, the number of
	connected components is successively $n$, then $2$, then $1$,
	while the number of handles is successively $0$, $2$, $1$, $2$, $1$ and $0$.}
	\label{fig:rips}
\end{figure*}

We conclude this section with a geometric construction that is often used in the topological data analysis of point clouds:
the Rips complex of $\inputPointCloud$ with diameter threshold $\diameterThreshold\geq0$
is the simplicial complex featuring all the simplices $\sigma\subset
\inputPointCloud$ whose diameter
$\delta(\sigma)=\max\limits_{x,y\in\sigma}\lVert x-y\rVert_2$ is smaller than
$\diameterThreshold$ (i.e., every edge included in $\sigma$ must be shorter than
$\diameterThreshold$) (see \autoref{fig:rips}):
\begin{equation}
	\rips_r(\inputPointCloud)=\{\sigma\subset\inputPointCloud\mid \delta(\sigma)\leq
	\diameterThreshold\}.
\end{equation}
We write $\calk^\homologyDim=\{\sigma\in\calk\mid\dim\sigma=\homologyDim\}$ the set of $\homologyDim$-simplices of a simplicial complex $\calk$, and $\skeleton{\calk}{k}=\{\sigma\in\calk\mid\dim\sigma\leq\homologyDim\}$ its $\homologyDim$-\emph{skeleton}.
Then, $\rips_r^{(\homologyDim)}(\inputPointCloud)$ is the Rips complex of
$\inputPointCloud$ with diameter
$\diameterThreshold\geq0$ and maximum
simplex dimension $\homologyDim$.

\subsection{Persistent homology}
\label{sec:persistentHomology}

Persistent homology, introduced independently by several authors
\cite{B94,frosini99,robins99,edelsbrunner02}, focuses on the evolution of the
topological properties of an increasing sequence of $m$ nested
simplicial complexes $\bbk=(\filt{0},\filt{1},\ldots,\filt{m-1})$,
which is called a \textit{filtration}. Persistent homology can be applied on
various filtrations, like the sublevel sets of a scalar field on a mesh. However
a filtration of particular interest for point clouds is the Rips filtration
(\autoref{fig:rips}), which consists in the sequence of Rips complexes with an
increasing threshold $r$. More precisely, the Rips filtration of $X$ can be
written as
$\bigl(\rips_{0}(X),\rips_{r_1}(X),\ldots,\rips_{r_{m-1}}(X),
\rips_{\infty}(X)\bigr) $ with increasing $r_i$'s such that the Rips complex
changes between $r_i$ and $r_{i+1}$ (i.e.
$\rips_{r_i}(X)\subsetneq\rips_{r_{i+1}}(X)$).

\textit{Persistent homology groups} for a filtration $\bbk$ are algebraic constructions that contain \textit{persistent homology classes} and that are defined from homology groups (see \autoref{appendix:homology} for a brief summary or~\cite{edelsbrunner_computational_2010} for a more complete introduction).
Informally, these classes represent topological features (that can be
geometrically interpreted as clusters for \PH{0}, cycles for \PH{1}, voids for
the \PH{2}\ldots) which appear at some point in the filtration and may disappear
later. Formally, for $0\leq i\leq j\leq m$, the $\homologyDim$-th persistent
homology group $\PHG{i}{j}$ is the image of the $\homologyDim$-th homology group
$\HG{i}$ by the inclusion morphism $\morphism{i}{j}:\HG{i}\rightarrow\HG{j}$.
Its classes can be understood as those which already exist in $\HG{i}$ and that
still exist in $\HG{j}$. Therefore, we say that a class $\PHC\in\HG{i}$ is
\textit{born} entering $\filt{i}$ when $\gamma\not\in\PHG{i-1}{i}$, i.e., when
it is not the image of a pre-existing class by the inclusion morphism. We also
say that such a class born at $\filt{i}$ \textit{dies} entering $\filt{j}$ when
$\morphism{i}{j-1}(\gamma)\not\in\PHG{i-1}{j-1}$ but
$\morphism{i}{j}(\gamma)\in\PHG{i-1}{j}$, i.e., when $\gamma$ is merged into a
pre-existing class (possibly the 0 class) when going from $\filt{j-1}$ to
$\filt{j}$. In this case, the birth of $\gamma$ is associated with a
$\homologyDim$-simplex $\sigma_i\in\filt{i}$ and its death with
$(\homologyDim+1)$-simplex $\sigma_j\in\filt{j}$. Such a pair $(\sigma_i,
\sigma_j)$ is called a \textit{persistence pair}.
If the filtration is defined by scalar values $f$ on the simplices
(e.g., the diameter $\delta$ for the Rips filtration), the \textit{persistence}
of such a class is the positive value $f(\sigma_j)-f(\sigma_i)$.

In the specific case of a Rips filtration, we can say more simply that for any $\homologyDim$, a \PH{\homologyDim} pair of positive persistence $(\sigma_i,\sigma_j)$ is created (resp. killed) by the longest edge of $\sigma_i$ (resp. $\sigma_j$).
Indeed, a Rips complex is entirely determined by its 1-skeleton since the diameter of any simplex is the length of the longest edge it features (it is a \textit{flag complex}). In the following we refer to these edges as $\homologyDim$-\textit{critical edges}.

\begin{figure}
	\centering
	\includegraphics[width=.95\linewidth]{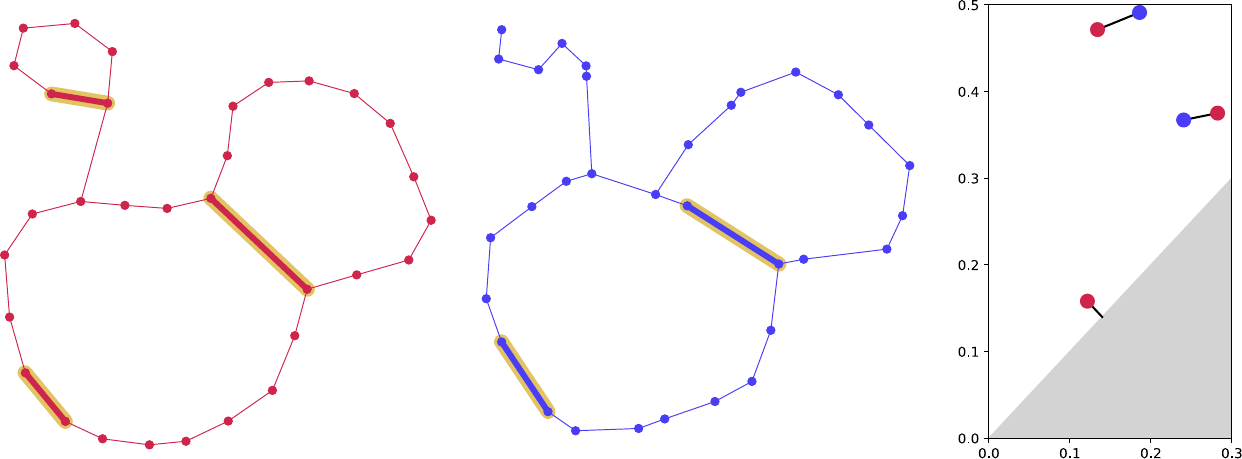}
	\caption{Two point clouds (red and blue) in the plane represented
	with their $\rng$, along with their respective 1-dimensional
	persistence diagrams (right). The edges in
	$\rng\setminus\emst$ are highlighted in bold and yellow.
	The number of non-diagonal points in each diagram is exactly the number of
	$\rng$-polygons in the associated point cloud (3 for the red one, 2 for the blue
	one), and also the number of $\rng\setminus\emst$ edges. The optimal assignment inducing
	the Wasserstein distance $\calw_2$ between them is shown in black.}
	\label{fig:wasserstein}
\end{figure}

\textit{Persistence diagrams} (\autoref{fig:wasserstein}) are concise
topological representations that summarize the persistent homology groups.
More precisely, the $\homologyDim$-th persistence diagram is a 2D
multiset where each $\homologyDim$-dimensional persistence pair $(\sigma_i,
\sigma_j)$ is represented by a point
$\bigl(f(\sigma_i),f(\sigma_j)\bigr)\in\bbr^2$ above
the diagonal.
We write $\dgm{k}(f)$ for the $\homologyDim$-th persistence diagram of the filtration given by a scalar function $f:\calk\rightarrow\bbr$, or directly $\dgmrips{k}(\inputPointCloud)$ for the $\homologyDim$-th persistence diagram of the Rips filtration of $\inputPointCloud$.
Many points clouds have the same persistence diagrams:
in particular, $\dgmrips{k}(\inputPointCloud)$ is invariant to point permutations in $\inputPointCloud$.

Persistence diagrams (of equal homology dimension) are usually compared with metrics born of optimal transport that come with stability properties \cite{cohen2006vines,skraba_wasserstein_2023}.
Let $\cald, \cald'\subset\bbr^2$ be two persistence diagrams to be
compared. We define the \textit{augmented} diagrams as
$\augmented{\cald}=\cald\cup\Delta\cald'$ and
$\augmented{\cald'}=\cald'\cup\Delta\cald$ where $\Delta$ is the map that
projects to
the diagonal, i.e., $\Delta:(b,d)\mapsto(\frac{b+d}{2},
\frac{b+d}{2})$. This ensures that $|\augmented{\cald}|=|\augmented{\cald'}|$.
The $p$-cost $c_p(x,y)$ between two points $x\in\augmented{\cald}$ and
$y\in\augmented{\cald'}$ is defined as 0 if both $x$ and $y$ are on the diagonal
and as $\bbr^2$'s $p$-norm $\|x-y\|_p$ otherwise.
Then, the $L_q$-Wasserstein distance $\calw_q^p$ is:
\begin{equation}
\calw_q^p(\cald,\cald')=
\min\limits_{\psi\in\Psi}\left(\sum\limits_{x\in\augmented{\cald}}c_p\bigl(x,
\psi(x)\bigr)^q\right)^{1/q},
\end{equation}
where $\Psi=\mathrm{Bij}(\augmented{\cald},\augmented{\cald'})$ (\autoref{fig:wasserstein}).
Because for any fixed $q$, all $\calw_q^p$ are bi-Lipschitz equivalent
\cite{skraba_wasserstein_2023}, we focus on the distances $\calw_p=\calw_p^p$
with $p<\infty$. In practice, the Wasserstein distances $\calw_p$ can be
approximated thanks to the \textit{auction} algorithm, an asymmetric and generic
approach introduced by Bertsekas~\cite{bertsekas_new_1981} that can be
specialized to persistence diagrams with dedicated data structures to leverage
their geometric structure~\cite{kerber2017geometry}.

\subsection{Persistence homology computation}
\label{sec:persistenceComputation}

Since the introduction of computational PH
\cite{edelsbrunner02}, numerous algorithms and implementations have been
proposed, based either on the boundary matrix reduction approach or on its
geometric interpretation known as the \texttt{PairCells} algorithm~\cite{zomorodian_computational_2010},
which we detail in the following as our
novel loss (introduced in \autoref{sec:newLoss}) is defined relatively to the
\textit{cascade}, an object computed by this algorithm.
Indeed, \texttt{PairCells} computes not only persistent pairs but also
representatives of each \PH{} class for any simplicial complex $\calk$ with its
simplices endowed with distinct scalar values $f(\sigma)$ that are increasing by
inclusion (which induces the filtration of sublevel sets).
For each $k$-simplex $\sigma\in\calk$, it stores a $k$-chain $\cascade(\sigma)$ that is initially equal to $\sigma$, and whose $(k-1)$-boundary $\partial(\cascade(\sigma))$ is, after the execution of the algorithm, a representative of the \PH{} class killed by $\sigma$.
In the following we will call this boundary a \emph{persistent generator}
of the \PH{} class~\cite{Iuricich22, guillou2023discrete}.
In addition, these cascades will enable us in \autoref{sec:newLoss} to find $2$-chains that fill the $1$-cycles.

\begin{figure*}
	\def\svgwidth{\linewidth}
	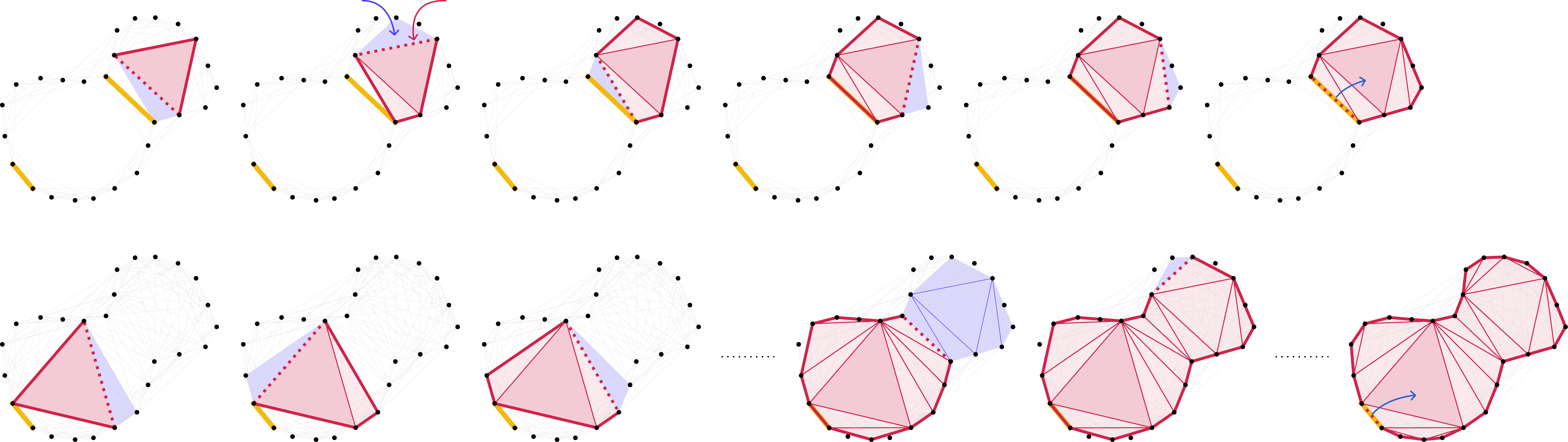
	\caption{Some steps of two executions of the \texttt{\texttt{EliminateBoundaries}} \autoref{algo:eliminateBoundaries} procedure run on the 2-simplices $\sigma_0$ (top line) and $\sigma_1$ (bottom line), corresponding to the creation of two \PH{1} pairs in the previous planar point cloud (with its smallest cycle removed for clarity).
	Unpaired edges -- which are in $\rng(\inputPointCloud)\setminus\emst(\inputPointCloud)$ --
	are depicted in yellow.
	At each step, $\cascade[\sigma]$ is shown in red, and its boundary
$\partial(\cascade[\sigma])$ is highlighted in bold, with its longest edge,
i.e., $\tau=\text{Youngest}\bigl(\partial(\cascade[\sigma])\bigr)$, dotted.
	When it exists, $\cascade\bigl[\partner[\tau]\bigr]$ is shown is blue.
	A persistent pair $(\tau, \sigma)$ is created when $\partner[\tau]=\varnothing$, i.e., when $\tau$ coincides with a yellow unpaired edge (rightmost images).
	In both executions, $\rips_{\delta(\sigma)}^1$ is shown in light gray.
	}
	\label{fig:eliminateBoundaries}
\end{figure*}

\texttt{PairCells} (\autoref{algo:pairCells}) takes one $k$-simplex
$\sigma\in\calk$ at a time, by increasing value $f(\sigma)$, and applies the
\texttt{EliminateBoundaries} procedure (\autoref{algo:eliminateBoundaries}) on
it.
This procedure runs a \textit{homologous propagation}\cite{guillou2023discrete} from $\sigma$: the $(k-1)$-boundary $\partial(\cascade(\sigma))$ is expanded step by step, each time selecting its \emph{youngest} $(k-1)$-simplex $\tau$ (i.e., the one with the highest value $f(\tau)$) and adding modulo~2 to $\cascade(\sigma)$ the cascade associated with the partner of $\tau$, if it exists (see \autoref{fig:eliminateBoundaries}).
The procedure stops either when an unpaired simplex $\tau$ is selected, in which
case $\sigma$ and $\tau$ are paired together since $\sigma$ kills the
$(k-1)$-dimensional homology class introduced by $\tau$; or when the boundary
$\partial\bigl(\cascade(\sigma)\bigr)$ becomes empty, in which case we know
that $\sigma$ creates a $k$-dimensional homology class.
This algorithm can be understood as a geometric interpretation of the persistence algorithms based on boundary matrix reduction.

In practice, if we are not interested in the cascades themselves, it can be
faster to only store the boundary of each simplex's cascade
$\partial\bigl(\cascade(\sigma)\bigr)$
(instead of the
whole cascade itself) and to perform a modulo~2 addition operation on those
boundaries when expanding a cascade
(\autoref{algo:eliminateBoundaries:propagation} of
\autoref{algo:eliminateBoundaries})~\cite{guillou2023discrete}. Another
acceleration is the preliminary pairing of the \textit{apparent pairs}, pairs of
simplices with the same filtration
value~\cite{guillou2023discrete,bauer_ripser_2021}
(some of them appear in \autoref{fig:eliminateBoundaries} as these dotted red edges whose partner's cascade is a single triangle).

\begin{algorithm}
	\DontPrintSemicolon
	\label{algo:pairCells}\caption{\texttt{PairCells}}
	\KwIn{$\calk$}
	\For{$\sigma\in\calk$ by increasing $f(\sigma)$}{
		$\partner[\sigma]\gets\varnothing$\;
		$\cascade[\sigma]\gets\sigma$\;
		\texttt{EliminateBoundaries}($\sigma$)\;
		\If{$\partial(\cascade[\sigma])\neq\varnothing$} {
			$\tau\gets\text{Youngest}\bigl(\partial(\cascade[\sigma])\bigr)$\;
			$\partner[\sigma]\gets\tau$\;
			$\partner[\tau]\gets\sigma$\;
		}
	}
\end{algorithm}

\begin{algorithm}
	\DontPrintSemicolon
	\label{algo:eliminateBoundaries}\caption{\texttt{EliminateBoundaries}}
	\KwIn{$\sigma\in\calk$}
	\While{$\partial(\cascade[\sigma])\neq\varnothing$}{
		$\tau\gets\text{Youngest}\bigl(\partial(\cascade[\sigma])\bigr)$\;
		\eIf{$\partner[\tau]=\varnothing$} {
			return\;
		}{
$\cascade[\sigma]\gets\cascade[\sigma]+\cascade\bigl[\partner[\tau]\bigr]$\;
\label{algo:eliminateBoundaries:propagation}
		}
	}
\end{algorithm}

\subsection{Geometric interpretation of persistent homology}
\label{sec:geometricPH}

Recent boundary matrix reduction-based open-source implementations include
\textit{Gudhi}~\cite{maria2014gudhi}, \textit{PHAT}~\cite{bauer2017phat} or
\textit{Ripser}~\cite{bauer_ripser_2021}. In particular, \textit{Ripser} is
specialized in the PH computation of Rips filtrations (given by point clouds or
distance matrices) yielding much faster and more memory-efficient execution than
previous implementations.
However, it is sometimes possible to leverage the Euclidean embedding
of point clouds for which we want to compute the PH of the Rips filtration to
speed up the computation. In this section, we highlight some known results that
link \PH{0} and \PH{1} of the Rips filtration to some geometric structures
defined in \autoref{sec:geometry}.

The first result is that the set of edges that destroy
a \PH{0} class of the Rips filtration (i.e., that reduce the
number of connected components by 1) is exactly the set of edges of the
Euclidean minimum spanning tree (see
\cite{doraiswamy2020topomap} for example for a proof). It allows to compute
the \PH{0} of the Rips filtration by constructing the minimum spanning tree of
the input point cloud, e.g., with
Kruskal's algorithm based on the union-find
data structure~\cite{cormen}, in time complexity
$\calo\left(n^2\alpha(n^2)\right)$ where $\alpha$ is the inverse of the
Ackermann function. This approach is largely used in practice, e.g. in
\textit{Ripser}~\cite{bauer_ripser_2021}.

On the contrary, the edges in the complementary of $\emst(X)$ are the edges that
create a \PH{1} class~\cite{skraba2017randomly}, possibly with zero
persistence -- in which case this class dies immediately with a triangle of
same diameter.
More recently, Koyama et al.~\cite{koyama2023reduced} showed that the edges of
$\rng(X)\setminus\emst(X)$ are exactly the edges that create a \PH{1}
class of positive persistence.
A corollary
is that the quantity $|\rng(X)\setminus\emst(X)|$ is exactly the number of \PH{1} classes with positive persistence (see
\autoref{fig:wasserstein}).
Koyama et al. also introduced \textit{Euclidean PH1}, a method that specializes
the \PH{1} computation for Rips filtrations of
point clouds in $\bbr^2$ and $\bbr^3$, using the above geometric result.
Compared to other Rips PH algorithm like \textit{Ripser}, they improve both
running time and memory consumption in this specific case thanks to the
construction from the $\rng$ of a reduced Vietoris--Rips complex that features
far fewer triangles than the original one, but that has the same \PH{0} and
\PH{1}. This permits to get a much smaller boundary matrix to be reduced.

\subsection{Topological Autoencoders}
\label{sec:topologicalAutoencoders}

Our method builds on \textit{Topological Autoencoders}
(TopoAE)~\cite{moor2020topological}, a dimension reduction technique that
combines an autoencoder with a topological feature preservation loss.
We therefore briefly summarize this approach below.

Let $\calx$ be the input, high-dimensional space and $\calz$ the latent, low-dimensional space. For visualization purposes, in the following we take $\calz=\bbr^2$. An \textit{autoencoder}~\cite{hinton_reducing_2006} is a composition $\dec\circ\enc$ of an \textit{encoder} $\enc:\calx\rightarrow\calz$ and a \textit{decoder} $\dec:\calz\rightarrow\calx$. We write $Z=\enc{X}\in\calz$ the representation of an input vector $X\in\calx$ and $\Tilde{X}=\dec(\enc{X})$ its reconstruction.
In practice, $\enc$ and $\dec$ are neural networks, e.g., with one or several fully connected layer(s), possibly with one or several additional convolutional layer(s) if $\calx$ represents an image space.
\begin{figure}
	\centering
	\def\svgwidth{.9\linewidth}
	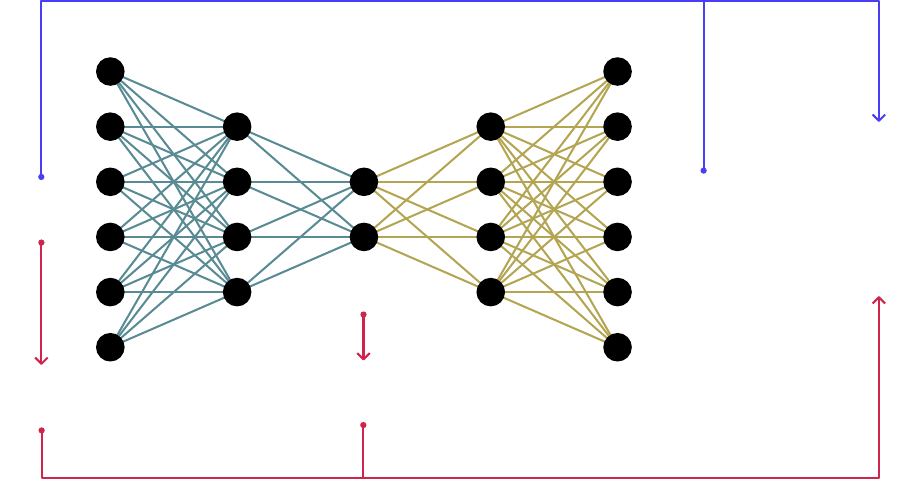
	\caption{An autoencoder with reconstruction term $\call_r$ and topological regularization term $\call_t$. Here both the encoder and the decoder are fully-connected networks with 1 hidden layer.}
	\label{fig:autoencoder}
\end{figure}
The
reconstruction loss
of an autoencoder is simply the MSE between an
input and its reconstruction:
\begin{equation}
	\call_r=\|\dec(\enc X)-X\|^2.
\end{equation}

For constraining autoencoders to preserve topological structures, one can add to
this reconstruction term a topological regularization term $\call_t$ which
compares some topological abstraction of the input and its low-dimensional
representation (\autoref{fig:autoencoder}).
Moore et al.~\cite{moor2020topological} discussed
a regularization term considering directly the lengths of
carefully selected edges:
\begin{equation} \label{eq:topoae}
	\begin{split}
\ltopoae{d}=&\|A^\inputPointCloud[\critical{d}(\inputPointCloud)]
-A^\latentPointCloud[\critical{d}(\inputPointCloud)]\|_2^2 +\\
		&\|A^\latentPointCloud[\critical{d}(\latentPointCloud)]-A^\inputPointCloud[\critical{d}(\latentPointCloud)]\|_2^2
	\end{split}
	,
\end{equation}
where $A^\inputPointCloud[\cdot]$ and $A^\latentPointCloud[\cdot]$ denote
evaluating the distance matrix of $\inputPointCloud$ and $\latentPointCloud$
respectively, and where $\critical{d}(\inputPointCloud)$ (resp.
$\critical{d}(\latentPointCloud)$) denotes all $\homologyDim$-critical edges,
for any $0\leq\homologyDim\leq d$, in $\inputPointCloud$ (resp.
$\latentPointCloud$), i.e., the edges that either create or kill a
\PH{\homologyDim} class of positive persistence (see
\autoref{sec:persistentHomology}).

In practice however, the method is only documented
for \PH{0}, so that
$\critical{0}(\inputPointCloud)=\emst(\inputPointCloud)$ and
$\critical{0}(\latentPointCloud)=\emst(\latentPointCloud)$ (see
\autoref{sec:geometricPH}).
In that case, the regularization term rewrites as:
\begin{equation} \label{eq:topoae0}
	\begin{split}
		\ltopoae{0}=&\|A^\inputPointCloud[\emst(\inputPointCloud)]-A^\latentPointCloud[\emst(\inputPointCloud)]\|_2^2+\\
		&\|A^\latentPointCloud[\emst(\latentPointCloud)]-A^\inputPointCloud[\emst(\latentPointCloud)]\|_2^2
	\end{split}.
\end{equation}
The idea of this loss function is to force the length of topologically
critical edges in the input to be preserved in the representation (first line of 
\autoref{eq:topoae0}), and reciprocally (second line). More precisely, the 
second term tries to push away in $\calz$ the ends of the edges in 
$\emst(\latentPointCloud)\setminus\emst(\inputPointCloud)$ that are, 
intuitively, too short in $\calz$. The hope is that at the end of the 
optimization, the length of both $\emst$'s edges match in high and low 
dimension, inducing the same 0-th persistence diagrams 
(which we show in \autoref{sec:topoae-wasserstein}).

Finally, the weights of the autoencoder are optimized iteratively to minimize the value of the loss function. Note that this involves computing the required critical edges of the current representation $\latentPointCloud$ at each iteration.

\section{Topological Autoencoder Analysis}
\label{sec:analysis}

In this section we provide a novel theoretical
analysis of the TopoAE loss function~\cite{moor2020topological} and its
link with the Wasserstein distance between the persistence diagrams in high and
low dimensions. Our aim is to better understand its limitations and the
configurations that may cause them.

\subsection{Relation to the Wasserstein distance}
\label{sec:topoae-wasserstein}

We first show below that the topological regularization term $\ltopoae{0}$ (\autoref{eq:topoae0})
upper bounds the $L_2$-Wasserstein distance $\calw_2$ between both 0-dimensional persistence diagrams.
\begin{lemma}
	\label{lemma:TopoAE0_bound}
	For any point clouds $\inputPointCloud$ and $\latentPointCloud$ of equal 
	size, we have the following inequality:
	\[\calw_2\bigl(\dgmrips{0}(\inputPointCloud),
	\dgmrips{0}(\latentPointCloud)\bigr)^2\leq\ltopoae{0}(
	\inputPointCloud,\latentPointCloud).\]
	Besides, under general position hypothesis (unique pairwise distances), if $\ltopoae{0}(\inputPointCloud,\latentPointCloud)=0$, then $\dgmrips{0}(\inputPointCloud)=\dgmrips{0}(\latentPointCloud)$ and the \PH{0} pairs are the same, i.e. $\emst(\inputPointCloud)=\emst(\latentPointCloud)$.
	\end{lemma}
	\begin{proof}
	Let $\calk$ be the (abstract) 1-dimensional simplicial complex
	$\calk=\emst(\inputPointCloud)\cup\emst(\latentPointCloud)$ (see \autoref{fig:wassersteinBound}). Let
	$\delta_\inputPointCloud:\calk\rightarrow\bbr$ (resp.
	$\delta_\latentPointCloud:\calk\rightarrow\bbr$) be
	the simplex diameter function in $\inputPointCloud$ (resp. in
	$\latentPointCloud$).
	\begin{figure}
		\centering
		\def\svgwidth{\linewidth}
		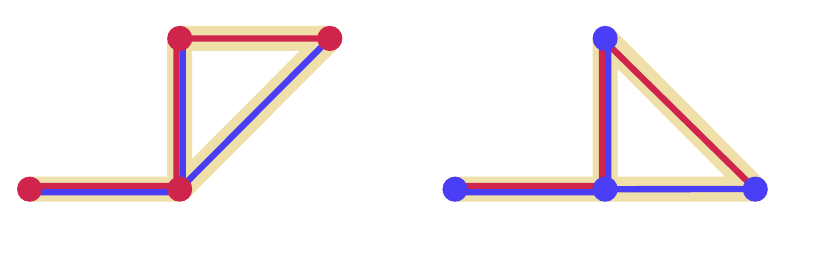
		\caption{Notations for the proof of \autoref{lemma:TopoAE0_bound}. In this example
		$\emst(\inputPointCloud)=\bigl\{\{v_1,v_2\},\{v_2,v_3\},\{v_3,v_4\}\bigr\}$ (in red) and
		$\emst(\latentPointCloud)=\bigl\{\{v_1,v_2\},$\allowbreak$\{v_2,v_3\},$\allowbreak$\{v_2,v_4\}\bigr\}$ (in blue). Therefore
		$\calk=\bigl\{\{v_1,v_2\},$\allowbreak$\{v_2,v_3\},$\allowbreak$\{v_2,v_4\},$\allowbreak$\{v_3,v_4\}\bigr\}$ (in yellow). Here $\ltopoae{0}(\inputPointCloud, \latentPointCloud)$ evaluates as
		$\bigl(\delta_X(\{v_2,v_4\})-\delta_Z(\{v_2,v_4\})\bigr)^2+
		 \bigl(\delta_X(\{v_3,v_4\})-\delta_Z(\{v_3,v_4\})\bigr)^2
		=2(\sqrt{2}-1)^2$.
		Note that this provides an example where
		$\ltopoae{0}(\inputPointCloud, \latentPointCloud)>0$ but
		$\dgmrips{0}(\inputPointCloud)=\dgmrips{0}(\latentPointCloud)$, i.e. the
		converse of the stated implication (\autoref{lemma:TopoAE0_bound})
is not true.}
		\label{fig:wassersteinBound}
	\end{figure}
	Because the minimum spanning tree contains exactly all the edges that kill a 0-dimensional persistent homology class, and because $\emst(\inputPointCloud)\subset\calk$ and $\emst(\latentPointCloud)\subset\calk$, we have $\dgmrips{0}(\inputPointCloud)=\dgm{0}(\delta_\inputPointCloud)$ and $\dgmrips{0}(\latentPointCloud)=\dgm{0}(\delta_\latentPointCloud)$.
	Hence, the following inequalities:
	\begin{align*}		
		&\calw_2\bigl(\dgmrips{0}(\inputPointCloud),\dgmrips{0}(\latentPointCloud)\bigr) ^ 2 = \calw_2\bigl(\dgm{0}(\delta_\inputPointCloud),
		\dgm{0}(\delta_\latentPointCloud)\bigr)^2 \\
		&\leq\sum\limits_{\substack{\sigma\in\calk \\ \dim(\sigma)\in\{0,1\}}}|\delta_\inputPointCloud(\sigma)-\delta_\latentPointCloud(\sigma)|^2 \text{\begin{tabular}{ll}
				(by Cellular Wasserstein \\
				stability theorem~\cite{skraba_wasserstein_2023})
		\end{tabular}}\\
		&=\sum\limits_{\substack{\sigma\in\calk \\ \dim(\sigma)=1}}|\delta_\inputPointCloud(\sigma)-\delta_\latentPointCloud(\sigma)|^2 \text{ (Rips filtration)}\\
		&\leq\sum\limits_{\substack{\sigma\in\emst(\inputPointCloud) \\ \dim(\sigma)=1}}|\delta_\inputPointCloud(\sigma)-\delta_\latentPointCloud(\sigma)|^2 + \sum\limits_{\substack{\sigma\in\emst(\latentPointCloud) \\ \dim(\sigma)=1}}|\delta_\inputPointCloud(\sigma)-\delta_\latentPointCloud(\sigma)|^2\\
		&=\text{\small{$\|A^\inputPointCloud[\critical{0}(\inputPointCloud)]-A^\latentPointCloud[\critical{0}(\inputPointCloud)]\|_2^2+\|A^\latentPointCloud[\critical{0}(\latentPointCloud)]-A^\inputPointCloud[\critical{0}(\latentPointCloud)]\|_2^2$}}\\
		&=\ltopoae{0}(\inputPointCloud,\latentPointCloud).
	\end{align*}
	In particular, when $\ltopoae{0}(\inputPointCloud,\latentPointCloud)=0$, then $\dgmrips{0}(\inputPointCloud)=\dgmrips{0}(\latentPointCloud)$. In that case, general position hypothesis (unique pairwise distances) on $\inputPointCloud$ and $\latentPointCloud$ -- that guarantees the uniqueness of their respective $\emst$ -- implies $\emst(\inputPointCloud)=\emst(\latentPointCloud)$, hence the same \PH{0} pairs.
\end{proof}

This result is, to our knowledge, the first theoretical
guarantee that justifies
using the original, $0$-dimensional version of the TopoAE
loss function (\autoref{eq:topoae0}) as a regularization term for preserving
\PH{0}. In particular, if this term is zero, we know that the representation
$\latentPointCloud$ has the same \PH{0} -- and therefore the same 0-th
persistence diagram -- than the input $\inputPointCloud$.

\subsection{Counter-example for higher dimensional PH}
Unfortunately, \autoref{lemma:TopoAE0_bound} does not generalize to higher-dimensional \PH{}.
Indeed, as soon as we add the edges that create or destroy \PH{1}
classes (\autoref{sec:topologicalAutoencoders}), a
zero $\ltopoae{1}$ loss (\autoref{eq:topoae}) no longer implies
equality of the persistence diagrams:
\begin{equation}
	\label{eq:counter-ex}
	\ltopoae{1}(\inputPointCloud,\latentPointCloud)=0 \centernot\implies\left\{
	\begin{array}{l}
		\dgmrips{0}(\inputPointCloud)=\dgmrips{0}(\latentPointCloud)\\
		\dgmrips{1}(\inputPointCloud)=\dgmrips{1}(\latentPointCloud)
	\end{array}
	\right..
\end{equation}

\begin{figure}
	\centering
	\def\svgwidth{\linewidth}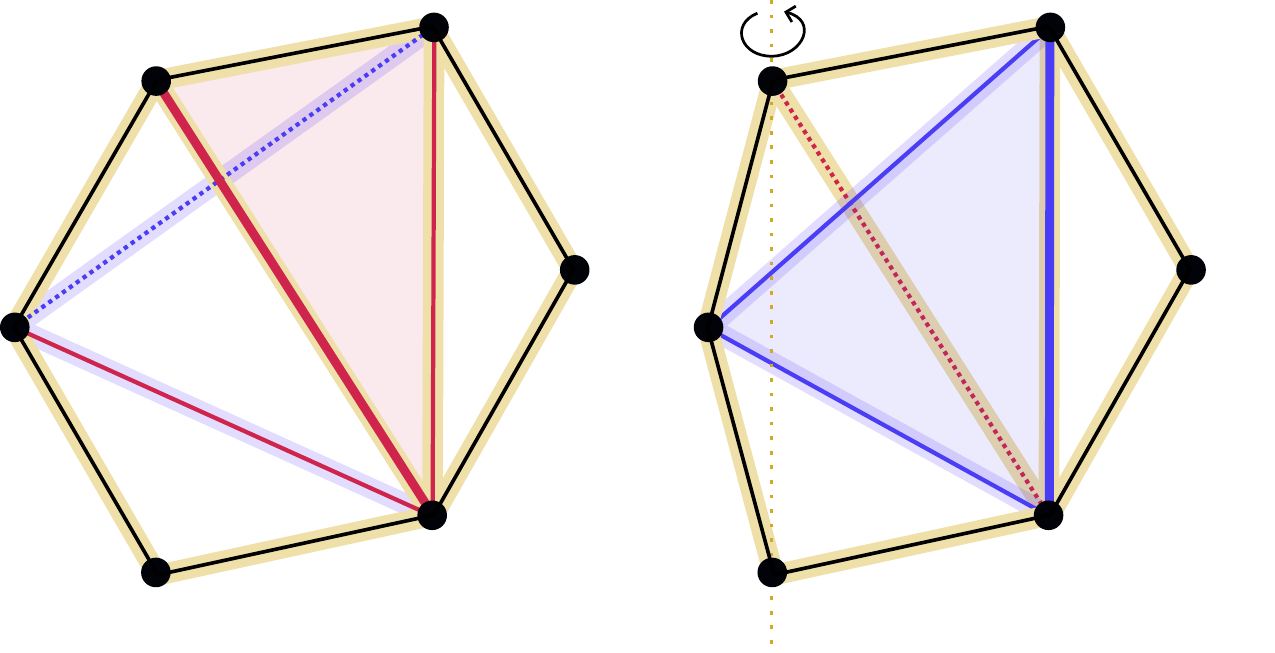
	\caption{An example for the non-implication of
	\autoref{eq:counter-ex}. The
	left point cloud $\latentPointCloud$ is in $\bbr^2$ and is a deformed regular
	hexagon; the right one $\inputPointCloud$ is in $\bbr^3$ and its leftmost point
	has been rotated with respect to the dotted yellow axis with angle $\pi/3$ (see \autoref{appendix:counter-example-data} for raw coordinates).
	Both $\rng$s are pictured in black and the length of their edges are 1.
	The triangle that kills the cycle in $\inputPointCloud$ (resp.
	$\latentPointCloud$) is shown in blue (resp. red) with its longest edge
	highlighted in bold. For $\latentPointCloud$, the remaining cascade edge
	(see \autoref{eq:GCS}) is also shown (its length is 1.623).
	The 0-th persistence diagrams are the same: 
	$\dgmrips{0}(\inputPointCloud)=\dgmrips{0}(\latentPointCloud)=\{(0,1)\text{ with
	multiplicity 5}, (0,+\infty)\}$; but not the 1-th persistence diagrams:
	$\dgmrips{1}(\latentPointCloud)=\{(1,1.819)\}$ and
	$\dgmrips{1}(\inputPointCloud)=\{(1,1.732)\}$. However,
	the critical edges considered in $\ltopoae{1}$ (i.e., $\rng$ edges
	and 1-cycle killing edges, highlighted in yellow) have the
	same length in both $\inputPointCloud$ and $\latentPointCloud$, therefore
	$\ltopoae{1}(\inputPointCloud,\latentPointCloud)=0$. On the contrary, the two
	additional edges considered in the 1-dimensional \emph{cascade distortion}
	(highlighted in purple, see \autoref{eq:cascae})
	are longer in $\latentPointCloud$ than in $\inputPointCloud$,
	hence $\lcascae{1}(\inputPointCloud,\latentPointCloud)>0$.}
	\label{fig:counter-example}
\end{figure}

\autoref{fig:counter-example} provides a detailed counter-example for two point
clouds $\inputPointCloud\subset\bbr^3$ and $\latentPointCloud\subset\bbr^2$
which illustrates this non-implication.
This example shows that preserving the length
of topologically critical edges is not enough to guarantee the preservation
of $\PH{\homologyDim}$ when $\homologyDim\geq1$.

\section{Cascade Distortion}
\label{sec:newLoss}

This section presents a new penalty term called \emph{cascade distortion} (CD) which addresses the above counter-example.
This term favors an isometric embedding of the $2$-chains filling persistent
$1$-cycles
(i.e., that are computed thanks to the \texttt{PairCells}
algorithm), leading to a more accurate geometrical representation of
the persistent $1$-cycles in low dimensions.

\subsection{Formulation}
\label{sec:newLoss:formulation}

Let $\calp^k$ be the set of \PH{k} pairs with a positive persistence, i.e.,
$\calp^k=\{(\tau,\sigma)\in\simplicesofdim{\calk}{k}\times\simplicesofdim{\calk}{k+1}\mid
\partner[\tau]=\sigma\text{ and }\delta(\tau)<\delta(\sigma)\}$.
We now define
the $d$-dimensional \textit{global cascade skeleton} ($\skelcasc^d$) as the set
of edges that appear in the cascade of any \PH{k} pair 
with $0\leq k\leq d$:
\begin{equation} \label{eq:GCS}
\skelcasc^d(\inputPointCloud)=
\bigcup\limits_{k=0}^d\bigcup\limits_{(\tau,\sigma)\in\calp^k}\cascade[
\sigma]^{(1)}.
\end{equation}
In \autoref{fig:eliminateBoundaries},
$\skelcasc^1(\inputPointCloud)$ appears in the bottom-right image as the set of
red edges.
This leads us to propose
the following $d$-dimensional \textit{Cascade Distortion} (CD) loss function,
which looks similar to the original TopoAE
loss (\autoref{eq:topoae}), but which constrains the length of all 
the edges appearing within the global cascade of $\inputPointCloud$, instead of its critical edges only:

\begin{equation} \label{eq:cascae}
\begin{split}
	\lcascae{d}(\inputPointCloud,\latentPointCloud)=&\|A^\inputPointCloud[\skelcasc^d(\inputPointCloud)]-A^\latentPointCloud[\skelcasc^d(\inputPointCloud)]\|_2^2+\\&\|A^\latentPointCloud[\critical{d}(\latentPointCloud)]-A^\inputPointCloud[\critical{d}(\latentPointCloud)]\|_2^2
\end{split}.
\end{equation}
In practice, we focus our study on $\lcascae{1}$ for visualization purposes.
Intuitively, the role on the first term (top line of \autoref{eq:cascae})
is to favor an isometric embedding of the input cycles and the 2-chains that
fill them, so that their shapes are preserved at best.
The second term (bottom line of \autoref{eq:cascae}), like in
\tAE, involves pushing away vertices that are too close in $\latentPointCloud$,
hence penalizing the cycles that exist in $\latentPointCloud$ but not in
$\inputPointCloud$.
This way, the first term favors a faithful projection of the high-dimensional cycles, while the second term removes spurious low-dimensional cycles from the planar projection.
In addition, this loss enables the usage of a dedicated algorithm for the fast
computation of $\critical{1}(\latentPointCloud)$ at each
optimization iteration (\autoref{sec:algorithms}).

In practice, $\skelcasc^1(\inputPointCloud)$ is computed using a modified
version of \texttt{PairCells}, which, similarly to~\cite{Iuricich22},
uses a two-step approach.
First, the persistence pairs $\calp^1$ with a positive persistence are computed
with an efficient, dedicated algorithm, e.g., \textit{Ripser}~\cite{bauer_ripser_2021}.
Then, for each pair $(\tau,\sigma)\in\calp^1$ in ascending order, we run the
\texttt{EliminateBoundary} procedure (\autoref{algo:eliminateBoundaries}) from $\sigma$ until the youngest
edge of the boundary $\partial(\cascade[\sigma])$ is $\tau$, storing the cascade
edges encountered in the meantime. At each step,
$\text{Youngest}\bigl(\partial(\cascade[\sigma])\bigr)$ forms either an
apparent pair (with a triangle that is searched at this moment), or a positive persistence
pair that was handled previously.
Note that it is mandatory to choose the same order as \textit{Ripser} on the
triangles with identical diameter (i.e., the reverse colexicographic order).
This is more efficient than a naive execution of \texttt{PairCells},
since it benefits from the accelerations of \textit{Ripser}.

\subsection{Relation to TopoAE}
\label{sec:relationTopoAECascadeAE}

The above \textit{Cascade Distortion} term (\autoref{eq:cascae}) can be understood as a generalization of the original TopoAE topological regularization term in that when restricting to \PH{0}, both formulations are equivalent.
\begin{lemma}
	For any point cloud $\inputPointCloud$, $\skelcasc^0(\inputPointCloud)=\emst(\inputPointCloud)$.
	Therefore, the 0-dimensional CD and TopoAE regularization
	terms are equal:
	$\ltopoae{0}(\inputPointCloud, \latentPointCloud)=\lcascae{0}(\inputPointCloud, \latentPointCloud)$.
\end{lemma}
\begin{proof}
	Let $\inputPointCloud$ be a point cloud. First note that for any $\sigma$, we have $\sigma\in\cascade[\sigma]$, hence:
	\[
		\underbrace{\emst(\inputPointCloud)
		= \bigcup\limits_{(\tau,\sigma)\in\calp^0}\{\sigma\}}
	    _{\text{see \autoref{sec:geometricPH}}}
		\subset\underbrace{\bigcup\limits_{(\tau,\sigma)\in\calp^0}\cascade[\sigma]
		=\skelcasc^0(\inputPointCloud)}
		_{\text{by definition (\autoref{eq:GCS})}}.
	\]
	Reciprocally, note that for any vertex $\tau$, $\partner[\tau]$ is an edge that kills a \PH{0} class, hence is an $\emst$ edge.
	Then, with \autoref{algo:eliminateBoundaries:propagation}
	(\autoref{algo:eliminateBoundaries}), $\cascade[\sigma]$ writes as the
	sum of $\sigma$ and cascades of edges of the form $\partner[\tau]$.
	Therefore, by recurrence on the edges by increasing length, for any $\sigma\in\emst(X)$, i.e., for any $(\tau,\sigma)\in\calp^0$, we have $\cascade[\sigma]\subset\emst(\inputPointCloud)$.
	Hence $\skelcasc^0(\inputPointCloud)\subset\emst(\inputPointCloud)$.
	In the end, $\skelcasc^0(\inputPointCloud)=\emst(\inputPointCloud)$.
\end{proof}
For higher-dimensional \PH{} however, we can only say that
$\ltopoae{d}(\inputPointCloud,
\latentPointCloud)\leq\lcascae{d}(\inputPointCloud, \latentPointCloud)$, i.e.,
the latter is more restrictive than the former.
For instance, for $d=1$, in the counter-example of \autoref{fig:counter-example} where we had $\dgmrips{1}(\inputPointCloud)\neq\dgmrips{1}(\latentPointCloud)$ but $\ltopoae{1}(\inputPointCloud,\latentPointCloud)=0$, the 1-dimensional cascade distortion is here indeed positive : $\lcascae{1}(\inputPointCloud,\latentPointCloud)>0$.
More generally, $\lcascae{1}$ additionally constrains the length of the edges of the 2-chains that fill
the persistent 1-cycles, hence favoring an isometric embedding of those chains.
Intuitively, this can be understood as trying to preserve the whole
\textit{shape} of the 1-cycles, instead of only their birth and death.
\revision{Although $\lcascae{2}$ could be similarly understood as a way to preserve the \emph{shape} of the cavities when projecting to $\bbr^{3}$, this paper does not further investigate the use of cascade distortion for $d>1$, as it would be much more computationally expensive.}

In the following, we call TopoAE++ the dimensionality reduction
technique obtained by replacing in TopoAE the topological regularization term
$\ltopoae{0}$ with $\lcascae{1}$, and by including the acceleration
presented in the next section.

\section{Fast algorithm for Rips PH in the plane}
\label{sec:algorithms}

Because the computation of the persistence diagram of the low-dimensional representation is executed at each step of the optimization process, it is a critical stage in terms of computation time. Therefore, we propose in this section a new fast algorithm that computes \PH{} for the Rips filtration of a two-dimensional point cloud, which is useful when our embedding space is $\bbr^2$, e.g., for visualization purpose.
This algorithm is purely geometric and involves no matrix reduction step,
contrary to other algorithms (e.g. \cite{maria2014gudhi, bauer2017phat,
bauer_ripser_2021, koyama2023reduced}),
which allows it to be significantly
faster and less memory-intensive (see \autoref{sec:performance:rips2d} for timing results).

\subsection{Geometrical results}
\label{sec:algorithm:geometry}

\begin{figure}
	\centering
	\includegraphics[width=.99\linewidth]{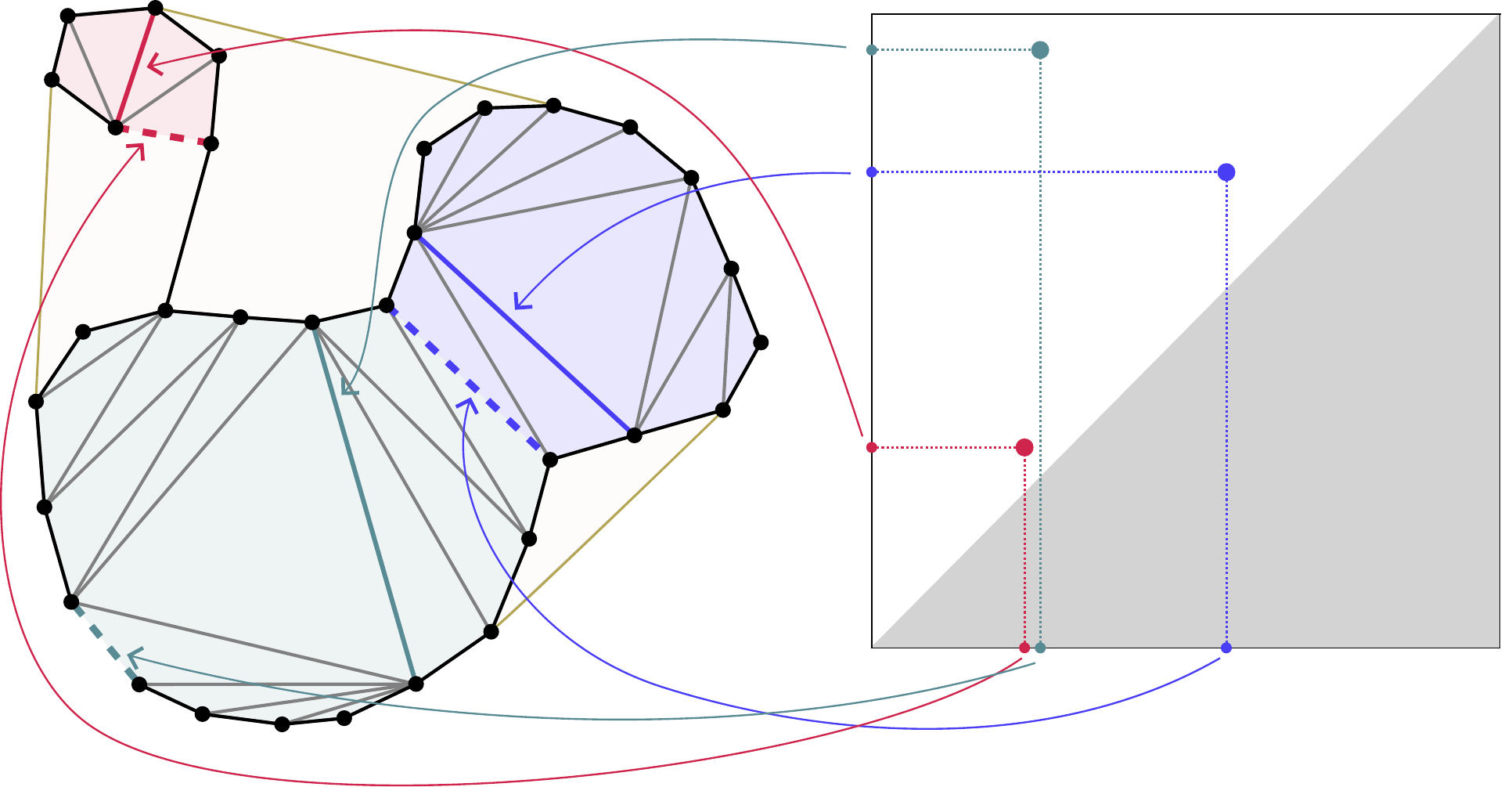}
	\caption{$\rng$-polygons (green, blue and red polygons) with their minmax length triangulations (gray edges). The arrows show to which critical edges (birth: dashed; death: continuous) is associated each point of the 1-dimensional persistence diagram. Each birth edge is in the $\rng$ and each death edge is the longest edge of the minmax length triangulations of its $\rng$-polygon. Missing edges from the convex hull are shown in yellow, as well as the remaining polygons that the convex hull forms with the $\rng$.}
	\label{fig:MMLT-PD}
\end{figure}

We rely on the fact that the edges that introduce a cycle are exactly the edges in $\rng(\inputPointCloud)\setminus\emst(\inputPointCloud)$ (as previously exploited in the related work~\cite{koyama2023reduced}, see \autoref{sec:geometricPH}). Consequently, the number of points in the 1-th persistence diagram is the number of polygons formed by $\rng(\inputPointCloud)$, which is also the cardinal of $\rng(\inputPointCloud)\setminus\emst(\inputPointCloud)$. In contrast to previous work~\cite{koyama2023reduced} (which also exploited boundary matrix reduction), we additionally claim that the critical edges that kill these \PH{1} classes
can be found efficiently through geometric -- rather than algebraic -- considerations, namely with a minmax length triangulation of each $\rng$-polygons (see \autoref{fig:MMLT-PD}).

\begin{figure}
	\centering
	\def\svgwidth{\linewidth}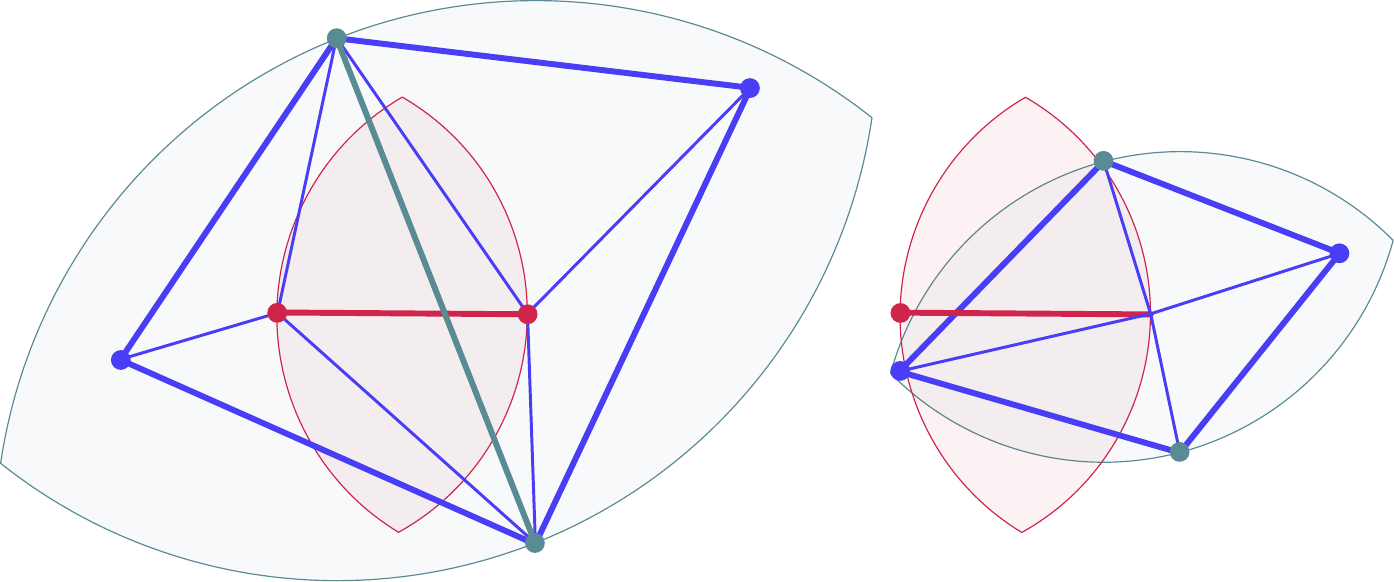
	\caption{Notations for proof of \autoref{lemma:interpolygonal-edges}.
	\revision{$pq$} (in red) is supposed to be an $\rng$ edge, hence an empty lens. An edge $ab$ (light
	green) that crosses \revision{$pq$} is shown to not be able to kill any \PH{1} class of positive persistence.}
	\label{fig:interpolygonal-edges}
\end{figure}

\begin{lemma}
	\label{lemma:interpolygonal-edges}
	In the plane, an edge that intersects an $\rng$ edge kills no \PH{1}
class of positive persistence.
\end{lemma}
\begin{proof}[\revision{Proof outline}]
	\revision{Let $ab$ be an edge that intersects an $\rng$ edge $pq$.
	Suppose that $ab$ kills a \PH{1} class $\gamma$ of positive persistence.
	Then $ab$ is the longest edge of two triangles $abc$ and $abd$, where $acbd$ is a representative 1-cycle of $\gamma$, and such that $c$ and $d$ are on both sides of $\lens(a,b)$.
	We show that it is always possible to construct a 2-chain whose boundary is $acbd$ and that contains only triangles of diameter $<|ab|$.
	This 2-chain depends on whether both $p$ and $q$ belong to $\lens(a,b)$ (\autoref{fig:interpolygonal-edges}, left) or only one of them does (\autoref{fig:interpolygonal-edges}, right).
	Hence, the class $\gamma$ is killed before the value $|ab|$, therefore not by $ab$.
	See Appendix~\ref{appendix:proof} for the full proof.}
\end{proof}

Therefore, the edges that kill a \PH{1} class of positive persistence can only
lie inside either an $\rng$-polygon, or a polygon formed by the $\rng$ and the
convex hull of $\inputPointCloud$ (\autoref{fig:MMLT-PD}).
Then, we have the following lemma for each $\rng$-polygon.

\begin{lemma}
	\label{lemma:MML-edges}
	Let $\Pi$ be a $\rng$-polygon. Under general position hypothesis (unique pairwise distances), the longest edge $\edgeMML$, of length noted $\delta_{\mml}$, of any minmax length triangulation of $\Pi$ kills a \PH{1} class $\gamma$ with positive persistence for which $\Pi$ is a representative.
\end{lemma}
\begin{proof}
	Let $\Pi$ be a $\rng$-polygon with point set $P$. Let $\edgeMML$ be the
	longest edge of the minmax length triangulations of this polygon, which is well-defined with the general position hypothesis. We have the
	following:
	\begin{itemize}
		\item for diameter threshold $r=\delta_{\mml}$, $\rips_{r}(P)$ contains a $\mml$ triangulation of $\Pi$, which can be seen as a 2-chain whose boundary is $\Pi$;
		\item for any diameter threshold $r<\delta_{\mml}$, $\rips_{r}(P)$ contains no 2-chain whose boundary is $\Pi$: otherwise, one could construct
 		a triangulation of $\Pi$ whose longest edge is of length at most $r$, hence smaller than $\delta_{\mml}$.
	\end{itemize}
	Hence, $\edgeMML$ kills a \PH{1} class $\gamma$ with positive persistence for which $\Pi$ is a representative.
\end{proof}

Therefore, in the plane, as the number of $\rng$-polygons is exactly the number of \PH{1} classes with a positive persistence, the edges that destroy the latter are exactly the edges found in the above lemma, i.e., that are the longest of the $\mml$ triangulations of an $\rng$-polygon (see \autoref{fig:MMLT-PD}).

Finally, for any $\rng$-polygon, we state a relation between $\deathValue$
and the length $\delta_{\dr}$ of the longest Delaunay edge $\DRKiller$ inside this polygon,
enabling the search of
$\ripsKiller$ among a smaller subset of the polygon's
diagonals (see \autoref{sec:algorithm:description}).

\begin{figure}
	\centering
	\def\svgwidth{.8\linewidth}
\begingroup%
  \makeatletter%
  \providecommand\color[2][]{%
    \errmessage{(Inkscape) Color is used for the text in Inkscape, but the package 'color.sty' is not loaded}%
    \renewcommand\color[2][]{}%
  }%
  \providecommand\transparent[1]{%
    \errmessage{(Inkscape) Transparency is used (non-zero) for the text in Inkscape, but the package 'transparent.sty' is not loaded}%
    \renewcommand\transparent[1]{}%
  }%
  \providecommand\rotatebox[2]{#2}%
  \newcommand*\fsize{\dimexpr\f@size pt\relax}%
  \newcommand*\lineheight[1]{\fontsize{\fsize}{#1\fsize}\selectfont}%
  \ifx\svgwidth\undefined%
    \setlength{\unitlength}{476.70562744bp}%
    \ifx\svgscale\undefined%
      \relax%
    \else%
      \setlength{\unitlength}{\unitlength * \real{\svgscale}}%
    \fi%
  \else%
    \setlength{\unitlength}{\svgwidth}%
  \fi%
  \global\let\svgwidth\undefined%
  \global\let\svgscale\undefined%
  \makeatother%
  \begin{picture}(1,0.99181982)%
    \lineheight{1}%
    \setlength\tabcolsep{0pt}%
    \put(0,0){\includegraphics[width=\unitlength,page=1]{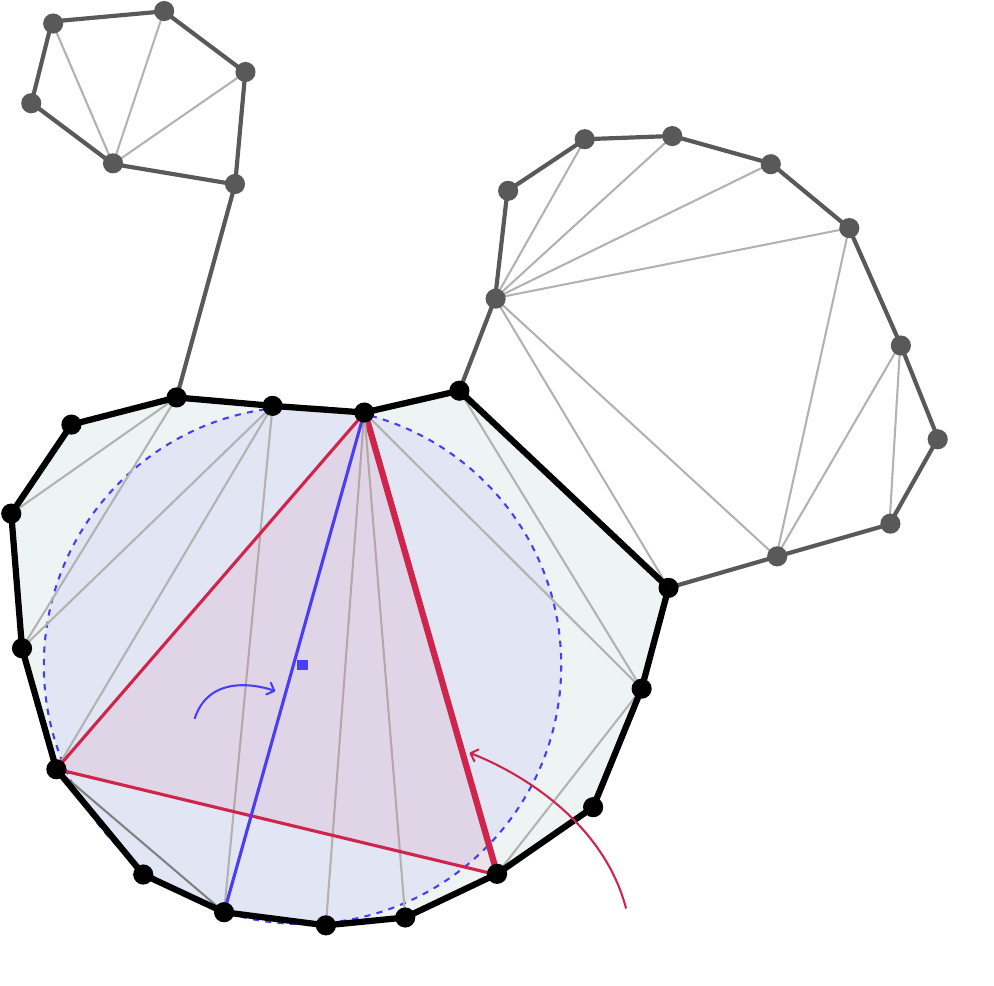}}%
    \put(0.16217264,0.47421874){\color[rgb]{0.28627451,0.23921569,0.96078431}\makebox(0,0)[t]{\lineheight{1.25}\smash{\begin{tabular}[t]{c}$\circumcircle$\end{tabular}}}}%
    \put(0.68098003,0.22217565){\color[rgb]{0,0,0}\makebox(0,0)[t]{\lineheight{1.25}\smash{\begin{tabular}[t]{c}$\Pi$\end{tabular}}}}%
    \put(0.18873553,0.22988344){\color[rgb]{0.28627451,0.23921569,0.96078431}\makebox(0,0)[t]{\lineheight{1.25}\smash{\begin{tabular}[t]{c}$\DRKiller$\end{tabular}}}}%
    \put(0.4911976,0.36424313){\color[rgb]{0.81176471,0.14117647,0.29411765}\makebox(0,0)[t]{\lineheight{1.25}\smash{\begin{tabular}[t]{c}$\ripsKiller$\end{tabular}}}}%
    \put(0.36645529,0.31154744){\color[rgb]{0.28627451,0.23921569,0.96078431}\makebox(0,0)[t]{\lineheight{1.25}\smash{\begin{tabular}[t]{c}$c(\circumcircle)$\end{tabular}}}}%
    \put(0.71355971,0.02166614){\color[rgb]{0.81176471,0.14117647,0.29411765}\makebox(0,0)[t]{\lineheight{1.25}\smash{\begin{tabular}[t]{c}$|\DRKiller|>|\ripsKiller|\geq\frac{\sqrt{3}}{2}|\DRKiller|$\end{tabular}}}}%
  \end{picture}%
\endgroup%

	\caption{Notations for the proof of \autoref{lemma:bounded_rips_delrips}.
	The $\rng$-polygon $\Pi$ (bold black), is such that
	its Delaunay triangulation (in gray) is not a $\mml$ triangulation. Hence,
	its cycle killing edge $\ripsKiller$ (bold red) is shorter than its
	longest Delaunay edge $\DRKiller$ (blue), i.e., $\deathValue < \delta_{\dr}$.
	The circumcircle $\circumcircle$ of the longest Delaunay edge $\DRKiller$ contains
	no points of $\inputPointCloud$ in its interior.}
	\label{fig:bounded_rips_delrips}
\end{figure}

\begin{lemma}
	\label{lemma:bounded_rips_delrips}
	In the plane,
	the values $\deathValue$ and $\delta_{\dr}$ of an $\rng$-polygon verify:
	\[\frac{\sqrt{3}}{2}\delta_{\dr}\leq\deathValue\leq\delta_{\dr}.\]
\end{lemma}
\begin{proof}
	Let $\Pi$ be an $\rng$-polygon and $P$ its point set. Let $\DRKiller\in\del(P)$ be the
	longest Delaunay edge inside $\Pi$, of length $\delta_{\dr}=|\DRKiller|$.
	The right inequality results from the Delaunay triangulation inside $\Pi$ being a 2-chain whose boundary is $\Pi$ and whose maximum diameter is $\delta_{\dr}$: hence the associated Rips \PH{1} class $\gamma$ is killed at value $\deathValue\leq\delta_{\dr}$.
	For the left inequality, because $\DRKiller$ is a Delaunay edge, there exists
	a circumcircle $\circumcircle$ to $\DRKiller$ of radius
	$r(\circumcircle)\geq|\DRKiller|/2$ that contains no other point of $P$ (see
	\autoref{fig:bounded_rips_delrips}). Now any 2-chain with $\Pi$ as boundary has
	to contain the center $c(\circumcircle)$ of that circle. However, the smallest
	triangles with no vertex within $\circumcircle$ and that contain $c(\circumcircle)$ are exactly the equilateral
	triangles circumscribed by $\circumcircle$, and their side length is
	$\sqrt{3}r(\circumcircle)\geq\frac{\sqrt{3}}{2}|\DRKiller|$.
	Therefore, the associated Rips \PH{1} class
	$\gamma$ is killed by a triangle $\tau$ of diameter at least
	$\frac{\sqrt{3}}{2}|\DRKiller|$, hence
	$\frac{\sqrt{3}}{2}\delta_{\dr}\leq\deathValue$.
\end{proof}

\subsection{Algorithm description}
\label{sec:algorithm:description}

Leveraging the above results, we introduce a fast algorithm that computes
\PH{} for the Rips filtration of a 2-dimensional point cloud. It is
summarized in \autoref{algo:2dpersistence} and we describe below more precisely
the different stages.

\begin{algorithm}
	\DontPrintSemicolon
	\label{algo:2dpersistence}\caption{2D Rips persistence method overview}
	\KwIn{Point cloud $X\subset\bbr^2$}
	\KwOut{Persistence diagrams $\dgmrips{0}(\inputPointCloud)$ and $\dgmrips{1}(\inputPointCloud)$ with associated simplices}
	Compute the Delaunay triangulation $\del(\inputPointCloud)$\;
	Compute the Urquhart graph $\ug(\inputPointCloud)$\;
	Apply Kruskal's algorithm on $\ug(\inputPointCloud)$ to compute
$\emst(\inputPointCloud)$ and obtain $\dgmrips{0}(X)$\;
	Check the emptiness of the lens of the edges in $\ug(\inputPointCloud)\setminus\emst(\inputPointCloud)$ to compute $\rng(\inputPointCloud)$\;
	For each $\rng$-polygon, find its local minmax length triangulation\revision{'s} longest edge (\autoref{algo:findMMLedge})\;
	Apply Kruskal's algorithm on the dual $\rng(\inputPointCloud)$ graph
to obtain $\dgmrips{1}(X)$\;
\end{algorithm}

\begin{algorithm}
	\DontPrintSemicolon
	\label{algo:findMMLedge}\caption{Find the longest edge of a minmax length triangulation of an $\rng$-polygon of $\inputPointCloud$}
	\KwIn{Points $P\subset\inputPointCloud$ of an $\rng$-polygon}
	\KwIn{Length $\delta_{\dr}$ of the longest Delaunay edge}
	\KwOut{Edge that destroy the associated 1-cycle}
	$\text{candidates}\gets\{\}$\;
	\For{$e=v_1v_2\in\binom{P}{2}$ s.t. $|e|\in\left[\frac{\sqrt{3}}{2}\delta_{\dr},\delta_{\dr}\right]$}{
		\tcp{check if $e$ is a 2-edge}
		\If{$\llens(e)\neq\varnothing$ and $\rlens(e)\neq\varnothing$} {
			\tcp{check if $e$ is expandable}
			\For{$x\in\llens(e)$}{
				\If{$\rlens(v_1x)=\varnothing$ and $\rlens(xv_2)=\varnothing$}{
					$\text{LeftExpandable}\gets$ true\;
					break\;
				}
			}
			\For{$y\in\rlens(e)$}{
				\If{$\rlens(v_2y)=\varnothing$ and $\rlens(yv_1)=\varnothing$}{
					$\text{RightExpandable}\gets$ true\;
					break\;
				}
			}
			\If{$\text{\normalfont{LeftExpandable}}$ and $\text{\normalfont{RightExpandable}}$} {
				$\text{candidates}\gets\text{candidates}\cup\{e\}$\;
			}
		}
	}
	\Return{$\min(\text{candidates}, |\cdot|)$}
\end{algorithm}

\paragraph{Delaunay triangulation} We use
CGAL~\cite{fabri_cgal_2009} to compute the Delaunay triangulation,
with identifiers on vertices and
triangles.
This implementation deals with the boundary (simplices on the convex hull) with a
virtual point at infinity.

\paragraph{Relative neighborhood graph} To compute $\rng(\inputPointCloud)$, we first compute
$\ug(\inputPointCloud)$ by keeping Delaunay edges whose link (made of two points) is outside their
lens. We then apply Kruskal's algorithm (which uses a
union-find data structure~\cite{cormen}) on $\ug(\inputPointCloud)$ to compute
$\emst(\inputPointCloud)$,
of which $\dgmrips{0}(X)$ can be deduced
(see \autoref{sec:geometry} and \autoref{sec:geometricPH}). Then, checking the
emptiness of the lens of the edges in $\ug(\inputPointCloud)\setminus\emst(\inputPointCloud)$ using disk queries on
a precomputed k-d tree, we determine whether they belong to $\rng(\inputPointCloud)$, and
construct a data structure that represents $\rng(\inputPointCloud)$ and its polygons (see next
paragraph).

\paragraph{Data structure for $\rng$-polygons} During the computation of $\ug(\inputPointCloud)$ and $\rng(\inputPointCloud)$, we maintain a union-find data structure on Delaunay triangles that permits to deduce efficiently the $\rng$-polygons. More precisely, as soon as we encounter a non-$\rng$ edge (either during $\ug(\inputPointCloud)$ computation or during $\rng(\inputPointCloud)$ refinement), we merge (if not already merged) the two classes associated with its two adjacent triangles. We also keep for each current class the longest deleted Delaunay edge, which needs to be maintained at each merge. Besides, we store $\ug(\inputPointCloud)$ and $\rng(\inputPointCloud)$ edges as quad-edges (i.e., an edge is stored as the identifiers of its two extremities and the identifiers of its two adjacent triangles). This allows an implicit representation of the $\rng$-polygons, each of which is associated with a root of the union-find data structure over triangles. For the next stage, we also recover the vertices associated with each polygon.

\begin{figure}
	\centering
	\def\svgwidth{.75\linewidth}
\begingroup%
  \makeatletter%
  \providecommand\color[2][]{%
    \errmessage{(Inkscape) Color is used for the text in Inkscape, but the package 'color.sty' is not loaded}%
    \renewcommand\color[2][]{}%
  }%
  \providecommand\transparent[1]{%
    \errmessage{(Inkscape) Transparency is used (non-zero) for the text in Inkscape, but the package 'transparent.sty' is not loaded}%
    \renewcommand\transparent[1]{}%
  }%
  \providecommand\rotatebox[2]{#2}%
  \newcommand*\fsize{\dimexpr\f@size pt\relax}%
  \newcommand*\lineheight[1]{\fontsize{\fsize}{#1\fsize}\selectfont}%
  \ifx\svgwidth\undefined%
    \setlength{\unitlength}{454.27053833bp}%
    \ifx\svgscale\undefined%
      \relax%
    \else%
      \setlength{\unitlength}{\unitlength * \real{\svgscale}}%
    \fi%
  \else%
    \setlength{\unitlength}{\svgwidth}%
  \fi%
  \global\let\svgwidth\undefined%
  \global\let\svgscale\undefined%
  \makeatother%
  \begin{picture}(1,0.98697319)%
    \lineheight{1}%
    \setlength\tabcolsep{0pt}%
    \put(0,0){\includegraphics[width=\unitlength,page=1]{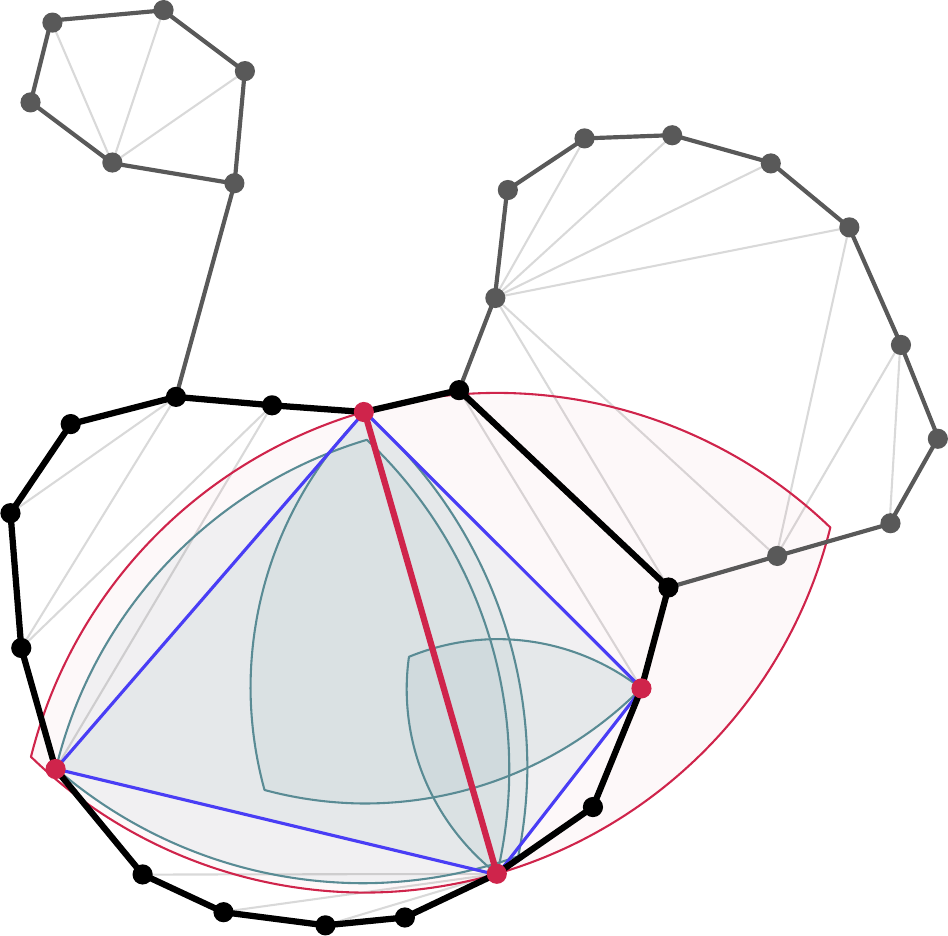}}%
    \put(0.09807649,0.58686482){\color[rgb]{0,0,0}\makebox(0,0)[t]{\lineheight{1.25}\smash{\begin{tabular}[t]{c}$\Pi$\end{tabular}}}}%
    \put(0.38443897,0.57952511){\color[rgb]{0.81176471,0.14117647,0.29411765}\makebox(0,0)[t]{\lineheight{1.25}\smash{\begin{tabular}[t]{c}$v_2$\end{tabular}}}}%
    \put(0.54548885,0.01888209){\color[rgb]{0.81176471,0.14117647,0.29411765}\makebox(0,0)[t]{\lineheight{1.25}\smash{\begin{tabular}[t]{c}$v_1$\end{tabular}}}}%
    \put(0.72624676,0.24308233){\color[rgb]{0.81176471,0.14117647,0.29411765}\makebox(0,0)[t]{\lineheight{1.25}\smash{\begin{tabular}[t]{c}$y$\end{tabular}}}}%
    \put(0.0352256,0.12346433){\color[rgb]{0.81176471,0.14117647,0.29411765}\makebox(0,0)[t]{\lineheight{1.25}\smash{\begin{tabular}[t]{c}$x$\end{tabular}}}}%
  \end{picture}%
\endgroup%

	\caption{Illustration of the expandability -- tested by
	\autoref{algo:findMMLedge} -- of the 2-edge $e=v_1v_2$. Indeed there exists
	$x,y\in\lens(e)$ (red area) on either side of $e$ such that the
	"inner" half-lenses (green areas) of the edges $v_1x$, $xv_2$, $v_2y$,
	$yv_1$ are empty.
	Thus, there exists a planar triangulation of $\Pi$ whose longest edge is $e$.
	Besides, $e$ happens to be the smallest of the expandable 2-edges in
	$\Pi$: $e$ is therefore the longest edge of the $\mml$ triangulations of $\Pi$.}
	\label{fig:expandability}
\end{figure}

\paragraph{Minmax length triangulations} Now all we need is to find the  edges that kill persistent cycles: using \autoref{lemma:MML-edges}, we search the longest edge $\edgeMML$ of a $\mml$ triangulation of each polygon. For that we partially follow the algorithm described in~\cite{edelsbrunner_quadratic_1993} that computes $\mml$ triangulations in quadratic time, even though there is no need here to compute a full triangulation since we only need its longest edge. Therefore for each $\rng$-polygon, we enumerate the polygon diagonals with length between $\frac{\sqrt{3}}{2}\simeq0.866$ and 1 times the length of the longest deleted Delaunay edge within the polygon (see \autoref{lemma:bounded_rips_delrips}). Among them, we search the smallest that is an \emph{expandable 2-edge} (a property introduced in~\cite{edelsbrunner_quadratic_1993} that allows an edge to be the longest of a triangulation of the polygon, see \autoref{appendix:MMLTs} for more details)
using disk queries on a k-d tree on the vertices of the polygon (see \autoref{fig:expandability} and \autoref{algo:findMMLedge}).

\paragraph{Persistent pairs} The last step consists in pairing edges that create cycles with those that kill them (a cycle-creating edge is not necessarily associated with a killer edge of an adjacent polygon, in case of nested cycles).
We solve this relying on duality, an overall strategy also employed in
the seminal persistence algorithm~\cite{edelsbrunner02}.
Specifically, for our setup, we compute \PH{0} of the dual of $\rng(\inputPointCloud)$ with opposite filtration values, which is
the
\PH{1} of the input. In other words, we apply Kruskal's algorithm on a
dual graph where dual nodes are $\rng$-polygons $\Pi$ with filtration value
$-\deathValue(\Pi)$, and where dual edges are cycle-creating edges
$e\in\rng(\inputPointCloud)\setminus\emst(\inputPointCloud)$ with filtration value $-|e|$.

\subsection{Parallelization}
\label{sec:algorithm:parallelization}

Several stages can be parallelized for accelerating the running time. The most
critical stage is the computation of minmax length triangulations
(it represents around 50\% of the computation time on average in sequential for uniform point clouds). It can be
naively parallelized on the different $\rng$-polygons. However, these polygons
can be of varying sizes, leading to a poor speed-up in some cases.
Instead, we develop a slightly more efficient parallelization, which consists in storing in a single array every candidate edge (of length
in $\left[\frac{\sqrt{3}}{2}\delta_{\dr},\delta_{\dr}\right]$) of every polygon, and to
parallelize the expandability checks on this array.

Finally, other stages can be easily parallelized with a less noticeable
effect, such as using a parallel sorting implementation for sorting the
$\ug(\inputPointCloud)$ edges by length (before applying Kruskal's
algorithm on them to get $\emst(\inputPointCloud)$).

\revision{
\subsection{Obstacles to generalization}
\label{sec:algo_generalize}
Generalizing this approach to higher-dimensional ambient space (e.g., $\bbr^3$) and/or higher-dimensional homology (e.g., \PH{2}) is difficult. Indeed, the fact that edges introducing a cycle are Delaunay edges does not hold for higher-dimensional homology: a triangle that introduces a cavity (i.e., a \PH{2} class) is not necessarily a Delaunay triangle. Furthermore, we rely heavily on duality to efficiently compute "codimension~1" persistent homology (i.e., \PH{1} in $\bbr^2$), which cannot be used as is to compute \PH{1} in $\bbr^3$, for example.
}

\section{Results}
\label{sec:results}

This section presents both qualitative and quantitative experimental results.
All results were obtained on a laptop computer with 64 GB of RAM, a Core
i7-13850HX CPU (8 cores at 5.3 GHz and 12 cores at 3.8 GHz), and a RTX 2000 GPU.
We implemented our method in C++ (with the libraries OpenMP, LibTorch
\cite{pytorch} and CGAL
\cite{fabri_cgal_2009}) as modules for TTK~\cite{tierny2017topology, masood2021overview}.
We used as encoder (resp. decoder) a fully connected network with two hidden layers of size 128 and 32 (resp. 32 and 128), with ReLU activation functions, and batch normalization.
For each experiment, we performed 1000 iterations of the Adam optimizer~\cite{kingma2017adam} with a learning rate fixed at 0.01.
\revision{See \julien{Appendix \ref{appendix:ablation}} for a more detailed 
study of hyperparameter selection.}

\subsection{Test data}
\label{sec:test_data}

We present results on synthetic, three-dimensional datasets in
\autoref{fig:tableSyntheticData}, which have been specifically designed
to include clearly salient topological features. The first one consists in 3
Gaussian clusters,
and features 3 significantly persistent \PH{0} pairs. The second one consists in
points sampled around an ellipse twisted along its major axis, and features one
significantly persistent \PH{1} pair. The third one consists in points sampled
around the edges of a tetrahedron (i.e., the complete $K_4$ graph which is
planar) and features 3 significantly persistent \PH{1} pairs.
We also consider a stress case in \autoref{fig:tableK5}, with a dense point cloud sampling the complete $K_5$ graph embedded in $\bbr^3$ (i.e., $5$ vertices in 3D, all pairwise connected by an edge).
Since $K_5$ is a non-planar graph, the resulting 3D
persistent generators (which follow the edges of $K_5$) cannot
be projected to the plane without intersections.
Then, it is not possible to project this stress case to the plane faithfully
with respect to \PH{1}, as the intersections of the projected 3D
generators create additional, short cycles in 2D.

We present results on real-life, acquired or simulated, high-dimensional, cycle-featuring
datasets in \autoref{fig:tableRealData}. COIL-20~\cite{nene_columbia_1996} is an
image dataset with 20 classes, each one containing 72 grayscale $64\times64$
images. Each class consists in a single object (e.g., a rubber duck)
viewed from 72 evenly spaced angles, so that the images within each class live
on a manifold that is homeomorphic to a circle.
This leads to significantly persistent pairs when computing the \PH{1} of the input endowed with the Euclidean metric between the images.
Cyclic features may also appear in motion capture
data~\cite{bei_wang_branching_2011}. Hence, we
also evaluate our approach on a dataset extracted from the
CMU motion capture database~\cite{noauthor_carnegie_nodate},
which records a human subject doing some periodic movement
(e.g., walking or running), each frame consisting in 62 skeleton joint
angles measurements.
Finally, data from biological processes can also feature meaningful cycles like in single-cell omics~\cite{saelens2019comparison}, a field that aims at sequencing single cells.
This enables the observation of gene expression within
individual cells at different stage of the cell cycle (resting, growth, DNA
replication, mitosis).
Specifically, we used the simulated dataset \emph{"cyclic\_2"}~\cite{cannoodt_single-cell_2018}, which features a significantly persistent
\PH{1} pair.
See \autoref{table:datasetSummary} for a summary of the presented datasets.
\begin{table}
	\centering
	\begin{tabular}{|c|c|c|c|c|}
		\hline
		Dataset & Nature & Dimension ($\highDim$) & Size ($n$) & Fig. \\
		\hline
		\datathreeblobs & Synthetic & 3 & 800 & \multirow{3}{*}{\autoref{fig:tableSyntheticData}} \\
		\datatwist & Synthetic & 3 & 100 & \\
		\datakfour & Synthetic & 3 & 300 & \\
		\hline
		\datakfive & Synthetic & 3 & 500 & \autoref{fig:tableK5} \\
		\hline
		\datacoil & Acquired & 4096 & 72 & \multirow{3}{*}{\autoref{fig:tableRealData}} \\
		\datamocap & Acquired & 62 & 138 & \\
		\datasinglecell & Simulated & 1170 & 243 & \\
		\hline
	\end{tabular}
	\vspace{2mm}
	\caption{Summary of datasets.}
	\label{table:datasetSummary}
\end{table}

\subsection{Quantitative criteria}
\label{sec:criteria}

In order to evaluate the topological accuracy of the representations generated by
our approach, we compute the
$L_2$-Wasserstein distance (see \autoref{sec:persistentHomology}) between the
1-dimensional persistence diagrams of the input $\inputPointCloud$ and the
low-dimensional representation $\latentPointCloud$:
\[\metwasser(\inputPointCloud,\latentPointCloud)=\calw_2\bigl(\dgmrips{1}(\inputPointCloud),\dgmrips{1}(\latentPointCloud)\bigr).\]
This quantity, which we want to be as low as possible, summarizes in some measure the extent to which similar cycles exist in the two point clouds.
However, as persistence diagrams lose information -- they are in particular invariant to permutations in the point clouds --
it does not convey whether the set of vertices involved in these cycles are
the same, in contrast to TopoAE-like loss functions.

We also compute the metric distortion $\metdistor$, which should also be as low as possible, expressed as the root mean squared error ($\RMSE$) between the pairwise distances in $\inputPointCloud$ and those in $\latentPointCloud$ -- which is what MDS-like methods typically minimize:
\[\metdistor(\inputPointCloud,\latentPointCloud)=\sqrt{\frac{1}{n}\sum\limits_{i
<j}\bigl(\lVert X_i-X_j\rVert_2-\lVert Z_i-Z_j\rVert_2\bigr)^2}.\]

We refer the reader to \revision{Appendix \ref{appendix:quantitativeDescription}} for a
description of other quality scores often considered in the DR literature
(see \revision{Appendix \ref{appendix:quantitativeObservations}} for experiments
with these indicators).

\subsection{Competing approaches}
We compare our approach to global methods (PCA~\cite{pearson1901liii},
MDS~\cite{torgerson1952multidimensional}) that do
not take topology into account. We also compare to several locally
topology-aware methods (Isomap~\cite{tenenbaum_global_2000},
t-SNE~\cite{van2008visualizing}, UMAP~\cite{mcinnes2018umap}) that are able to
consider the local manifold structure of the data but that might fail to project
faithfully the overall structure.
Regarding previously documented
globally topology-aware methods, although TopoMap~\cite{doraiswamy2020topomap}
preserves exactly \PH{0} and TopoAE~\cite{moor2020topological} preserves \PH{0}
in the sense of \autoref{lemma:TopoAE0_bound}, both methods have no guarantee
about \PH{1}. In addition, we show the results of the method suggested
in~\cite{carriere2021optimizing}, which we denote $\carriereMethod$ and which
consists in adding $\metwasser(\inputPointCloud,\latentPointCloud)$
to the TopoAE loss $\ltopoae{0}$. Both terms in the loss are given the same weights.
It is then supposed to give the best results in terms
of $\metwasser(\inputPointCloud,\latentPointCloud)$ values and therefore make the cycles in $\inputPointCloud$ and $\latentPointCloud$ as similar (i.e., in number, in size) as possible.
However, as explained in \autoref{sec:criteria}, $\metwasser(\inputPointCloud,\latentPointCloud)$ is not sensible to which simplices are involved in
the cycles, and the cycles in $\latentPointCloud$ might be created between simplices that do not belong to a cycle in $\inputPointCloud$.

For PCA, MDS, Isomap and t-SNE, we used the implementations in
scikit-learn~\cite{pedregosa2011scikit} (with default parameters
except Isomap for which the number of neighbors is set to 8); for
UMAP, the Python package umap-learn
provided by the authors; and for TopoMap, the existing implementation in TTK.
For TopoAE-like approaches, we used our own implementation, with
Ripser~\cite{bauer_ripser_2021} as latent space \PH{1} algorithm for \tAE{} and
\carriereMethod{}, and \autoref{algo:2dpersistence} for \tAE++. For
non-deterministic methods (i.e., \tAE, \carriereMethod{} and \tAE++
which rely on a stochastic optimization of a neural network with a random
initialization), we performed 10 runs and kept the result with the best
$\metwasser(\inputPointCloud,\latentPointCloud)$
\revision{(see 
\julien{Appendix \ref{appendix:failures}} for more information on this 
stochastic behavior)}.

\subsection{Result analysis}
\label{sec:results:analysis}
\begin{figure*}
	\adjustbox{width=1.013\linewidth,center}{
		\scriptsize{
		\begin{tabular}{|p{0cm}r||rr|rrr|rrr||r|}
			\hline

			\multicolumn{2}{|c||}{\multirow{2}{*}{Input}}&
			\multicolumn{2}{c|}{Global methods}&
			\multicolumn{3}{c|}{Locally topology-aware methods}&
			\multicolumn{4}{c|}{Globally topology-aware methods}\\

			\cline{3-11}
			\multicolumn{2}{|c||}{}
			&\multicolumn{1}{c}{PCA\cite{pearson1901liii}}
			&\multicolumn{1}{c|}{MDS\cite{torgerson1952multidimensional}}
			&\multicolumn{1}{c}{Isomap\cite{tenenbaum_global_2000}}
			&\multicolumn{1}{c}{t-SNE\cite{van2008visualizing}}
			&\multicolumn{1}{c|}{UMAP\cite{mcinnes2018umap}}
			&\multicolumn{1}{c}{TopoMap\cite{doraiswamy2020topomap}}
			&\multicolumn{1}{c}{\tAE\cite{moor2020topological}}
			&\multicolumn{1}{c||}{\carriereMethod\cite{carriere2021optimizing}}
			&\multicolumn{1}{c|}{\tAE++}\\
			
			\hline \raisebox{1.1cm}{\multirow{4}{*}{\rotatebox{90}{\datakfive{} $(n=500)$}}} &
			\raisebox{-3mm}{\includegraphics[width=.1\linewidth]{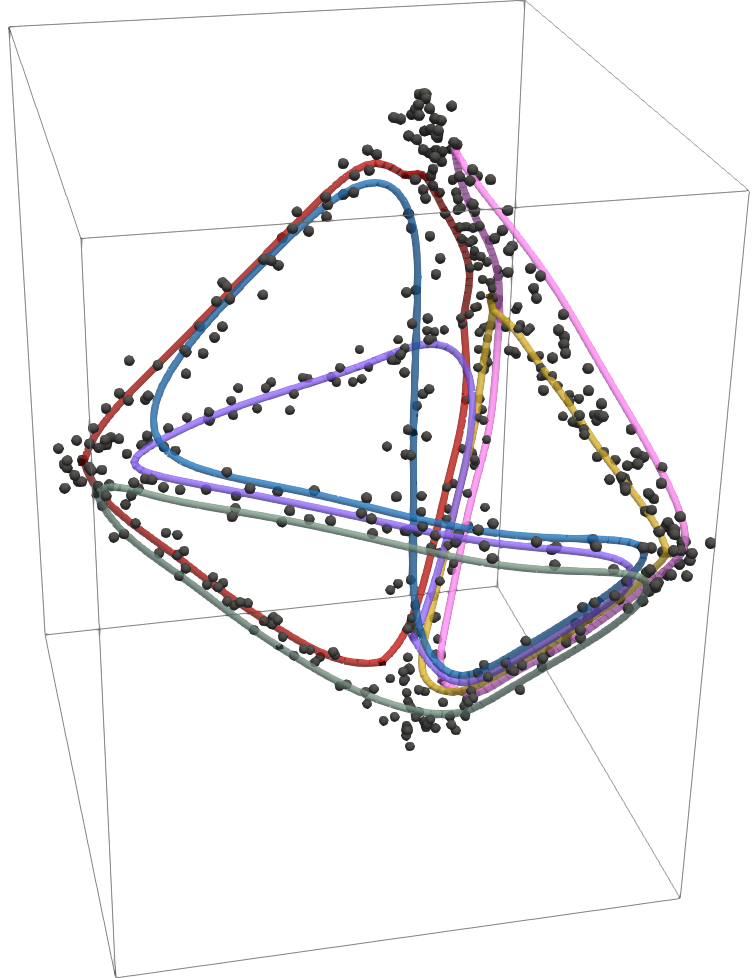}} &
			\includegraphics[width=.1\linewidth]{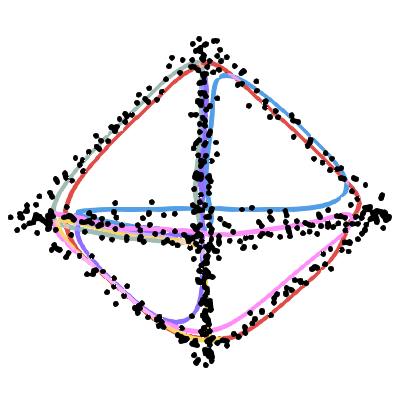}&
			\includegraphics[width=.1\linewidth]{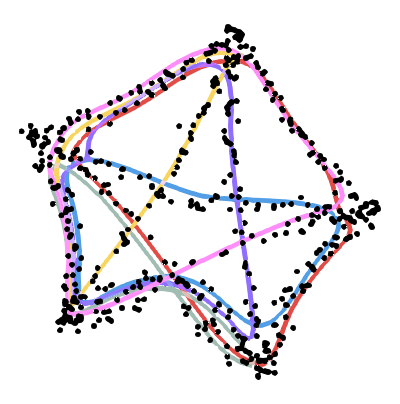}&
			\includegraphics[width=.1\linewidth]{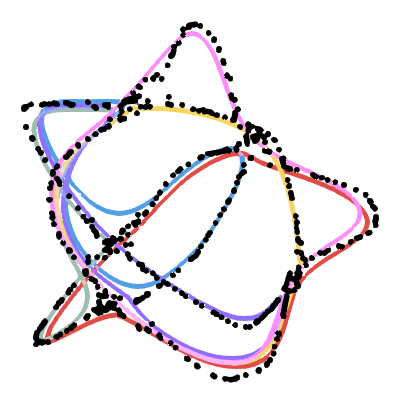}&
			\includegraphics[width=.1\linewidth]{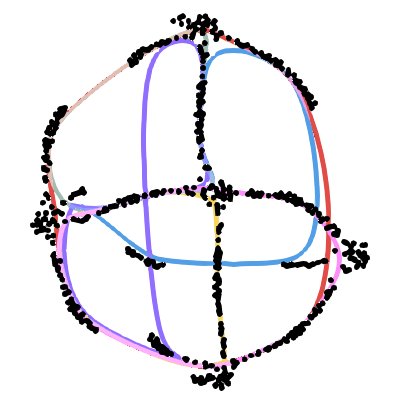}&
			\includegraphics[width=.1\linewidth]{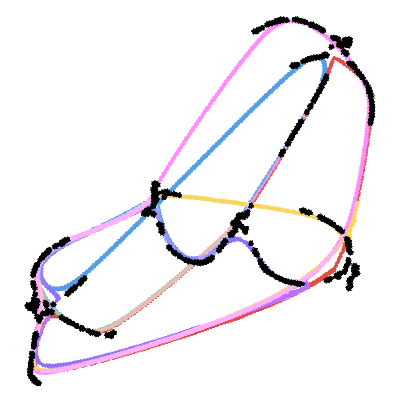}&
			\includegraphics[width=.1\linewidth]{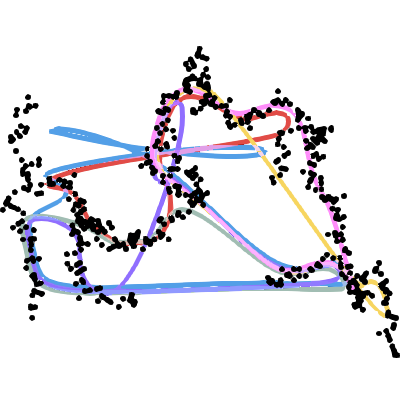}&
			\includegraphics[width=.1\linewidth]{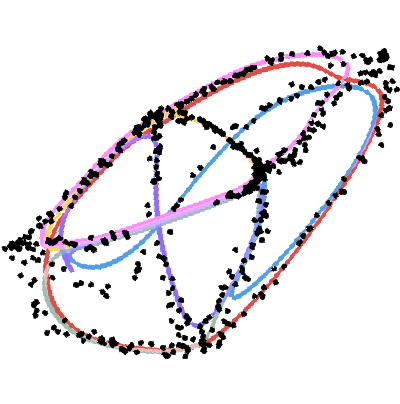}&
			\includegraphics[width=.1\linewidth]{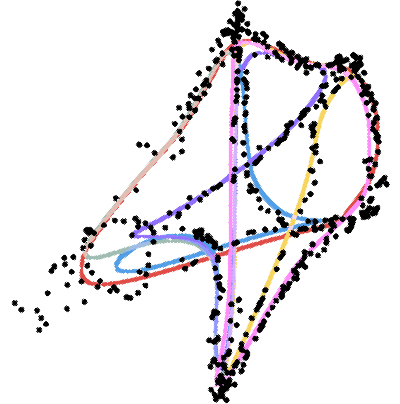}&
			\includegraphics[width=.1\linewidth]{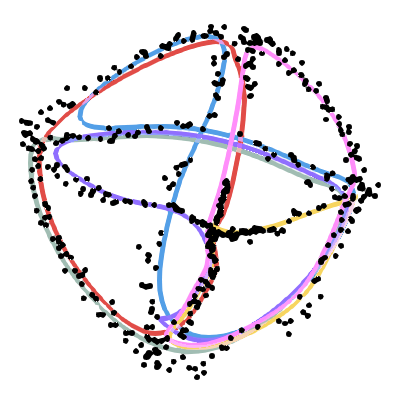}\\

			& $\metwasser(\inputPointCloud,\latentPointCloud)$ & 2.8e-01 & 3.5e-01 & 1.3e-01 & 4.7e+01 & 9.7e-01 & 4.8e-01 & 1.6e-01 & \textbf{2.2e-03} & \underline{8.0e-02}\\
			& $\metdistor(\inputPointCloud,\latentPointCloud)$ & \underline{3.1e-01} & \textbf{2.4e-01} & 4.7e-01 & 2.1e+01 & 9.8e+00 & 1.2e+00 & 9.2e-01 & 8.6e-01 & 3.4e-01\\
			& $\timing$ & 0.00043 & 0.59 & 0.088 & 0.63 & 0.14 & 0.063 & 29 & 170 & 6.0\\
			\hline
		\end{tabular}
		}
	}
	\caption{Comparison of DR methods on a stress case, i.e., a point cloud
  	sampled around the edges of the complete $K_5$ graph embedded in 3D.
	It features 6 significantly persistent \PH{1} pairs, and a generator for each
	is represented with a curve of distinct color.
	As $K_5$ is a non-planar graph, this point cloud cannot be projected
	faithfully in 2D in a topological sense (i.e., the high-dimensional persistent
	generators cannot be projected to the plane without intersections). Despite
	this, our method (TopoAE++) achieves a competitive Wasserstein distance.}
	\label{fig:tableK5}
\end{figure*}

\begin{figure*}
	\adjustbox{width=1.013\linewidth,center}{
		\scriptsize{
		\begin{tabular}{|p{0cm}r||rr|rrr|rrr||r|}
			\hline

			\multicolumn{2}{|c||}{\multirow{2}{*}{Input}}&
			\multicolumn{2}{c|}{Global methods}&
			\multicolumn{3}{c|}{Locally topology-aware methods}&
			\multicolumn{4}{c|}{Globally topology-aware methods}\\

			\cline{3-11}
			\multicolumn{2}{|c||}{}
			&\multicolumn{1}{c}{PCA\cite{pearson1901liii}}
			&\multicolumn{1}{c|}{MDS\cite{torgerson1952multidimensional}}
			&\multicolumn{1}{c}{Isomap\cite{tenenbaum_global_2000}}
			&\multicolumn{1}{c}{t-SNE\cite{van2008visualizing}}
			&\multicolumn{1}{c|}{UMAP\cite{mcinnes2018umap}}
			&\multicolumn{1}{c}{TopoMap\cite{doraiswamy2020topomap}}
			&\multicolumn{1}{c}{\tAE\cite{moor2020topological}}
			&\multicolumn{1}{c||}{\carriereMethod\cite{carriere2021optimizing}}
			&\multicolumn{1}{c|}{\tAE++}\\

			\hline \raisebox{1.6cm}{\multirow{4}{*}{\rotatebox{90}{\datacoil{} $(n=72)$}}} &
			\includegraphics[width=.11\linewidth]{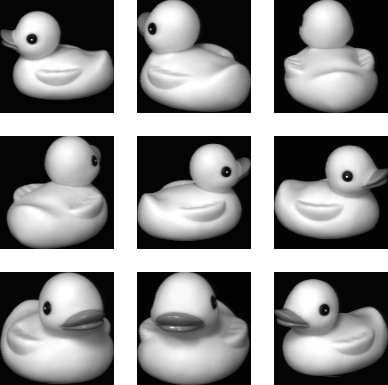} &
			\includegraphics[width=.1\linewidth]{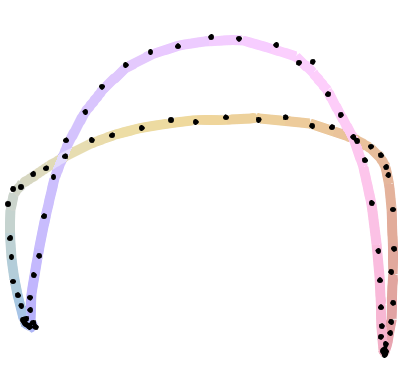}&
			\includegraphics[width=.1\linewidth]{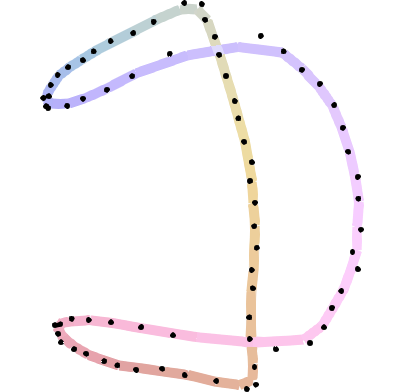}&
			\includegraphics[width=.1\linewidth]{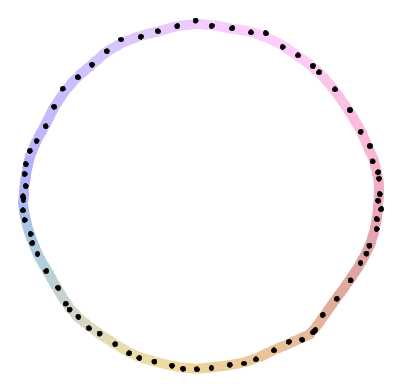}&
			\includegraphics[width=.1\linewidth]{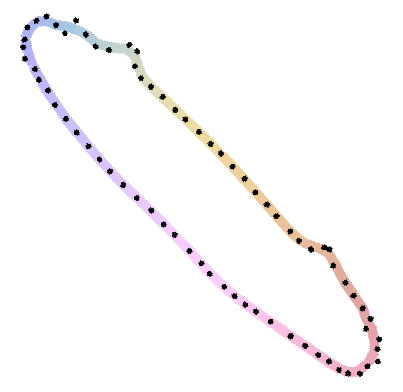}&
			\includegraphics[width=.1\linewidth]{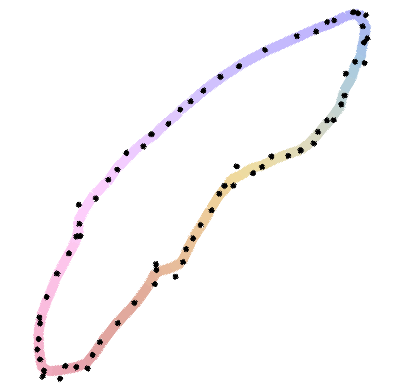}&
			\includegraphics[width=.1\linewidth]{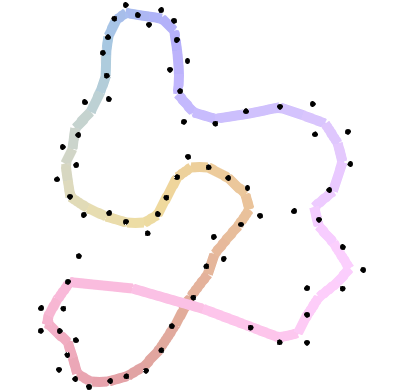}&
			\includegraphics[width=.1\linewidth]{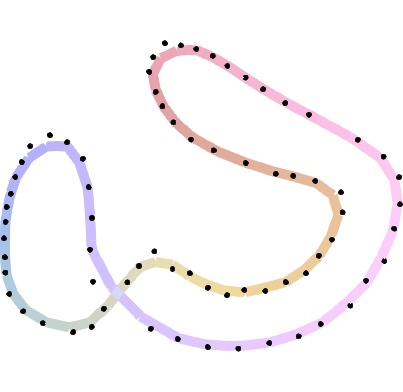}&
			\includegraphics[width=.1\linewidth]{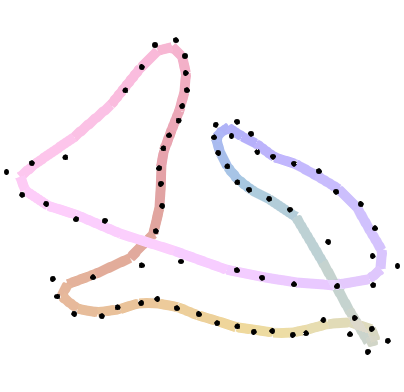}&
			\includegraphics[width=.1\linewidth]{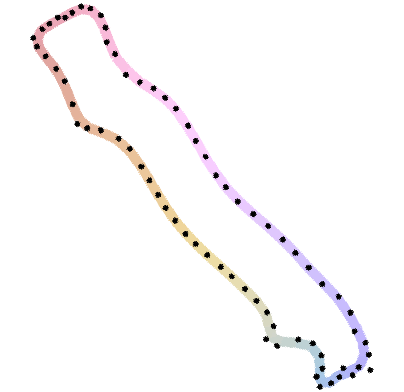}\\

			& $\metwasser(\inputPointCloud,\latentPointCloud)$ & 1.2e+01 & 1.0e+01 & 6.0e+02 & 1.3e+01 & 1.2e+01 & 2.7e+01 & 7.5e+00 & \underline{9.1e-02} & \textbf{1.6e-02}\\
			& $\metdistor(\inputPointCloud,\latentPointCloud)$ & \underline{2.9e+00} & \textbf{2.0e+00} & 2.1e+01 & 4.9e+00 & 6.2e+00 & 7.2e+00 & 8.2e+00 & 6.7e+00 & 1.9e+01\\
			& $\timing$ & 0.0071 & 0.086 & 0.083 & 0.30 & 2.5 & 0.14 & 2.0 & 4.0 & 2.4 \\

			\hline \raisebox{1.4cm}{\multirow{4}{*}{\rotatebox{90}{\datamocap{} $(n=138)$}}} &
			\includegraphics[width=.11\linewidth]{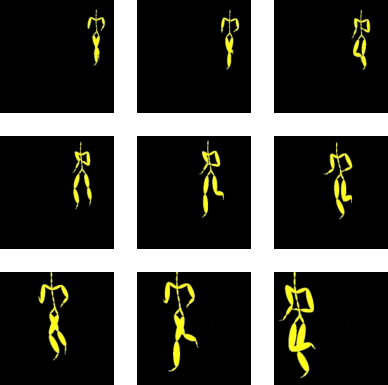} &
			\includegraphics[width=.1\linewidth]{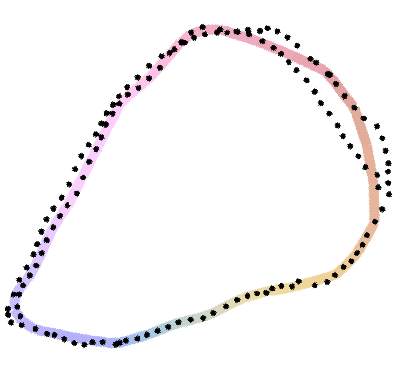}&
			\includegraphics[width=.1\linewidth]{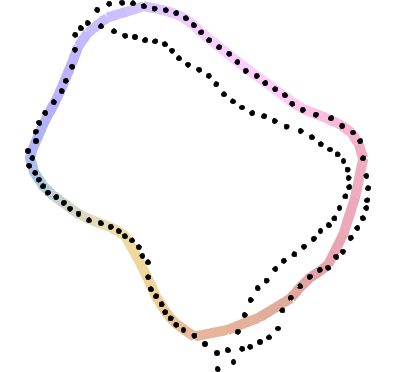}&
			\includegraphics[width=.1\linewidth]{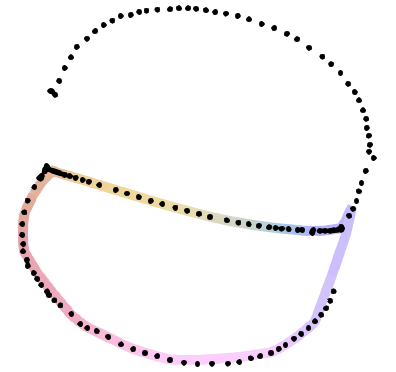}&
			\includegraphics[width=.1\linewidth]{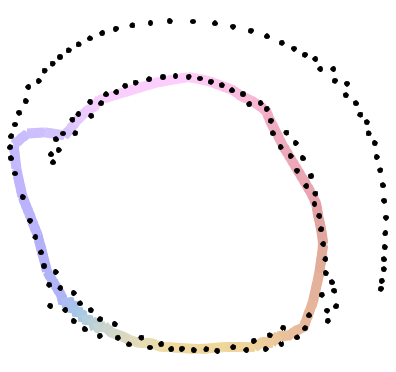}&
			\includegraphics[width=.1\linewidth]{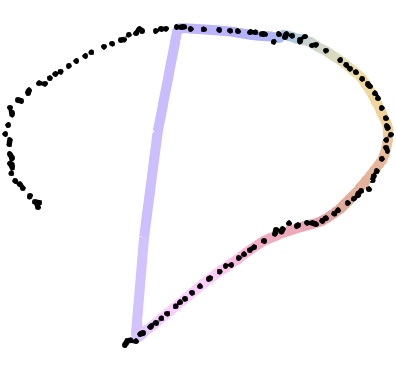}&
			\includegraphics[width=.1\linewidth]{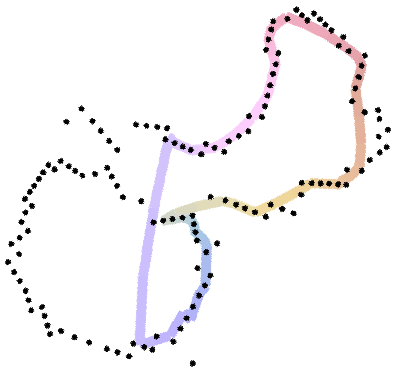}&
			\includegraphics[width=.1\linewidth]{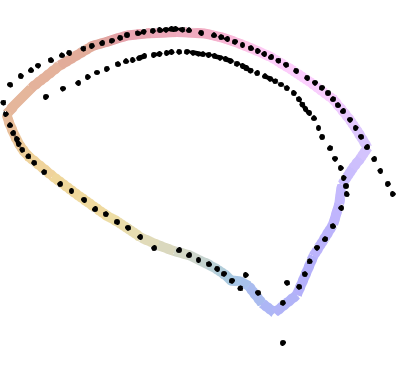}&
			\includegraphics[width=.1\linewidth]{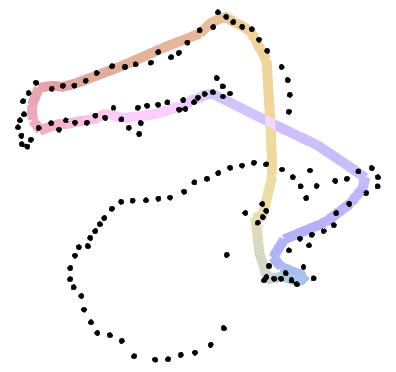}&
			\includegraphics[width=.1\linewidth]{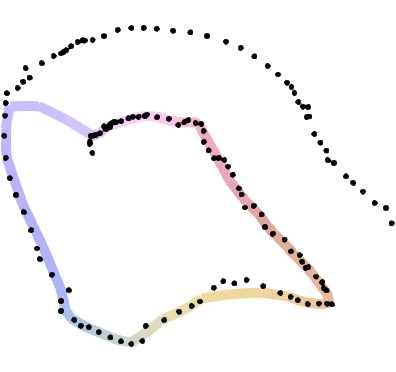}\\

			& $\metwasser(\inputPointCloud,\latentPointCloud)$ & 2.1e+03 & 1.9e+03 & 1.0e+04 & 1.6e+03 & 1.5e+03 & 2.1e+03 & 1.2e+03 & \textbf{9.4e+00} & \underline{1.2e+01}\\
			& $\metdistor(\inputPointCloud,\latentPointCloud)$ & \underline{2.4e+01} & \textbf{1.7e+01} & 1.3e+02 & 9.9e+01 & 1.0e+02 & 8.5e+01 & 2.6e+01 & 5.6e+01 & 2.9e+01\\
			& $\timing$ & 0.041 & 0.25 & 0.11 & 0.33 & 2.6 & 0.056 & 3.1 & 12 & 3.4\\

			\hline \raisebox{1.62cm}{\multirow{4}{*}{\rotatebox{90}{\datasinglecell{} $(n=243)$}}} &
			&
			\includegraphics[width=.1\linewidth]{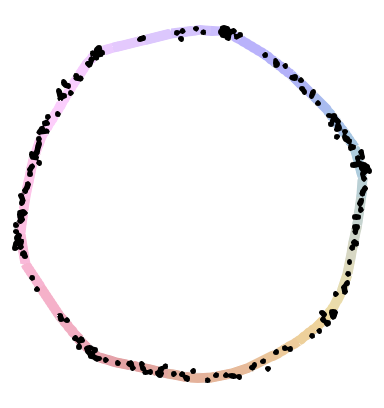}&
			\includegraphics[width=.1\linewidth]{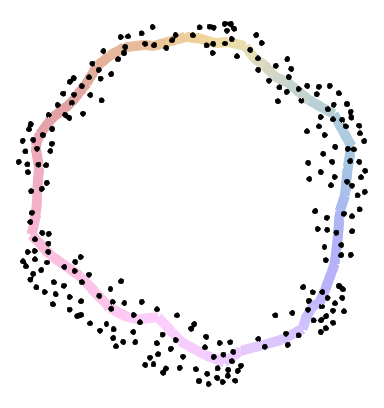}&
			\includegraphics[width=.1\linewidth]{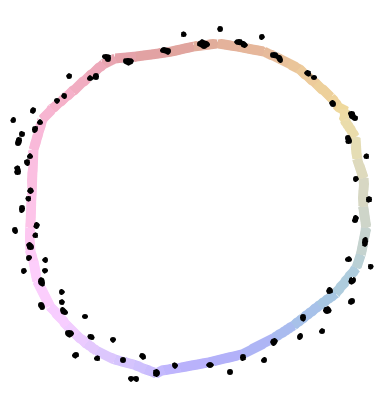}&
			\includegraphics[width=.1\linewidth]{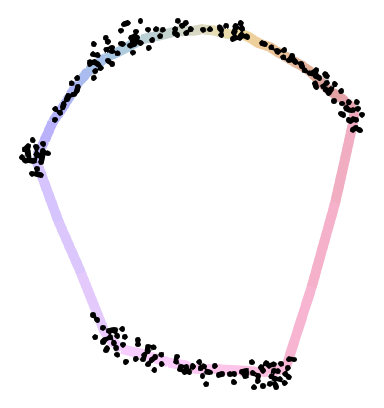}&
			\includegraphics[width=.1\linewidth]{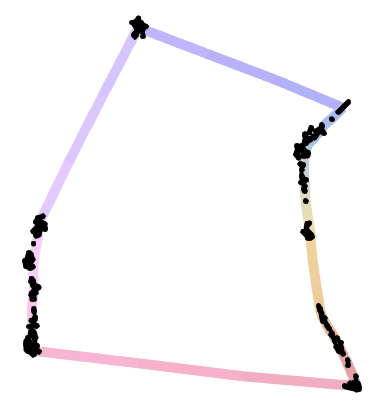}&
			\includegraphics[width=.1\linewidth]{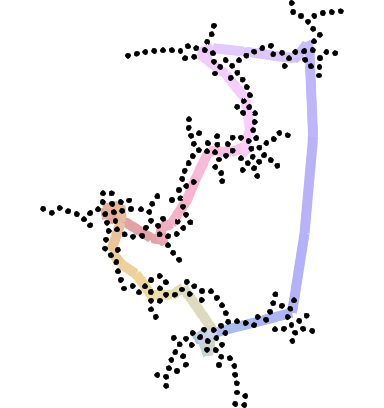}&
			\includegraphics[width=.1\linewidth]{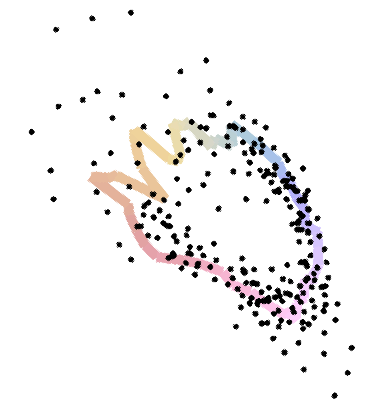}&
			\includegraphics[width=.1\linewidth]{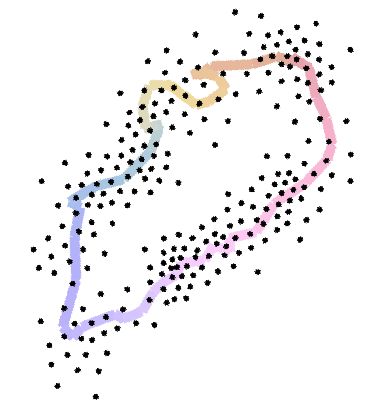}&
			\includegraphics[width=.1\linewidth]{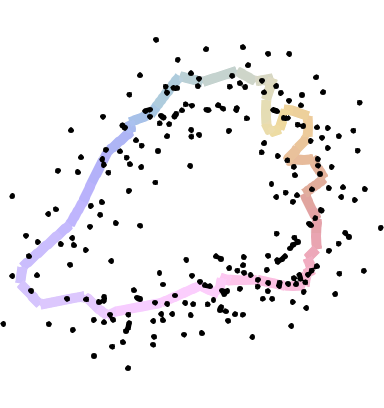}\\

			& $\metwasser(\inputPointCloud,\latentPointCloud)$ & 3.0e+03 & 3.1e+03 & 1.8e+05 & 1.8e+03 & 1.8e+03 & 2.8e+04 & 1.8e+03 & \textbf{6.4e+02} & \underline{8.6e+02}\\
			& $\metdistor(\inputPointCloud,\latentPointCloud)$ & \underline{3.1e+01} & \textbf{2.3e+01} & 4.8e+02 & 9.6e+01 & 1.0e+02 & 1.2e+03 & 9.8e+01 & 3.2e+02 & 1.5e+02\\
			& $\timing$ & 0.0089 & 0.77 & 0.019 & 0.33 & 2.5 & 0.023 & 7.6 & 27 & 6.0\\
			\hline
		\end{tabular}
		}
	}
	\caption{Comparison of DR methods on three real-life, acquired or simulated, high-dimensional
	datasets (see \autoref{sec:test_data} for a description) that all feature a
	single significantly persistent \PH{1} pair. A generator for this pair
	is projected down to 2D,
	in transparency and slightly smoothed (with a color
	map depicting its arc-length parameterization). Quantitatively, our approach
	(TopoAE++) achieves a competitive Wasserstein distance. Qualitatively, it generates
	fewer crossings of the projected high-dimensional persistent generator.}
	\label{fig:tableRealData}
\end{figure*}

\subsubsection{Qualitative analysis}
Our visual validation approach involves computing a generator for each
significantly persistent pair in the input and observing their low-dimensional
projections. These generators should ideally be projected without
self-intersection and coincide with the cycles in the low-dimensional
representation. In such a case, the output would be considered to faithfully
represent the high-dimensional cycles.

In the depicted examples, \tAE++ succeeds in projecting the cycles from the
input in this visual sense, whenever possible
(with the exception of \datakfive{}, for which a faithful planar projection
is not possible, see \autoref{sec:test_data}).
It also tends to preserve the \textit{shape} of the cycles.
For example, \datatwist{} (\autoref{fig:tableSyntheticData}) is projected with
the original ellipse shape preserved, but without torsion.
Similarly, in \datacoil{} (\autoref{fig:tableRealData}), the
high-dimensional cycle has a stretched shape (since the rubber duck has a
similar silhouette when viewed from the front and back, but not when viewed from
the left and right), which \tAE++ takes into account (views from the left and
right corresponds to the "ends" of the stretched planar cycle).
\datamocap{} (\autoref{fig:tableRealData}) comes from the capture of a subject performing a
translation in addition to its periodic movement (running). It is therefore
projected to a spiral shape, where the longest edge of the projection of the
input cycle (appearing in lavender) corresponds to the same body configuration
during the running cycle, but at two different locations. Finally,
\datasinglecell{} (\autoref{fig:tableRealData}) is difficult to project in the plane as it presents points
that lie on a cycle but are distant from each other due to the high
dimensionality: \tAE++ takes that into account and its result can be
interpreted as a compromise between preserving \PH{0} and \PH{1}.

On contrary, global methods (PCA, MDS) do not guarantee the preservation of
these cycles (see, e.g., their results on \datatwist{},
\autoref{fig:tableSyntheticData}, and
\datacoil{}, \autoref{fig:tableRealData}), while locally
topology-aware methods sometimes fail to project
them correctly (e.g., t-SNE and UMAP break apart the cycles in
\datakfour{}, \autoref{fig:tableSyntheticData}, and
\datasinglecell{}, \autoref{fig:tableRealData}, when
projecting). In addition, even if the
cycle is visually correctly projected, these methods tend to forget the
\textit{shape} of the cycle (e.g., Isomap projects \datatwist{},
\autoref{fig:tableSyntheticData}, and
\datacoil{}, \autoref{fig:tableRealData}, as almost perfect circles).
Finally, as expected, previously documented
globally topology-aware methods generally fail to
faithfully project the cycles, as they only incorporate constraints on \PH{0}
(TopoMap, \tAE) or on the Wasserstein distance $\metwasser$ (\carriereMethod),
but not on \PH{1} (see the next paragraph).

\begin{figure}[b]
	\def\svgwidth{.95\linewidth}
	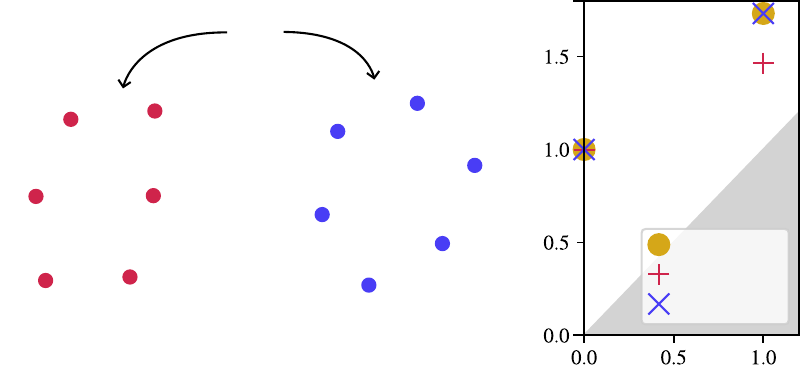
	\caption{Topological accuracy comparison between projections obtained
	with the loss $\ltopoae{1}$ (a naive extension of \cite{moor2020topological}
to
	\PH{1}, \autoref{eq:topoae}) and our cascade distortion ($\lcascae{1}$,
	\autoref{eq:cascae}) on the counter-example
	$\inputPointCloud\in\bbr^3$ of \autoref{fig:counter-example}, producing the 2D
	embeddings $\latentPointCloud_\mathrm{TAE}$ and
$\latentPointCloud_\mathrm{CD}$
	respectively (left).
	In this example, the optimization achieved a virtually zero value for
	both losses. However, in the space of persistence diagrams,
	$\dgmrips{1}(\latentPointCloud_\mathrm{TAE})$ does not match
	$\dgmrips{1}(\inputPointCloud)$, while
	$\dgmrips{1}(\latentPointCloud_\mathrm{CD})$ coincides with it (right).}
	\label{fig:counter-example-diags}
\end{figure}

\subsubsection{Quantitative analysis}

From a more quantitative point of view, in the examples shown, \tAE++ generally produces the second best $\metwasser$ after \carriereMethod{}, both significantly better than all other methods for this metric.
However, as \carriereMethod{} directly optimizes $\metwasser$ and forgets the
simplices involved in \PH{1} pairs, the vertices involved in planar cycles
(i.e., corresponding to the points of $\dgmrips{1}(\latentPointCloud)$)
do not match those involved in input cycles (i.e., corresponding to the
points of $\dgmrips{1}(\inputPointCloud)$). This leads to a good $\metwasser$
metric but a poor projection.
For instance, in \datatwist{} (\autoref{fig:tableSyntheticData}),
\carriereMethod{} creates a cycle (the top one) that has the right birth and
death -- while the bottom loop has a persistence close to zero -- to cancel out
$\metwasser$; yet the original cycle is poorly projected, with a
self-intersection.
Similarly, on \datakfive{} (\autoref{fig:tableK5}), \carriereMethod{} places the points in order to have 6 persistent cycles with the right birth and death to cancel out $\metwasser$, while \tAE++ tries to preserve \PH{1}, which is impossible here due to the non-planarity of $K_5$.

The distortion metric is higher with TopoAE++, than in global methods
(PCA, MDS) which directly minimize it (by design for MDS).
It can be seen as a necessary compromise to faithfully project cycles
(e.g., \datatwist{}, \autoref{fig:tableSyntheticData},
requires distortion to be untwisted in 2D;
the cycle in \datacoil, \autoref{fig:tableRealData}, which is
folded up in high dimension, requires
distortion to be unfolded).
See \autoref{table:quantitative} of 
\julien{Appendix \ref{appendix:quantitative}} for
additional \revision{common DR indicators (measuring the preservation of both global and local distances)}, along with a companion
discussion.

\begin{figure}[b]
	\centering
	\scriptsize{
	\revision{
	\begin{tabular}{c|c|c|c|}
		\cline{2-4}
		 & \datacoil & \datamocap & \datasinglecell \\
		\hline
		\multicolumn{1}{|c|}{\raisebox{.25cm}{$\call_t=\ltopoae{1}$}} &
		\includegraphics[width=.1\linewidth]{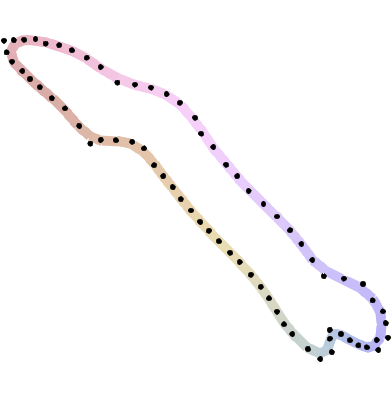} &
		\includegraphics[width=.1\linewidth]{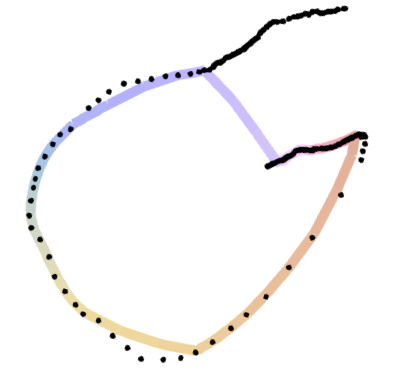} &
		\includegraphics[width=.1\linewidth]{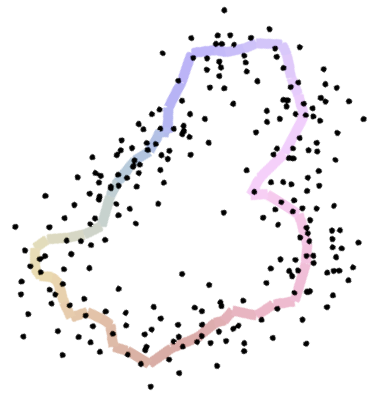} \\
		\multicolumn{1}{|c|}{$\metwasser(\inputPointCloud,\latentPointCloud)$} & 3.2e-02 & 3.0e+01 & 1.3e+03 \\
		\multicolumn{1}{|c|}{$\timing$} & 5.9 (2.4) & 15 (3.3) & 26 (5.9) \\
		\hline
		\multicolumn{1}{|c|}{\raisebox{.25cm}{$\call_t=\lcascae{1}$}} &
		\includegraphics[width=.1\linewidth]{results/figures_cropped/coil1_TopoAE++} &
		\includegraphics[width=.1\linewidth]{results/figures_cropped/mocap_09_07_TopoAE++} &
		\includegraphics[width=.1\linewidth]{results/figures_cropped/cyclic_2_TopoAE++}\\
		\multicolumn{1}{|c|}{$\metwasser(\inputPointCloud,\latentPointCloud)$} & 1.6e-02 & 1.2e+01 & 8.6e+02 \\
		\multicolumn{1}{|c|}{$\timing$} & 2.4 & 3.4 & 6.0 \\
		\hline
	\end{tabular}
	}
	}
	\caption{Topological accuracy comparison
	($\metwasser(\inputPointCloud,\latentPointCloud)$)
	between projections obtained with the loss $\ltopoae{1}$
	and our cascade distortion $\lcascae{1}$, for our real-life datasets.
	For $\ltopoae{1}$, we report the runtime
	with the gold standard for Rips persistence computation~\cite{bauer_ripser_2021}
	as well as, in parentheses, with our novel planar Rips
	persistence computation algorithm (\autoref{sec:algorithms}).
	On average, projections using our novel loss ($\lcascae{1}$), combined with our fast planar Rips
	persistence algorithm, improve topological accuracy by $48\%$, with a runtime improvement of $71\%$.}
	\label{table:comparisonTAEwithCasc}
\end{figure}

\autoref{fig:counter-example-diags} provides further evaluations of the
internal aspects of our approach. Specifically, for the counter-example of
\autoref{fig:counter-example}, it compares projections obtained by minimizing
a naive extension of~\cite{moor2020topological} to \PH{1} ($\ltopoae{1}$,
\autoref{eq:topoae}) and our novel loss ($\lcascae{1}$,
\autoref{eq:cascae}). In this example, while the optimization achieves a
virtually zero value for both losses, a clear gap occurs between the
persistence diagrams
$\dgmrips{1}(\latentPointCloud_\mathrm{TAE})$ and
$\dgmrips{1}(\inputPointCloud)$, while
$\dgmrips{1}(\latentPointCloud_\mathrm{CD})$ coincides with
$\dgmrips{1}(\inputPointCloud)$.
This indicates that $\lcascae{1}$ enables a better preservation of the
persistent homology of the input $\latentPointCloud$. This gain in topological
accuracy is further evaluated in \autoref{table:comparisonTAEwithCasc}, which
compares the resulting Wasserstein distances
($\metwasser(\inputPointCloud,\latentPointCloud)$) for our real-life datasets.
In particular, we observe in this figure that,
while both losses do not produce
intersections in the projection of the high-dimensional generator (colored
curve), projections based on
our novel loss improve topological
accuracy by $48\%$ on average. This illustrates that our novel loss
produces projections which are more faithful to the input data, in
terms of the number, size and shape of its cyclic patterns.

%

\subsection{Time performance}
\label{sec:performance}

\subsubsection{Fast 2D persistence computation}
\label{sec:performance:rips2d}

\begin{figure}
	\centering
	\includegraphics[width=\linewidth]{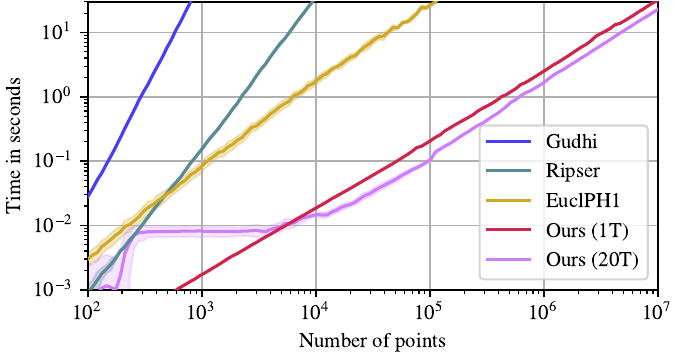}
	\caption{Comparison of Rips persistence computation times
	between \textit{Gudhi}, \textit{Ripser}, \textit{Euclidean PH1} and our algorithm
	(on a single thread, 1T, and on 20 threads, 20T), on point clouds randomly,
	uniformly distributed over $[0,1]^2$.}
	\label{fig:benchmark-rips-uniform}
\end{figure}

\begin{figure}
	\centering
	\includegraphics[width=\linewidth]{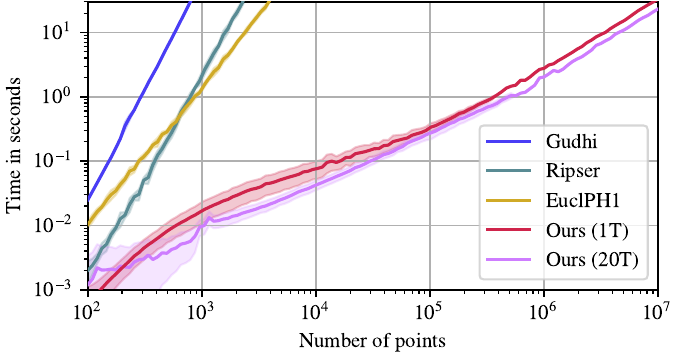}
	\caption{Comparison of Rips persistence computation times
	between \textit{Gudhi}, \textit{Ripser}, \textit{Euclidean PH1} and our algorithm
	(on a single thread, 1T, and on 20 threads, 20T), on a stress case (for
	\emph{Ripser} and \emph{Euclidean PH1}): a point cloud randomly
	distributed over the unit circle, with Gaussian noise ($\sigma=0.1$).}
	\label{fig:benchmark-rips-ring}
\end{figure}

We compared our persistence algorithm (\autoref{algo:2dpersistence}) for point
clouds in $\bbr^2$ to the following more generic Rips \PH{0} and \PH{1}
implementations: \textit{Gudhi}~\cite{maria2014gudhi} (a generic implementation
that handles various filtrations), \textit{Ripser}~\cite{bauer_ripser_2021} (a
specialization to Rips filtrations that can be applied to any point
cloud in $\bbr^d$ or distance matrix), and \textit{Euclidean PH1}
\cite{koyama2023reduced} (that works theoretically for any point cloud in
$\bbr^d$, but implemented only for $d=2$ or $d=3$).
We observe that on uniformly sampled point clouds, the running time asymptotic
behavior is roughly cubic (with regard to the number of input points)
for \textit{Gudhi} and quadratic for \textit{Ripser},
while it remains close to being linear for \textit{Euclidean PH1} and our
algorithm. However, our approach offers a speedup of about 2
orders of magnitudes compared to \textit{Euclidean PH1} (see
\autoref{fig:benchmark-rips-uniform}).
The speedup seems even more noticeable with points clouds sampled next to a
circle which creates one high-persistence pair in the 1-dimensional persistence
diagram and particularly slows down \textit{Euclidean PH1} (see
\autoref{fig:benchmark-rips-ring}).
Finally, we have observed empirically that the use of
\autoref{lemma:bounded_rips_delrips} in \autoref{algo:findMMLedge}
(i.e., checking expandability only for the edges of length in
$\left[\frac{\sqrt{3}}{2}\delta_{\dr},\delta_{\dr}\right]$)
enables discarding $84\%$ of the possible polygon diagonals on average,
which significantly contributes to the reported speedup.

\subsubsection{Overall DR approach}

\tAE++ can hardly compete with conventional DR methods (PCA, MDS, Isomap, t-SNE) in terms of computation time.
However, this must be put into perspective. First, we construct a complete
projection from the input space $\calx$ to the
latent space $\bbr^2$, and not just a projection of the point cloud
$\inputPointCloud$ (i.e., we can project a new point, at the low price of
one evaluation of the encoder network). Second,
we incorporate constraints from persistent homology, which is computationally
expensive by nature.

However, our fast 2D Rips \PH{} algorithm (\autoref{sec:algorithms})
manages to overcome the runtime overhead due to
the computation of \PH{1} of $\latentPointCloud$ at each iteration of the
optimization.
Indeed, comparing \tAE++ with \carriereMethod{} -- which also requires this
computation, implemented with \textit{Ripser} -- we observe a speedup
by a factor of $2$ to $20$ in the presented examples.
The experiments from \autoref{table:comparisonTAEwithCasc} (comparing
\tAE++ to a naive extension of \cite{moor2020topological} to \PH{1})
report compatible speedups ($3.7$ on average).
In addition, \tAE++ is sometimes even faster than \tAE{} although
the latter only computes \PH{0}, both initially for the
input $\inputPointCloud$ and at each optimization step for
$\latentPointCloud$.

\subsection{Limitations}
\label{sec:limitations}

The first obvious limitation is of a fundamental nature: there are
points clouds for which there is no planar embedding that is faithful with
regard to the \PH{1} (see \autoref{fig:tableK5} and its description,
\autoref{sec:test_data}).
Besides, like many other topological methods, we are limited in the size of the input due
to the initial \PH{} computation, if only in terms of memory. Specifically,
\textit{Ripser}, which is commonly considered as the gold
standard implementation for Rips filtrations in high dimensions, runs out
of memory on a 64 GB machine when computing the \PH{1} of around 50,000 points.
This limitation could be addressed by batch processing or subsampling approaches
(see, e.g.,~\cite{moor2020topological}). In addition, as with the original \tAE,
unsatisfactory results may occur when the minimization of the topological
regularization term is not successful (i.e., stuck in a "bad" local minimum).
In our experiments, as for \tAE{} and \carriereMethod{}, this was mitigated by
running 10 times the projection with a random initialization, and considering
as an output the projection minimizing
$\metwasser(\inputPointCloud,\latentPointCloud)$.
More sophisticated strategies could be considered in the future,
\revision{such as using a locally topology-aware method (e.g., UMAP) for initialization, or penalizing the self-crossings of the input persistent generators during optimization to help escape bad local minima}.
Finally,
\revision{the proof of \autoref{lemma:TopoAE0_bound} does not work with higher-dimensional homology, as taking
$\skelcasc^1(\inputPointCloud)\cup\skelcasc^1(\latentPointCloud)$ does not allow to recover $\dgmrips{1}(\inputPointCloud)$ and $\dgmrips{1}(\latentPointCloud)$ like in 0-dimensional homology.
Therefore,} we have no theoretical guarantee that our cascade distortion
loss function upper bounds the Wasserstein distance between the persistence diagrams, despite its
experimental ability to accurately preserve the cycles. However,
we have not found a counter-example yet, similar to the example
presented in \autoref{fig:counter-example}.
\revision{If such a bound exists, finding it would require tools beyond those used in this paper. Therefore, answering this question is left to future work.}

\section{Conclusion}
\label{sec:conclusion}

We have introduced \tAE++, a dimensionality reduction method
aiming at accurately visualizing the cyclic patterns present in high
diemnsional data. For this, we revisited TopoAE~\cite{moor2020topological},
provided a novel theoretical analysis of its original formulation
for \PH{0}, and introduced a new generalization to \PH{1}.
We have shown experimentally that our novel projection
provides an improved balance between the topological accuracy and the
visual preservation of the input $1$-cycles.
As a side benefit of our work, to overcome the computational overhead
due to the \PH{1} computation, we have presented
a novel, fast, geometric algorithm that computes Rips \PH{}
for planar point clouds, which may be of independent interest.

In future work, \revision{cleverer initializations and}
the possibility to handle automatically bad local minima during
the minimization, e.g., by penalizing
self-intersections of input generators, could be investigated.
Besides, an interesting extension would be to generalize the method to either
higher-dimensional latent spaces (e.g., $\calz=\bbr^3$, still
for visualization purpose\revision{s}), or \revision{to} constraints on higher-dimensional homology
(e.g., constraining \PH{2} to preserve the \emph{cavities}%
\revision{, by optimizing $\lcascae{2}$ instead of $\lcascae{1}$}).
However, \revision{this might prove computationally challenging, since
it would require computing \PH{2} in low dimension at each iteration while} our fast planar Rips \PH{1} algorithm does not seem \revision{easily generalizable} to such a context.
Finally, another interesting extension would be to consider
other filtrations of specific interest for the applications.

\section*{Acknowledgments}
This work is partially supported by the European Commission grant
ERC-2019-COG \emph{``TORI''} (ref. 863464, \url{https://erc-tori.github.io/}); and by ANR-23-PEIA-0004 (PDE-AI \url{https://pde-ai.math.cnrs.fr/}).

\bibliography{biblio}
\bibliographystyle{IEEEtran}

\clearpage
\appendices
\section{Additional background}
\label{appendix:background_appendix}

\subsection{Homology}
\label{appendix:homology}

\textit{Homology groups}, introduced by Henri Poincaré in the early 20th century
in his \textit{Analysis Situs}, are algebraic constructions that describe
"holes" in a topological space.
Here, we provide a brief summary of the notions used to define them.

Let $\calk$ be a simplicial complex. A \textit{$\homologyDim$-chain} $c$ is a formal sum of $\homologyDim$-simplices with \textit{modulo 2} coefficients\footnote{The coefficients can be chosen in another field or ring (which yields different homology groups), but $\bbz/2\bbz$ is the most common choice.}: $c=\sum\alpha_i\sigma_i$ with 
$\alpha_i\in\bbz/2\bbz$. Two $\homologyDim$-chains can be summed simplex by simplex, which permits to define the group of $\homologyDim$-chains, noted $\chain{\homologyDim}$.
The \textit{boundary} of a $\homologyDim$-simplex is the sum of its $(\homologyDim-1)$-dimensional faces, i.e. more formally, if $\sigma=(v_0,\ldots,v_k)$, then its boundary is \[\partial_\homologyDim\sigma=\sum\limits_{i=0}^k(v_0,\ldots,\hat{v_i},\ldots,v_k)\]
where the hat means that $v_i$ is omitted. Then the boundary of a $\homologyDim$-chain $c$ is defined as the modulo 2 sum of the boundaries of its simplices, i.e. $\partial_\homologyDim c=\sum\alpha_i\partial_\homologyDim\sigma_i$. This defines the \textit{$\homologyDim$-th boundary map} $\partial_\homologyDim$, which is a morphism $\partial_\homologyDim:\chain{\homologyDim}\rightarrow\chain{\homologyDim-1}$.

A \textit{$\homologyDim$-cycle} is a $\homologyDim$-chain whose boundary is 0. The group of $\homologyDim$-cycles is noted $\cycle{\homologyDim}=\ker\partial_\homologyDim$.
A \textit{$\homologyDim$-boundary} is a $\homologyDim$-chain which is the boundary of some $(\homologyDim+1)$-chain. The group of $\homologyDim$-boundaries is noted $\bound{\homologyDim}=\Ima\partial_{\homologyDim+1}$.
The fundamental lemma of homology states that
$\partial_{\homologyDim-1}\circ\partial_\homologyDim=0$, i.e., that the boundary
of a boundary is always an empty chain. In particular this implies the
inclusions
$\bound{\homologyDim}\subseteq\cycle{\homologyDim}\subseteq\chain{\homologyDim}$.

Two $\homologyDim$-cycles $a,b\in\cycle{\homologyDim}$ are \textit{homologous}
whenever $a=b+\partial c$ for some $c\in\chain{\homologyDim+1}$. The set of the
equivalence classes for this relation forms the
$\homologyDim$-th homology group of $\calk$:
$\homol{\homologyDim}=\cycle{\homologyDim}/\bound{\homologyDim}$. Its rank, i.e.
the maximal number of linearly independent classes (called \textit{generators}), is called the
\textit{$\homologyDim$-th Betti number} of $\calk$:
$\betti{\homologyDim}(\calk)=\rank\homol{\homologyDim}=\log_2\bigl(
|\homol{\homologyDim }|\bigr)$. The Betti numbers can be easily
interpreted geometrically for small values of $\homologyDim$: for a simplicial complex $\calk$ embedded in $\bbr^3$,
$\betti{0}(\calk)$ is the number of connected components, $\betti{1}(\calk)$ is the number of
cycles, and $\betti{2}(\calk)$ is the number of voids.

\subsection{Minmax length triangulations}
\label{appendix:MMLTs}
In this section, we give more details on the computation of minmax length ($\mml$) triangulations, that are all taken from~\cite{edelsbrunner_quadratic_1993}. General position is supposed, i.e., no two edges are equally long.
A \emph{2-edge} is an edge whose right and left lenses are both non-empty.
Conversely, a \emph{1-edge} is an edge such that one of its half-lens is empty but not the other one.
By definition, 1- and 2-edges are not $\rng$ edges.
Note that every triangulation of an
$\rng$ polygon has a 2-edge (its longest edge is always a 2-edge).

An \emph{expandable 2-edge} is a 2-edge $e=pq$ for which there exists $x\in\llens(e)$ and $y\in\rlens(e)$ such that the edges $px$, $xq$, $qy$, $yp$ are either $\rng$ edges, or 1-edges with an empty right lens (see \autoref{fig:expandability}).
Expandability is a property that somehow allows an edge to be the longest edge of a triangulation.
Indeed, let $e$ be an expandable 2-edge inside an
$\rng$-polygon $\Pi$.
Then it is possible to construct by recurrence a triangulation of $\Pi$ such that the only 2-edge -- and thus longest edge -- in that triangulation is $e$ itself.

Besides, the \emph{2-edge lemma} (Sec. 5.3 of~\cite{edelsbrunner_quadratic_1993}) states that there exists a $\mml$ triangulation of $\Pi$ that contains an expandable 2-edge.
Therefore, to find a $\mml$ triangulation of an
$\rng$-polygon, it suffices to find the shortest expandable 2-edge $e$, which
is the longest edge of any $\mml$-triangulation. This is the only information we
use in our application.
However, if a complete $\mml$ triangulation is wanted, $e=pq$ has to be found together with $x$ and $y$ defined as above.
Then, the incomplete polygons defined by $px$, $xq$, $qy$, $yp$ have to be triangulated (Sec. 5.1 of~\cite{edelsbrunner_quadratic_1993}).

\section{Raw data for \autoref{fig:counter-example}}
\label{appendix:counter-example-data}
The point clouds $X\subset\bbr^3$ and $Z\subset\bbr^2$ used in the counter-example depicted in \autoref{fig:counter-example} are given by the following coordinates:
\[X=
\begin{bmatrix}
	1     &  0     & 0     \\
	 0.5  &  0.866 & 0     \\
	-0.48 &  0.667 & 0     \\
	-0.73 & -0.199 & 0.433 \\
	-0.48 & -1.065 & 0     \\
	 0.5  & -0.866 & 0
\end{bmatrix}
\quad
Z=
\begin{bmatrix}
	1     & 0     \\
	0.5   & 0.866 \\
	-0.48 & 0.667 \\
	-0.98 &-0.199 \\
	-0.48 &-1.065 \\
	0.5   &-0.866
\end{bmatrix}.
\]

\revision{
\section{\revision{Detailed proof for \autoref{lemma:interpolygonal-edges}}}
\label{appendix:proof}
In this section, we restate \autoref{lemma:interpolygonal-edges} and then prove it.
\setcounter{lemma}{2}
\begin{lemma}
	In the plane, an edge that intersects an $\rng$ edge kills no \PH{1}
	class of positive persistence.
\end{lemma}
\begin{proof}
	Let $ab$ be an edge that intersects an $\rng$ edge $pq$.
	Suppose that $ab$ kills a \PH{1} class $\gamma$ of positive persistence.
	Then $ab$ is the longest edge of two triangles $abc$ and $abd$, where $acbd$ is a representative 1-cycle of $\gamma$, and such that $c$ and $d$ are on both sides of $\lens(a,b)$ (see \autoref{fig:interpolygonal-edges}).
	Thus, by definition, all distances $|ac|$, $|bc|$, $|ad|$ and $|bd|$ are $<|ab|$.
	Moreover, wlog, we can suppose that $c$ and $p$, and $d$ and $q$, are in the same half-space delimited by the line $(ab)$.

	We first show that at least $p$ or $q$ belongs to $\lens(a,b)$.
	Suppose that both $p$ and $q$ are outside $\lens(a,b)$.
	Then, both $pq$ and $ab$ are in $\rng(\{a,b,p,q\})\subset\del(\{a,b,p,q\})$.
	This implies that this Delaunay triangulation features two edges that 
intersect, which is impossible. 
	Therefore, we have two remaining cases:

	\begin{enumerate}[leftmargin=0.45cm]
		\item Both $p$ and $q\in\lens(a,b)$ (\autoref{fig:interpolygonal-edges}, left).
		In this case, the lengths $|ap|$, $|bp|$, $|aq|$, $|bq|$ are all $<|ab|$.
		In addition, since $a\notin\lens(p,q)$, either $|ap|\geq|pq|$ or $|aq|\geq|pq|$, which implies $|pq|<|ab|$ with the above inequalities.
		Finally, $c$ and $p$, and $d$ and $q$, are in the same half-lense of $\lens(a,b)$, hence $|cp|<|ab|$ and $|dq|<|ab|$.
		Therefore, the 2-chain $acp+bcp+apq+bpq+adq+bdq$, whose boundary is $acbd$, contains only triangles of diameter $<|ab|$.

		\item $p\notin\lens(a,b)$ and $q\in\lens(a,b)$ (\autoref{fig:interpolygonal-edges}, right;
		the symmetric case is handled similarly).
		In this case, $|aq|<|ab|$ and $|bq|<|ab|$.
		In addition, like in the previous case, $d$ and $q$ are in the same half-lense of $\lens(a,b)$, hence $|dq|<|ab|$.
		The last inequality to show is $|cq|<|ab|$.

		First, we show that $|ap|\geq|pq|$ and $|bp|\geq|pq|$ (i.e., we are in a configuration similar to the one depicted \autoref{fig:interpolygonal-edges}, right).
		To prove the first inequality, suppose by contradiction that $|ap|<|pq|$ (we can do a similar reasoning to prove $|bp|\geq|pq|$).
		Then, as $a\notin\lens(p,q)$, we have $|aq|\geq|pq|$, hence $|ap|<|pq|\leq|aq|<|ab|$.
		Therefore, as by hypothesis $p\notin\lens(a,b)$, we have $|bp|\geq|ab|$.
		Now, remind that $ab$ intersects $pq$ by hypothesis.
		Because of that,
		due to triangular inequalities, the sum of two opposite edges 
		of the quadrilateral $apbq$ is smaller than
		the sum of the two diagonals, i.e.,
		$|aq|+|bp|<|ab|+|pq|$.
		This gives $|bp|<|ab|+(|pq|-|aq|)\leq|ab|$ (because $|pq|\leq|aq|$), hence a contradiction.

		Now, notice that as $p\notin\lens(a,b)$, either $ac$ or $bc$ intersects $pq$.
		Indeed, if not, $p$ lies inside the interior of the triangle $abc$, which is included in $\lens(a,b)$.
		Suppose that $ac$ intersects $pq$ (the other case is handled similarly).
		Because of that,
		once again due to triangular inequalities, the sum of two 
		opposite edges of the quadrilateral $cqap$ is smaller than
		the sum of the two diagonals:
		$|cq|+|ap|<|ac|+|pq|$, hence $|cq|<|ac|+(|pq|-|ap|)<|ab|$ (because $|ac|<|ab|$ and $|ap|\geq|pq|$).

		Therefore, the 2-chain $acq+bcq+adq+bdq$, whose boundary is $acbd$, contains only triangles of diameter $<|ab|$.
	\end{enumerate}
	\begin{figure}
		\centering
		\def\svgwidth{\linewidth}\input{figures/interpolygonal-edges-review.pdf_tex}
	\end{figure}
	In both cases, there always exists a 2-chain of boundary $acbd$, that contains only triangles of diameter $<|ab|$.
	Hence, the \PH{1} class $\gamma$ is killed before the value $|ab|$, thus not by $ab$.
\end{proof}
}

\section{Additional quantitative analysis}
\label{appendix:quantitative}
\begin{figure*}
	\centering
	\scriptsize{
	\begin{tabular}{|c|r||r|r||r|r|r||r|r|r||r|}
		\hline
		\multirow{2}{*}{Dataset} & \multirow{2}{*}{Indicator} & \multicolumn{2}{c||}{Global methods} & \multicolumn{3}{c||}{Locally topology-aware methods} & \multicolumn{4}{c|}{Globally topology-aware methods} \\
		\cline{3-11}
		& &
		\methodText{PCA}{\cite{pearson1901liii}} &
		\methodText{MDS}{\cite{torgerson1952multidimensional}} &
		\methodText{Isomap}{\cite{tenenbaum_global_2000}} &
		\methodText{t-SNE}{\cite{van2008visualizing}} &
		\methodText{UMAP}{\cite{mcinnes2018umap}} &
		\methodText{TopoMap}{\cite{doraiswamy2020topomap}} &
		\methodText{\tAE}{\cite{moor2020topological}} &
		\methodText{\carriereMethod}{\cite{carriere2021optimizing}} &
		\methodText{\tAE++}{} \\
		\hline

		\multirow{7}{*}{\datasetText{\datathreeblobs}} & $\metwasserzero$ & 2.0e+01 & 1.5e+01 & 2.6e+01 & 6.7e+02 & 2.1e+02 & \textbf{4.2e-04} & 2.0e+01 & 1.2e+01 & \underline{1.0e+01}\\
		& $\metwasser$ & 4.5e-01 & 5.0e-01 & 4.7e-01 & 2.3e+01 & 7.2e-01 & 4.3e+00 & 5.8e-01 & \textbf{1.1e-01} & \underline{4.4e-01}\\
		& $\metdistor$ & \textbf{2.7e-01} & \underline{4.1e-01} & 1.1e+00 & 3.5e+01 & 8.5e+00 & 7.0e+00 & 1.3e+00 & 1.5e+00 & 6.4e-01\\
		& $\LC$ & \textbf{1.00} & \underline{0.99} & 0.98 & 0.85 & 0.89 & 0.86 & 0.95 & 0.93 & 0.97\\
		& $\TA$ & \textbf{0.89} & \underline{0.88} & 0.72 & 0.38 & 0.42 & 0.48 & 0.70 & 0.66 & 0.78\\
		& $\Trust$ & 0.93 & 0.94 & 0.92 & \textbf{0.99} & \underline{0.98} & 0.96 & 0.97 & 0.97 & 0.96\\
		& $\Cont$ & \textbf{0.99} & 0.98 & 0.98 & 0.98 & 0.98 & 0.92 & 0.98 & 0.97 & \underline{0.98}\\
		\hline
		\multirow{7}{*}{\datasetText{\datatwist}} & $\metwasserzero$ & 1.1e-01 & 1.3e-01 & 4.5e-01 & 2.9e+00 & 6.6e-01 & \textbf{7.5e-06} & 8.6e-02 & 2.8e-02 & \underline{5.2e-03}\\
		& $\metwasser$ & 1.1e+00 & 2.2e-01 & 1.9e+01 & 3.3e+00 & 1.5e+01 & 1.8e-01 & 1.6e-01 & \textbf{1.8e-05} & \underline{1.7e-03}\\
		& $\metdistor$ & \underline{2.2e-01} & \textbf{2.1e-01} & 2.4e+00 & 3.9e+00 & 2.8e+00 & 8.2e-01 & 3.8e-01 & 7.3e-01 & 7.8e-01\\
		& $\LC$ & \textbf{0.99} & \underline{0.99} & 0.86 & 0.99 & 0.58 & 0.86 & 0.98 & 0.99 & 0.99\\
		& $\TA$ & \textbf{0.91} & \underline{0.89} & 0.63 & 0.87 & 0.42 & 0.64 & 0.85 & 0.87 & 0.88\\
		& $\Trust$ & 0.98 & 0.99 & \textbf{1.00} & 0.99 & 1.00 & 0.99 & 0.99 & 0.96 & \underline{1.00}\\
		& $\Cont$ & 0.99 & 0.99 & \textbf{1.00} & 0.99 & 1.00 & 0.98 & 0.99 & 0.98 & \underline{1.00}\\
		\hline
		\multirow{7}{*}{\datasetText{\datakfour}} & $\metwasserzero$ & 1.2e-01 & 6.9e-02 & 1.1e-01 & 3.0e+01 & 2.0e+01 & \textbf{2.3e-07} & 1.9e-01 & 7.8e-02 & \underline{5.8e-02}\\
		& $\metwasser$ & 2.7e-01 & 2.8e-01 & 3.8e-01 & 4.7e+01 & 3.9e-01 & 1.4e-01 & 6.5e-02 & \textbf{7.9e-04} & \underline{2.4e-02}\\
		& $\metdistor$ & \underline{2.7e-01} & \textbf{1.8e-01} & 5.1e-01 & 1.6e+01 & 9.1e+00 & 5.7e-01 & 1.0e+00 & 6.8e-01 & 3.4e-01\\
		& $\LC$ & \underline{0.83} & \textbf{0.89} & 0.65 & 0.66 & 0.53 & 0.57 & 0.75 & 0.68 & 0.77\\
		& $\TA$ & \underline{0.61} & \textbf{0.63} & 0.39 & 0.48 & 0.34 & 0.38 & 0.51 & 0.47 & 0.56\\
		& $\Trust$ & 0.97 & 0.98 & 0.95 & \textbf{1.00} & 1.00 & 0.99 & 0.98 & 0.98 & \underline{1.00}\\
		& $\Cont$ & \underline{1.00} & 1.00 & 1.00 & 0.99 & 0.99 & 0.98 & 0.99 & 0.99 & \textbf{1.00}\\
		\hline
		\multirow{7}{*}{\datasetText{\datakfive}} & $\metwasserzero$ & 2.7e-01 & 1.5e-01 & 1.8e-01 & 9.3e+01 & 1.8e+01 & \textbf{7.3e-07} & 2.2e-01 & 1.3e-01 & \underline{9.0e-02}\\
		& $\metwasser$ & 2.8e-01 & 3.5e-01 & 1.3e-01 & 4.7e+01 & 9.7e-01 & 4.8e-01 & 1.6e-01 & \textbf{2.2e-03} & \underline{8.0e-02}\\
		& $\metdistor$ & \underline{3.1e-01} & \textbf{2.4e-01} & 4.7e-01 & 2.1e+01 & 9.8e+00 & 1.2e+00 & 9.2e-01 & 8.6e-01 & 3.4e-01\\
		& $\LC$ & \underline{0.81} & \textbf{0.86} & 0.67 & 0.76 & 0.58 & 0.41 & 0.73 & 0.72 & 0.81\\
		& $\TA$ & \textbf{0.63} & \underline{0.61} & 0.43 & 0.55 & 0.41 & 0.33 & 0.50 & 0.50 & 0.57\\
		& $\Trust$ & 0.93 & 0.97 & 0.96 & \textbf{1.00} & \underline{1.00} & 0.99 & 0.99 & 0.97 & 1.00\\
		& $\Cont$ & 0.99 & 0.99 & \textbf{1.00} & 0.99 & 0.98 & 0.97 & 0.99 & 0.99 & \underline{1.00}\\
		\hline
		\multirow{7}{*}{\datasetText{\datacoil}} & $\metwasserzero$ & 1.2e+02 & 8.1e+01 & 3.0e+00 & 1.5e+02 & 1.5e+02 & \textbf{2.3e-06} & 1.9e+00 & 4.5e+00 & \underline{1.2e+00}\\
		& $\metwasser$ & 1.2e+01 & 1.0e+01 & 6.0e+02 & 1.3e+01 & 1.2e+01 & 2.7e+01 & 7.5e+00 & \underline{9.1e-02} & \textbf{1.6e-02}\\
		& $\metdistor$ & \underline{2.9e+00} & \textbf{2.0e+00} & 2.1e+01 & 4.9e+00 & 6.2e+00 & 7.2e+00 & 8.2e+00 & 6.7e+00 & 1.9e+01\\
		& $\LC$ & \textbf{0.94} & \underline{0.93} & 0.83 & 0.88 & 0.85 & 0.74 & 0.81 & 0.62 & 0.88\\
		& $\TA$ & \textbf{0.81} & 0.74 & 0.47 & \underline{0.77} & 0.65 & 0.44 & 0.56 & 0.38 & 0.73\\
		& $\Trust$ & 0.97 & 0.97 & \underline{1.00} & \textbf{1.00} & 0.99 & 0.92 & 0.96 & 0.91 & 0.99\\
		& $\Cont$ & 0.96 & 0.97 & 0.98 & \textbf{1.00} & \underline{0.99} & 0.96 & 0.97 & 0.94 & 0.99\\
		\hline
		\multirow{7}{*}{\datasetText{\datamocap}} & $\metwasserzero$ & 4.3e+03 & 2.2e+03 & \underline{3.9e+02} & 6.8e+03 & 6.8e+03 & \textbf{2.4e-03} & 1.6e+03 & 7.9e+02 & 1.9e+03\\
		& $\metwasser$ & 2.1e+03 & 1.9e+03 & 1.0e+04 & 1.6e+03 & 1.5e+03 & 2.1e+03 & 1.2e+03 & \textbf{9.4e+00} & \underline{1.2e+01}\\
		& $\metdistor$ & \underline{2.4e+01} & \textbf{1.7e+01} & 1.3e+02 & 9.9e+01 & 1.0e+02 & 8.5e+01 & 2.6e+01 & 5.6e+01 & 2.9e+01\\
		& $\LC$ & \underline{0.94} & \textbf{0.95} & 0.67 & 0.90 & 0.47 & 0.43 & 0.89 & 0.61 & 0.83\\
		& $\TA$ & \underline{0.72} & \textbf{0.73} & 0.53 & 0.64 & 0.34 & 0.35 & 0.68 & 0.42 & 0.59\\
		& $\Trust$ & 0.98 & 0.98 & \underline{1.00} & 1.00 & \textbf{1.00} & 0.97 & 0.98 & 0.93 & 1.00\\
		& $\Cont$ & 0.98 & 0.99 & \textbf{1.00} & 1.00 & 0.99 & 0.98 & 0.99 & 0.98 & \underline{1.00}\\
		\hline
		\multirow{7}{*}{\datasetText{\datasinglecell}} & $\metwasserzero$ & 4.6e+05 & 4.6e+05 & 3.5e+05 & 4.6e+05 & 4.6e+05 & \textbf{1.2e-01} & 4.6e+05 & \underline{1.6e+05} & 3.8e+05\\
		& $\metwasser$ & 3.0e+03 & 3.1e+03 & 1.8e+05 & 1.8e+03 & 1.8e+03 & 2.8e+04 & 1.8e+03 & \textbf{6.4e+02} & \underline{8.6e+02}\\
		& $\metdistor$ & \underline{3.1e+01} & \textbf{2.3e+01} & 4.8e+02 & 9.6e+01 & 1.0e+02 & 1.2e+03 & 9.8e+01 & 3.2e+02 & 1.5e+02\\
		& $\LC$ & \textbf{0.99} & \underline{0.98} & 0.98 & 0.88 & 0.81 & 0.62 & 0.72 & 0.74 & 0.85\\
		& $\TA$ & \textbf{0.94} & \underline{0.86} & 0.84 & 0.68 & 0.54 & 0.40 & 0.55 & 0.52 & 0.61\\
		& $\Trust$ & \underline{0.99} & 0.99 & 0.99 & \textbf{0.99} & 0.99 & 0.96 & 0.97 & 0.96 & 0.97\\
		& $\Cont$ & \underline{0.99} & 0.99 & 0.99 & \textbf{0.99} & 0.99 & 0.95 & 0.94 & 0.96 & 0.96\\
		\hline

	\end{tabular}
	}
	\caption{\PH{}-based metrics and other common indicators used in DR (see \autoref{appendix:quantitative} for a description). The best value (before rounding) for an indicator is written in bold, the second best (before rounding) is underlined.}
	\label{table:quantitative}
\end{figure*}

\revision{\autoref{table:quantitative} shows the values of \PH{}-related metrics ($\metwasserzero$ and $\metwasser$) and some common DR indicators for the examples presented in the paper. These additional indicators measure the preservation of global distances (pairwise distances-based and triplets-based indicators) and the preservation of local distances (rank-based indicators).}

\subsection{Description of the indicators}
\label{appendix:quantitativeDescription}

\paragraph{Pairwise distances-based indicators}
In addition to the metric distortion $\metdistor$, one can measure the linear correlation $\LC\in[-1,1]$ between the pairwise distances in $\inputPointCloud$ and in $\latentPointCloud$.
Having $\LC=1$ means that these pairwise distances are perfectly correlated in high and low dimension.

\paragraph{Triplets-based indicator}
The \textit{triplet accuracy} $\TA\in[0,1]$ is the proportion of triangles whose three edges have the same relative order (in terms of length) both in high and low dimension~\cite{wang_understanding_2021}.
It measures to some extent the preservation of the global structure of
$\inputPointCloud$, which is preserved when $\TA=1$.

\paragraph{Rank-based indicators}
The rank of a point $\inputPointCloud_i$ relative to another point $\inputPointCloud_j$ is the integer $\rho_{ij}\in\bbn$ such that $\inputPointCloud_j$ is the $\rho_{ij}$-th nearest neighbor of $\inputPointCloud_i$.
There exists indicators based on these ranks~\cite{lee_quality_2009, venna2006local} that measure the preservation of the nearest neighbors.
In particular, the \emph{trustworthiness} ($\Trust\in[0,1]$) is penalized when a
high rank $\rho_{ij}$ in the input becomes low in the representation,
i.e., when faraway points in high-dimension become neighbors in
low-dimension.
On the contrary, the \emph{continuity} ($\Cont\in[0,1]$) is penalized when a low
rank $\rho_{ij}$ in the input becomes high in the representation, i.e.,
 when neighbors in high-dimension are projected to faraway points in
low-dimension.
These indicators are computed for a number $K$ of nearest neighbors
(in our experiments, $K=10$).

\subsection{Observations}
\label{appendix:quantitativeObservations}

TopoMap has the best results for $\metwasserzero$, which is expected since it preserves by design exactly \PH{0}.
Although in theory this quantity should be exactly 0, the approximation
performed by the auction algorithm when estimating
the Wasserstein distance (to compute $\metwasserzero$) makes it slightly positive.
Our approach (\tAE++) has competitive results for this metric as well, which was expected since, as \tAE, it constrains the preservation of \PH{0} in the sense of \autoref{lemma:TopoAE0_bound}.

Global indicators, namely the metric distortion $\metdistor$, the linear correlation $\LC$ between pairwise distances and the triplet accuracy $\TA$, are  best preserved by global methods, i.e., PCA and MDS.
Our approach (\tAE++) presents competitive results for these indicators when
comparing to other locally topology-aware methods, while it clearly
outperforms globally topology-aware methods.

Finally, \tAE++ also presents good results for neighborhood quality
indicators ($\Trust$ and $\Cont$). Indeed, in our
datasets, the correct embedding of the input cycle(s) favors the preservation of
the neighborhood.
More precisely, a self-intersection in the projection of an input cycle penalizes the trustworthiness (since faraway vertices would be projected as neighbors near this intersection), while a broken input cycle penalizes the continuity (since two neighbors in the input generator would be embedded faraway in low-dimension).

\revision{
	\section{Hyperparameters}
	\label{appendix:ablation}
	We provide a study of the hyperparameters describing the autoencoder's architecture (size of the hidden layers, \autoref{fig:study_architecture}) and its optimization (learning rate and weight applied to the topological regularization term, \autoref{fig:study_lr_w}), on \datacoil{} and \datasinglecell{} datasets. Except for very small networks (8-8 or 16-16), the size of the hidden layers has little impact on the performance of the autoencoder.
	The 128-32 choice gives an acceptable compromise for our datasets. We found that a learning rate of $10^{-2}$ gives good results for our datasets.
	Finally, the weight applied on $\lcascae{1}$ has to be chosen according to the dataset as it scales with its size. For instance, we used $10^{-2}$ for \datacoil{} and $10^{-4}$ for \datasinglecell{}.
	\begin{figure*}
		\begin{center}
			\begin{tabular}{ccc}
				\hspace{-.4cm}
				\raisebox{1.2cm}{\rotatebox{90}{\small\datacoil}}
				\includegraphics[width=.32\linewidth]{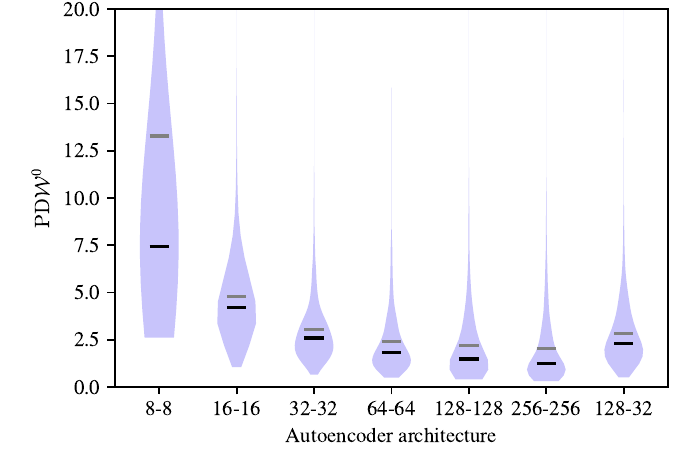}\hspace{-.3cm} &
				\includegraphics[width=.32\linewidth]{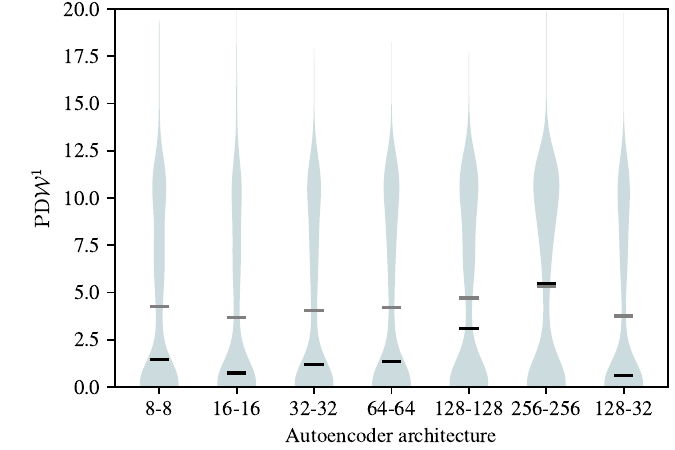}\hspace{-.3cm} &
				\includegraphics[width=.32\linewidth]{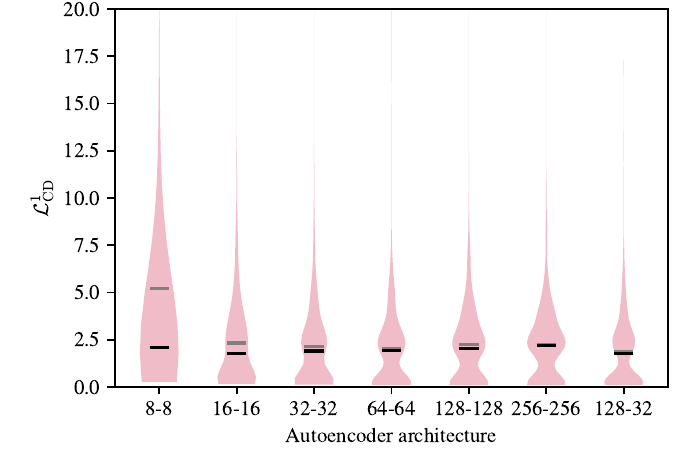}\hspace{-.3cm} \\
				\hspace{-.4cm}
				\raisebox{1.2cm}{\rotatebox{90}{\small\datasinglecell}}
				\includegraphics[width=.32\linewidth]{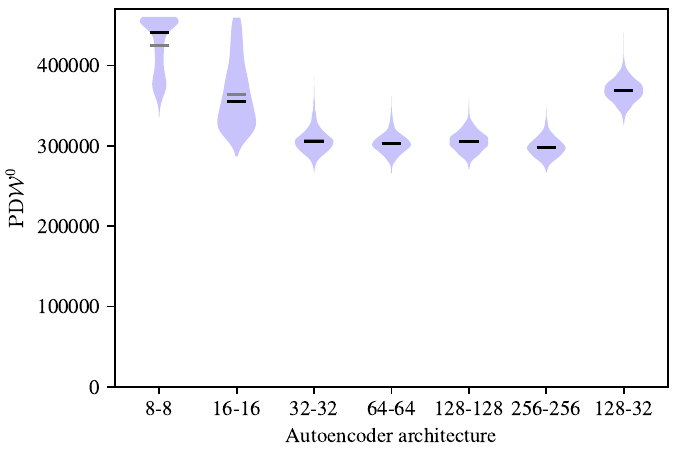}\hspace{-.3cm} &
				\includegraphics[width=.32\linewidth]{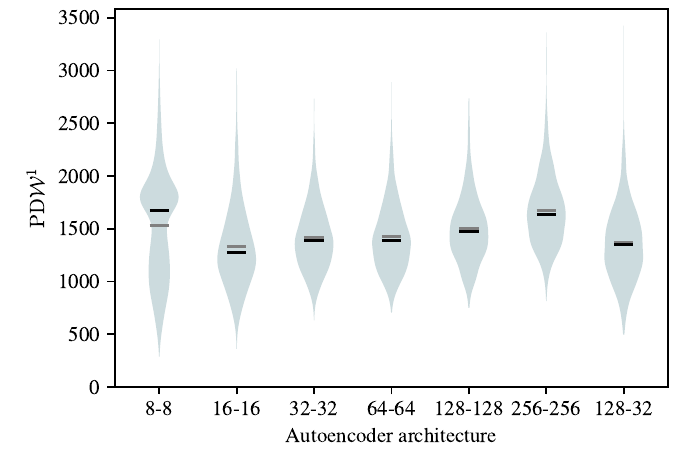}\hspace{-.3cm} &
				\includegraphics[width=.32\linewidth]{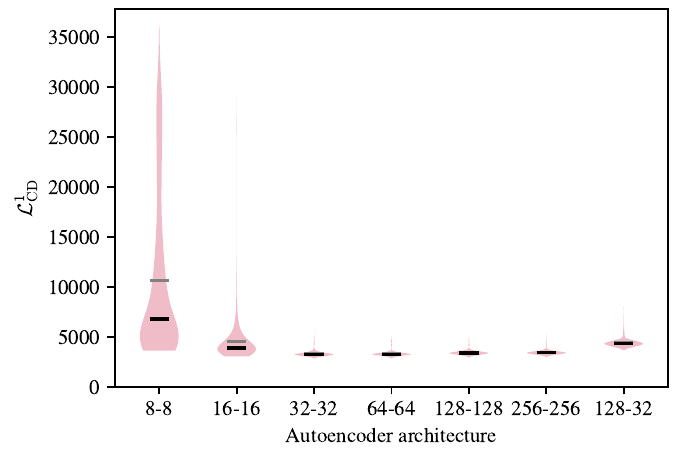}\hspace{-.3cm}
			\end{tabular}
		\end{center}
		\caption{\revision{Violin plots of the distribution of the final value 
		of $\lcascae{1}$ according to the size of the autoencoder's layers, computed for
		1000 runs of our method on \datacoil{} for the top line and \datasinglecell{}
		for the bottom line. We only consider here encoders with two fully connected
		hidden layers, whose size is indicated by the two number\julien{s} (the decoder
		is symmetrical). The mean value is shown in gray and the median in black.
		The rightmost violins of each chart in the top line correspond to the histogram
		in \autoref{fig:failures}.}}
		\label{fig:study_architecture}
	\end{figure*}
	\begin{figure}
		\begin{center}
			\includegraphics[width=.49\linewidth]{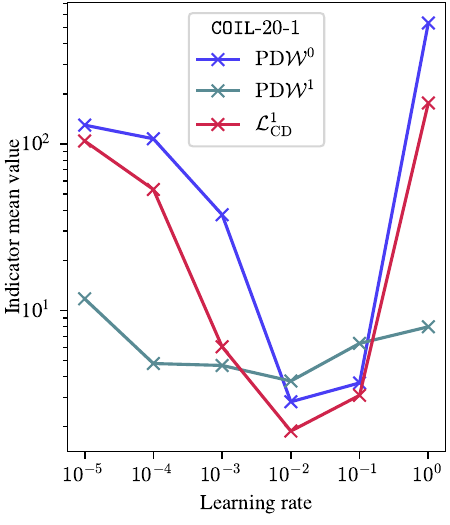}\hspace{-.1cm}
			\includegraphics[width=.49\linewidth]{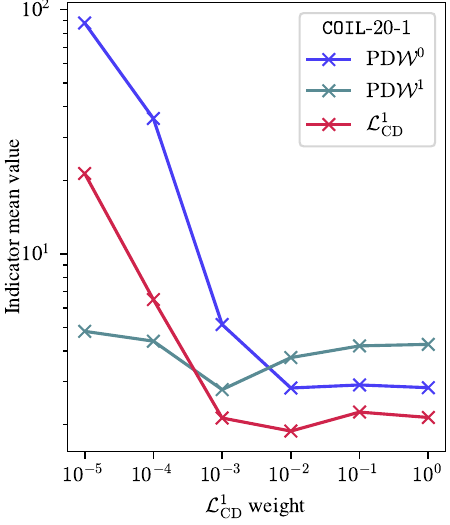}
			\includegraphics[width=.49\linewidth]{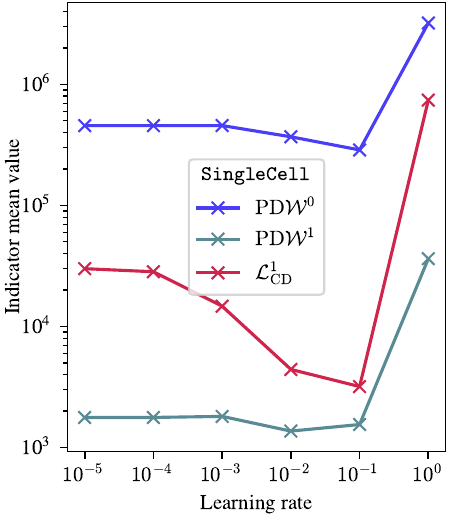}\hspace{-.1cm}
			\includegraphics[width=.49\linewidth]{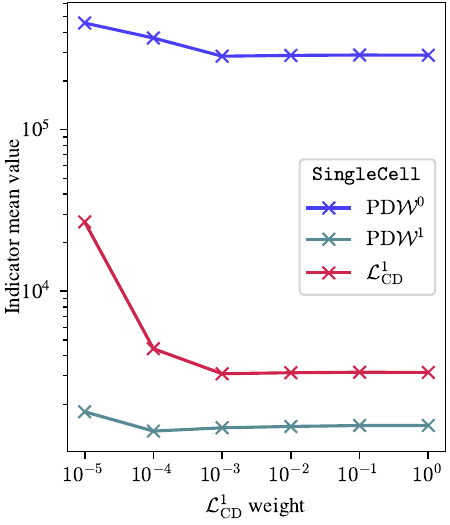}
		\end{center}
		\caption{\revision{Mean final values of $\metwasserzero$ (dark blue), 
$\metwasser$ 
		\julien{(teal)}
		and $\lcascae{1}$ (red) for varying learning rate\julien{s} (left) 
and varying weight\julien{s} applied to the $\lcascae{1}$ regularisation term, 
computed for 1000 runs of our method on \datacoil{} (top line) and 
\datasinglecell{} (bottom line).}}
		\label{fig:study_lr_w}
	\end{figure}
}

\revision{
	\section{Failures cases}
	\label{appendix:failures}
	We show in \autoref{fig:failures} a
	histogram of the final values of $\metwasserzero$, $\metwasser$ and 
$\lcascae{1}$ computed for 1000 runs of \tAE++ on \datacoil.
	For this dataset, the distributions of $\metwasser$ and $\lcascae{1}$ are typically bimodal (this is also visible in \autoref{fig:study_architecture}, top line).
	This reflects the existence of \emph{successful} runs (no input generator self-crossing, see left example) and \emph{unsuccessful} runs (trapped in a bad local minima featuring an input self-crossing, see right example).
	Even though we have curently mitigated this issue by keeping the best of several runs, this could be handled in future work with more elaborated initializations and by penalizing input generators' self-crossings during optimization.
	Finally, note that this phenomenon does not appear with all datasets (e.g., with \datasinglecell, see \autoref{fig:study_architecture}, bottom line, where distributions are not bimodal).
	\begin{figure}
		\centering
		\def\svgwidth{\linewidth}\tiny{
\begingroup%
  \makeatletter%
  \providecommand\color[2][]{%
    \errmessage{(Inkscape) Color is used for the text in Inkscape, but the package 'color.sty' is not loaded}%
    \renewcommand\color[2][]{}%
  }%
  \providecommand\transparent[1]{%
    \errmessage{(Inkscape) Transparency is used (non-zero) for the text in Inkscape, but the package 'transparent.sty' is not loaded}%
    \renewcommand\transparent[1]{}%
  }%
  \providecommand\rotatebox[2]{#2}%
  \newcommand*\fsize{\dimexpr\f@size pt\relax}%
  \newcommand*\lineheight[1]{\fontsize{\fsize}{#1\fsize}\selectfont}%
  \ifx\svgwidth\undefined%
    \setlength{\unitlength}{360.08877563bp}%
    \ifx\svgscale\undefined%
      \relax%
    \else%
      \setlength{\unitlength}{\unitlength * \real{\svgscale}}%
    \fi%
  \else%
    \setlength{\unitlength}{\svgwidth}%
  \fi%
  \global\let\svgwidth\undefined%
  \global\let\svgscale\undefined%
  \makeatother%
  \begin{picture}(1,1.00197239)%
    \lineheight{1}%
    \setlength\tabcolsep{0pt}%
    \put(0,0){\includegraphics[width=\unitlength,page=1]{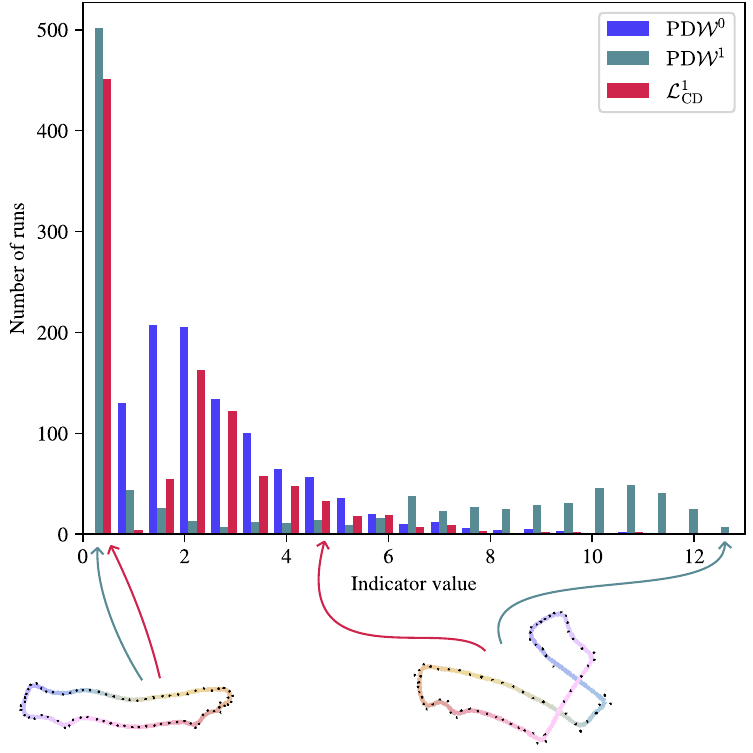}}%
    \put(0.40815313,0.03776984){\color[rgb]{0,0,0}\makebox(0,0)[t]{\lineheight{1.25}\smash{\begin{tabular}[t]{c}$\lcascae{1}=0.18$\end{tabular}}}}%
    \put(0.42061487,0.07455267){\color[rgb]{0,0,0}\makebox(0,0)[t]{\lineheight{1.25}\smash{\begin{tabular}[t]{c}$\metwasser=0.012$\end{tabular}}}}%
    \put(0.90038247,0.03522504){\color[rgb]{0,0,0}\makebox(0,0)[t]{\lineheight{1.25}\smash{\begin{tabular}[t]{c}$\lcascae{1}=4.9$\end{tabular}}}}%
    \put(0.91084343,0.07364169){\color[rgb]{0,0,0}\makebox(0,0)[t]{\lineheight{1.25}\smash{\begin{tabular}[t]{c}$\metwasser=12.4$\end{tabular}}}}%
  \end{picture}%
\endgroup%
}
		\caption{\revision{Histogram of the distributions of $\metwasserzero$, $\metwasser$ and $\lcascae{1}$ values after the optimization for 1000 runs of \tAE++ on \datacoil. Two examples of results are shown at the bottom: one successful on the left, the other unsuccessful on the right.}}
		\label{fig:failures}
	\end{figure}
}

\vfill

\end{document}